\documentclass[11pt]{article} 
\usepackage[margin=1in,footskip=0.25in]{geometry}

\usepackage{hyperref}
\newcommand{\HRule}{\rule{\linewidth}{0.5mm}}
\renewcommand*{\thefootnote}{\alph{footnote}}
\usepackage{LHCHWG_Public_Note_ttbbttW-def}
\usepackage{rotating}
\usepackage{booktabs}
\usepackage{multirow}

\makeatletter
\newcommand\footnoteref[1]{\protected@xdef\@thefnmark{\ref{#1}}\@footnotemark}
\makeatother

\begin{document}
\hypersetup{pageanchor=false}
\thispagestyle{empty}

	\begin{flushright}
	  {\large
            \textbf{\href{https://cds.cern.ch/record/xxxxxx}{LHCHWG-2022-003}}} \\[0.5cm]	
		{\large 	\textrm{December 15, 2022}} \\[2.0cm]
	\end{flushright}

	\begin{center}

	\textsc{\Large 	\href{https://twiki.cern.ch/twiki/bin/view/LHCPhysics/LHCHWG}{LHC Higgs Working Group}\footnote{\href{https://twiki.cern.ch/twiki/bin/view/LHCPhysics/LHCHWG}{\sl https://twiki.cern.ch/twiki/bin/view/LHCPhysics/LHCHWG}}} \\[0.5cm]
	\textsc{\Large 	Public Note} \\[1.5cm]
	
	\HRule \\[0.9cm]
	\textbf{\Large Study of \ttbb and \ttW background modelling for \ttH analyses} \\[0.7cm]
	\HRule \\[1.2cm]

	\textrm{\large
          Lars~Ferencz$\,^{1,}$\footnote{\href{mailto:lars.ferencz@desy.de}{lars.ferencz@desy.de}},
          Kirill~Grevtsov$\,^{1,}$\footnote{\href{mailto:kirill.grevtsov@desy.de}{kirill.grevtsov@desy.de}},
          Judith~Katzy$\,^{1,}$\footnote{\href{mailto:judith.katzy@desy.de}{judith.katzy@desy.de}},
          Andrea~Knue$\,^{2,}$\footnote{\href{mailto:andrea.knue@physik.uni-freiburg.de}{andrea.knue@physik.uni-freiburg.de}},
          Jan~van~der~Linden$\,^{3,}$\footnote{\href{mailto:jan.linden@kit.edu}{jan.linden@kit.edu}},
          Josh McFayden$\,^{4,}$\footnote{\href{mailto:joshua.angus.mcfayden@cern.ch}{joshua.angus.mcfayden@cern.ch}},
          Gianna Moenig$\,^{4,}$\footnote{\href{mailto:gianna.moenig@cern.ch}{gianna.moenig@cern.ch}},
          Emanuel~Pfeffer$\,^{3,}$\footnote{\href{mailto:emanuel.pfeffer@kit.edu}{emanuel.pfeffer@kit.edu}},
          Andrej~Saibel$\,^{5,}$\footnote{\href{mailto:andrej.saibel@cern.ch}{andrej.saibel@cern.ch}},
          Matthias~Schr\"oder$\,^{6,}$\footnote{\href{mailto:matthias.schroeder@uni-hamburg.de}{matthias.schroeder@uni-hamburg.de}},
          Joshuha~Thomas-Wilsker$\,^{7,}$\footnote{\href{mailto:joshuha.thomas-wilsker@cern.ch}{joshuha.thomas-wilsker@cern.ch}}
         } \\[0.3cm]	
	\textit{$^{1}$ DESY}\\
	\textit{$^{2}$ Universit\"at~Freiburg}\\
	\textit{$^{3}$ KIT}\\
	\textit{$^{4}$ University of Sussex} \\
        \textit{$^{5}$ Instituto de Física Corpuscular, Consejo Superior de Investigaciones Científicas}\\
	\textit{$^{6}$ Universit\"at~Hamburg}\\
        \textit{$^{7}$ Institute of High Energy Physics, Chinese Academy of Sciences}\\[1cm]
	\end{center} 
\begin{center}
{work done on behalf of the LHCHWG\\}
{\tiny{Reproduction of this article or parts of it is allowed as specified in the CC-BY-4.0 license.}}
\end{center}
        \newpage
        \thispagestyle{empty}



\mbox{}\vspace*{3em}
\begin{center}
	\textbf{Abstract}
\end{center}
This note presents Monte Carlo generator comparisons of the \ttbb and \ttW processes at particle level.
The aim is to compare the modelling of important backgrounds to \ttH
measurements in multi-lepton final states and in the \ttHbb decay channel and the
treatment of the associated theory uncertainties for a  combination of
the full Run-2 \ttH results  from  ATLAS and CMS. As a first step, modelling and theory uncertainties as used in ATLAS an CMS are compared in the relevant analysis regions. Significant differences in the treatment of systematic uncertainties between the experiments have been observed in \ttbb and $\ensuremath{t\bar{t}W}.$ As a first step, ATLAS and CMS agreed on a common reference value of the inclusive \ttW  cross section  to allow direct comparisons between experiments.
\clearpage

\thispagestyle{empty}
\setcounter{tocdepth}{2}
\tableofcontents
\clearpage

\hypersetup{pageanchor=true}
\renewcommand*{\thefootnote}{\arabic{footnote}}
\setcounter{footnote}{0}
\setcounter{page}{1}

\section{Introduction}
\label{sec:intro}

\begin{figure}[!htb]
  \centering
  \begin{tabular}{ccc}
    \includegraphics[width=0.35\textwidth]{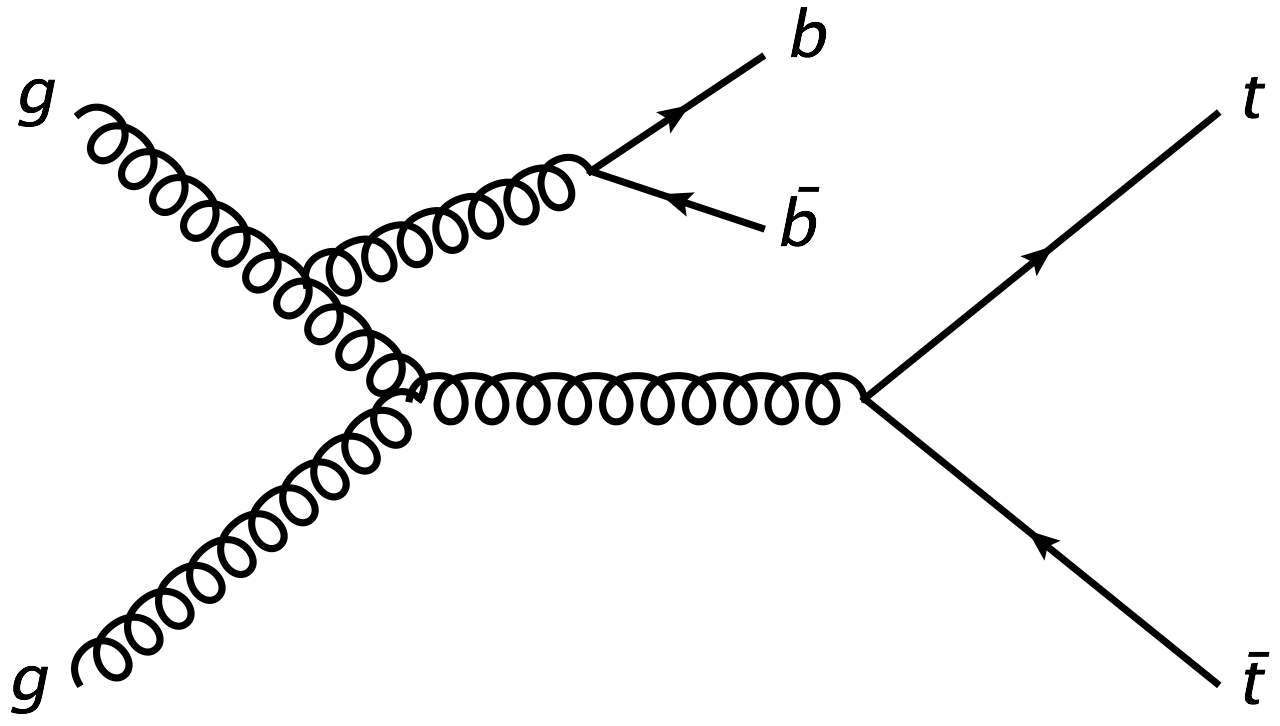}
    &&
    \includegraphics[width=0.35\textwidth]{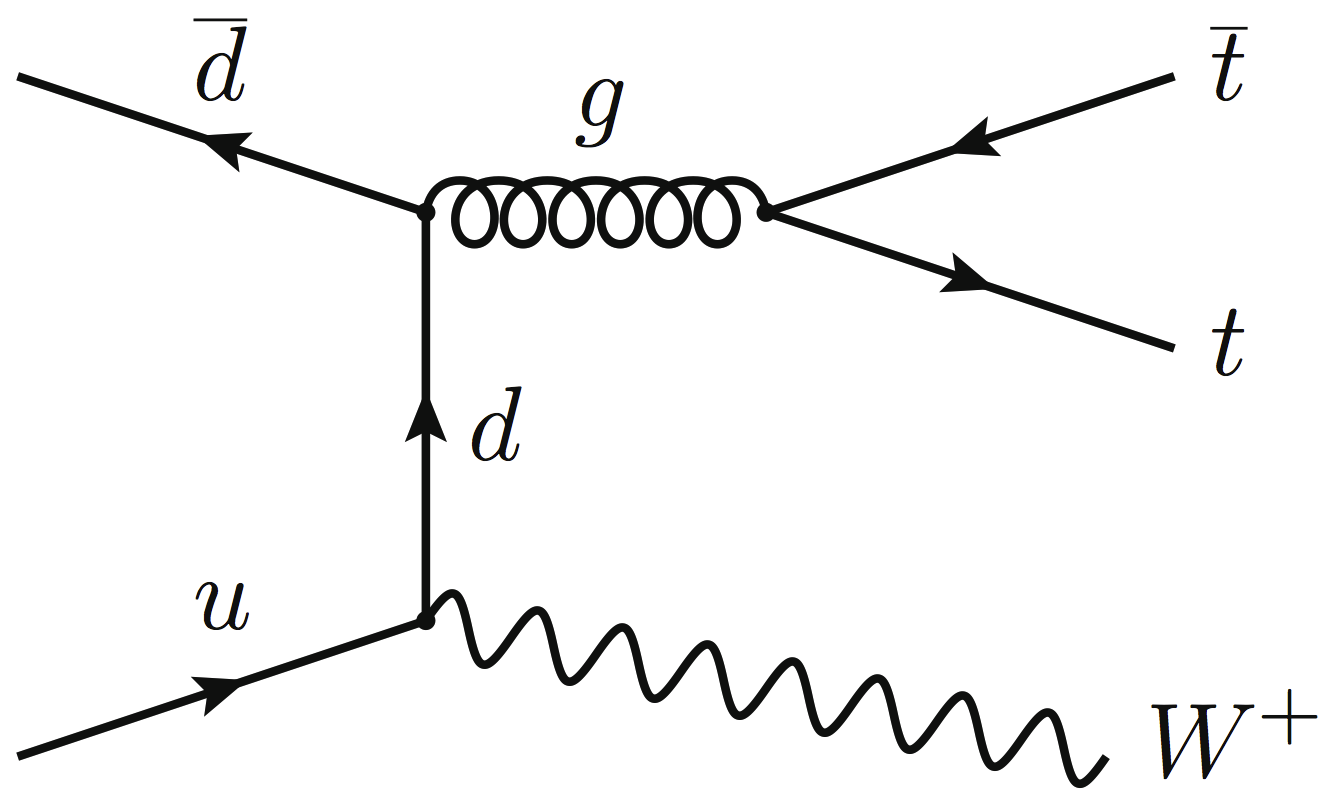} \\
  \end{tabular}
  \caption{Examples of tree-level Feynman diagrams for \ttbb (left) and \ttW (right). \label{intro:sig}}
\end{figure}

The search for Higgs boson production in association with a top quark pair (\ttH) has been performed in  the $H\rightarrow b\bar{b}$~\cite{HIGG-2020-23,HIGG-2017-03,CMS-PAS-HIG-18-030,CMS-HIG-17-026} decay channel and  in  multi-lepton final states~\cite{ATLAS-CONF-2019-045,CMS-HIG-19-008} which are primarily sensitive to the decays of H$\rightarrow WW^*$, H$\rightarrow \tau \tau$ and H$\rightarrow ZZ^*$.  These searches are limited by the modelling uncertainties of the main backgrounds, \ttbb and $t\bar{t}W,$ respectively. Examples of tree-level diagrams of the background processes are shown in Fig.~\ref{intro:sig}.

A comparison of Monte Carlo (MC) generators used by ATLAS and CMS is thus performed to compare the background modelling and the estimates of modelling uncertainties  in view of  future combinations of the experimental results. The goals is to provide input to a discussion between the experiments and between experiments and  theorists to define modelling uncertainties. Furthermore, the experiments aim to develop a common strategy  for combination of  the \ttHbb and $\ttH$(multi-lepton) analyses of the full Run-2 data set. Comparisons of observables relevant for the analyses are made at stable particle level, in a phase space similar to the reference measurements using the \rivet analysis toolkit~\cite{rivet}.

The note is structured as follows: comparisons of \ttbb distributions will be presented in Section~\ref{sec:ttbb} and comparisons of \ttW distributions in Section~\ref{sec:ttW}.
\clearpage
\section{Comparisons of Monte Carlo predictions for the \ttbb process}
\label{sec:ttbb}


In the following section \ttbb background predictions   and  variations considered to estimate their uncertainties  used by ATLAS and CMS in published and future analyses of \ttHbb are compared.
The first Run-2 \ttHbb analyses of both experiments~\cite{HIGG-2017-03,CMS-HIG-17-026}   based  on partial data sets predicted the $t\bar{t}+jets$ background  with  a  \ttbar  matrix element (ME) calculated at next-to-leading-order (NLO) accuracy in QCD in the five-flavour scheme (5FS) and matched to the \pythiaa parton shower (PS) \cite{Sjostrand:2014zea}  in the  \powheg framework ~\cite{Nason:2004rx,Frixione:2007vw,Frixione2007nw,Alioli:2010xd,Campbell:2014kua}. In this set-up, $b$-quarks not originating in the top quark decay chain  are produced by \pythiaa. 

The first predictions  using a \ttbb ME at NLO have been performed with stable top quarks in 5FS some time ago \cite{Bredenstein:2010rs, ttbbtheo1,Bevilacqua_2009}. They have been matched subsequently to parton shower programs \cite{ ttbbtheo2}. Very recently complete calculations for the \ttbb process in di-lepton top quark decay channel have been carried out in 5FS without matching to PS by two independent groups \cite{Bevilacqua_2021,Denner_2021,  ref21}. Such computations are  based on $e^+ \nu_e \mu^- \overline{\nu}_{\mu} b \overline{b} b \overline{b}$ matrix elements and include all resonant and non-resonant Feynman diagrams, interferences and off-shell effects of the top quark and the W gauge boson.  

The first  \ttHbb analysis based on the full Run-2 data set from  ATLAS~\cite{HIGG-2020-23} ("first full Run-2 analysis") used as nominal generator a calculation  where the  \ttbb ME is calculated at NLO with massive $b$-quarks\footnote{``quarks'' refers to both quarks and anti-quarks} in the four-flavour scheme (4FS)  \cite{Jezo:2018yaf} 
 and matched to \pythiaa in the \powhegboxres  framework  \cite{Jezo:2018yaf}, referred to as \ttbbpowheg  in the following.
 
  For  future analyses both experiments consider to use  the calculations of \ttbbpowheg   
  matched to \pythiaa as nominal generator  however with different settings of the  renormalisation and factorisation scale  compared to the original paper \cite{Jezo:2018yaf} and slightly different settings of the internal parameters based on more recent studies \cite{ttbbpubnote} as will be discussed below.

The estimation of systematic uncertainties differs significantly between the two experiments for the published analyses. ATLAS considered uncertainties   due the particular choice of matching algorithm and of the parton shower generator. For the  analyses  based on partial  and first full Run-2 data set, these differences were derived from  5FS \ttbar sample predicted by  \amcnp   \cite{Alwall:2014hca, Frederix:2012ps} matched to \pythiaa for the first and a sample where \powheg is  matched to \herwiga \cite{Bellm_2016} for the latter. Since the nominal generator in the first full Run-2 analysis was based on a \ttbbpowheg calculation,  the relative uncertainties derived from the 5FS \ttbar samples were used. Uncertainties due to higher order effects were estimated by varying the renormalisation and factorisation scales in the ME, \muR and \muF, simultaneously up and down by a factor of two. Correlations between the scale settings in the ME and  \alphas  in the PS ISR  were  considered by  simultaneous variation  with \muR and \muF to cover the effects of PS variations in the presence of matching~\cite{Cooper:2011gk}.

In the first Run-2 analysis, CMS considered the uncertainty due to the choice of generator settings by varying the \hdamp parameter in \powheg  which controls the transverse momentum (\pt) of the first additional emission beyond the leading-order Feynman diagram in the PS and therefore regulates the high-\pt emission against which the \ttbar system recoils. Comparisons with \sherpa were done internally but not added to the list of systematic uncertainties.  The renormalisation and factorisation scales  \muR and \muF as well as \alphas in both the PS ISR and FSR were varied independently, i.e. one parameter was changed at a time while keeping the other parameters at their nominal values.

For  future analyses, both experiments consider predictions with varied \muR and \muF scales and varied PS \alphas as well as different settings of \ttbbpowheg  internal parameters, however ATLAS studies  additional  uncertainties due to parton shower and matching.
To estimate the dependence on \ttbbpowheg  internal parameters, ATLAS varies the parameter \bornzero which regulates the splitting into the finite and the  singular part of the real emission in the \powheg framework. Variations of the parameter \hdamp  were studied in Ref.~\cite{ttbbpubnote} but no significant differences were found and  therefore this variation is  not further considered for uncertainty estimates.  Uncertainties due to the particular setting of PS  are estimated with set-ups of \ttbbpowheg  matched  to \herwiga and \pythiaa with a dipole recoil. The dependence on the particular choice of generator and the NLO matching algorithm is studied by comparing to NLO 4FS predictions of  \ttbb generated with   \sherpa 2.2.10~\cite{Sherpa,Cascioli:2013era, Cascioli:2011va}.
Details of the studies are given in Ref.~\cite{ttbbpubnote}.

In case of CMS, the dependence on \ttbbpowheg  internal parameters is estimated by varying the matching parameter \hdamp. 
  

Both experiments consider PDF uncertainties in the published and future analyses, however they are  neglected in the studies presented here due to the smallness of the effect.  Finally, in order to get comparable results,   the scale uncertainties are treated the same way for both experiments in all studies presented here,  i.e.\  \muR and \muF, PS ISR and PS FSR are changed individually by a factor 0.5 (2)  while keeping the other parameters at their nominal values.
All comparisons are performed using stable final-state particles in a fiducial phase space similar to the experimental measurements implemented in a dedicated routine in  the \rivet analysis toolkit~\cite{rivet, rivetroutine}.

The chapter is organised as follows.
Section\,\ref{sec:ttbb:samples} describes the samples used for the comparison and the technical set-up of their generation.
Section\,\ref{sec:ttbb:rivet} describes the observables and the fiducial phase space used for the comparison and finally, Sec.\,\ref{sec:ttbb:results} displays the resulting comparisons.

\subsection{MC generator set-ups}\label{sec:ttbb:samples}
The set-ups used to generate \ttbb  predictions with \ttbbpowheg,  \powheg,  \amcnp and \sherpa  are described in the following.
The generator configurations and version numbers are summarised in Table~\ref{tab:ttbbsamples} and their scale settings are given in Table~\ref{tab:ttbbscalechoices}.
The systematic uncertainty estimates due to scale and \alphas variations are summarised in Table~\ref{tab:ttbbvariations}.

The $b$-quark mass  is set to \SI{4.75}{\GeV} for CMS samples and for \sherpa, and to  \SI{4.95}{\GeV} for all other ATLAS samples.
The top quark mass is set to  \SI{172.5}{\GeV}.
The decay of the top quark is calculated by the corresponding generators (\powheg, \sherpa) respecting the spin correlation.
The PDF sets used in the ME calculation are selected from the NNPDF family for all samples, where ATLAS uses version 3.0 while CMS uses version 3.1.
The ATLAS \ttbbpowheg , \powheg and \amcnp samples use EvtGen \cite{evtgen} for simulation of the $B$-hadron decays, while the \sherpa sample and all CMS samples calculate the decays within the corresponding PS codes. All samples were produced  for final states with one or two leptons. 


\begin{description}
\item[\ttbbpowheg  samples:]~\\
Nominal \ttbb predictions are calculated using the \powhegboxres  framework at NLO with massive $b$-quarks~\cite{Jezo:2018yaf}  with the ``4FS NLO as 0118'' PDF sets. 
The renormalisation scale  is set to half of the geometric average of the transverse mass of top- and $b$-quarks  defined as
$\mT{i} = \sqrt{m^2_i + p^2_{\text{T},i}}$, where $m_i$ refers to the mass, $p_{\text{T},i}$
to the transverse momentum and $i$  to the top or $b$-quark. 
The factorisation scale is related to the average of the transverse mass of the outgoing partons in the ME calculation, see Table~\ref{tab:ttbbscalechoices}.
For ATLAS, it follows Ref.~\cite{Jezo:2018yaf}, while it is set to a factor two smaller in CMS following Ref.~\cite{Buccioni:2019plc}.  The \ttbbpowheg  internal parameters differ between the experiments:  \hbzd is set to 5 for ATLAS and to 2 for CMS, 
\hdamp is set to  \HT/2 for ATLAS and to 1.379 times the top quark mass for CMS. 
The \pythiaa parameters for PS and hadronisation modelling are set to the A14~\cite{ATL-PHYS-PUB-2014-021} and CP5~\cite{Sirunyan:2019dfx} tunes for ATLAS and CMS and the samples are referred to as ATLAS and CMS  PP8 \ttbb samples, respectively.

To vary \ttbbpowheg  internal parameters,  ATLAS sets the parameter \hbzd to 2. 
CMS varies in its set-up the \hdamp parameter to 2.305 times the top quark mass for the ``\hdamp up'' variation and to 0.8738 times the top quark mass for the  ``\hdamp down'' variation.

The ATLAS  \ttbbpowheg    calculation was performed using a special option where virtual corrections are switched off and then reweighted with virtual corrections switched on\footnote{steered via "for\_reweight 1"}, while the CMS samples used default calculation.

For the PS variations, ATLAS uses the set of LHE files  which store the results of the ME calculation by \ttbbpowheg  for the PP8 \ttbb  sample  and matches them  to a different PS prediction.
For the  prediction with the \pythiaa dipole shower only the treatment of the  recoil of the radiated parton  in the  shower  is changed and all other parameters are kept as the A14 tuned values.
Another sample is produced where \herwiga is used with the default tune provided with this generator version. 


\item[\sherpa \ttbb samples:]~\\
A \ttbb sample was generated using \sherpa version~2.2.10~\cite{Sherpa,Cascioli:2013era, Cascioli:2011va}.
The \ttbb MEs were calculated with massive $b$-quarks at NLO, using the \textsc{COMIX}~\cite{Gleisberg:2008fv} and \openloops~\cite{Cascioli:2011va} ME generators, and merged with the Sherpa PS, tuned by the authors~\cite{Schumann:2007mg}.
The same renormalisation and factorisation scales and PDFs are used as for the ATLAS PP8 \ttbb prediction.

\item[Inclusive  \ttbar samples:]~\\
The inclusive \ttbar samples are generated with the \powheg v2 NLO event generator~\cite{Nason:2004rx,Frixione:2007vw,Alioli:2010xd,Campbell:2014kua,Jezo:2015aia} and \amcnp  using a 5FS PDF set. 
The renormalisation and factorisation scales were set to the average transverse mass of the top quark and antiquark. 

For the \powheg samples of both experiments,  the PS and hadronisation is modeled by \pythiaa with the same versions and settings as for the PP8 \ttbb samples above.
The \hdamp parameter was set to the 1.5 times the top quark mass for ATLAS and to 1.379 times the top quark mass for CMS.
Another ATLAS sample is generated using \herwiga for the PS and hadronization. 
These samples are referred to as ATLAS (CMS) PP8 \ttbar and ATLAS PH7 \ttbar samples.

The inclusive \amcnp \ttbar sample uses the same scale settings and the same \pythiaa version as the ATLAS PP8 \ttbar sample and is referred to as ATLAS aMC+P8  \ttbar sample.

\end{description}


\begin{sidewaystable}
  \centering
  \caption{\label{tab:ttbbsamples}
    Configurations used for the event generation of the \ttbb process and the predicted total cross section for events with at least one lepton. 
  }
  \vskip6pt
  {\scriptsize
    \renewcommand{\arraystretch}{1.3}
    \begin{tabular}{llllllllrrr}
      \toprule
      & name & ME & Generator & ME order & Shower & Tune\footnote{ ``default'' refers to the generator's default tune}
      & NNPDF PDF set (ME) & \hdamp & \bornzero & $\sigma^{\geq1 \text{lep}}$ [\si{\pico\barn}]  \\
      \midrule
      ATLAS  & PP8 \ttbb & \ttbb  & \ttbbpowheg  & NLO & \pythia 8.224 & A14 & 4FS 3.0 NLO as 0118 & $H_T/2$ & 5 & 18.72 \\[3pt]
      CMS    & PP8 \ttbb & \ttbb  & \ttbbpowheg  & NLO & \pythia 8.230 & CP5 & 4FS 3.1 NLO as 0118 & $1.379 \cdot \mtop$ & 2 & 23.86 \\[3pt]
      \midrule
      ATLAS  & PP8 \ttbb \hbzd 2  & \ttbb  & \ttbbpowheg  & NLO & \pythia 8.224 & A14 & 4FS 3.0 NLO as 0118 & $\HT/2$ & 2 & 18.46  \\[3pt]
      
      ATLAS  & PP8 \ttbb dipole & \ttbb  & \ttbbpowheg  & NLO & \pythia 8.224 & A14, dipoleRecoil\footnote{called by \texttt{SpaceShower::dipoleRecoil} ``on''}& 4FS 3.0 NLO as 0118 & $\HT/2$ & 2 & 18.72 \\[3pt]
      ATLAS &PH7 \ttbb &  \ttbb &  \ttbbpowheg    & NLO & \herwig 7.1.6 & default & 4FS 3.0 NLO as 0118 & $\HT/2$ & 5 &18.47 \\[3pt]
      ATLAS & Sherpa \ttbb & \ttbb  & \sherpa 2.2.10 & NLO & \sherpa & default &  4FS 3.0 NNLO as 0118 & --- & --- & 20.24 \\ [3pt]
      CMS   & PP8 \ttbb \hdamp up & \ttbb  & \ttbbpowheg  & NLO & \pythia 8.230 & CP5 & 4FS 3.1 NLO as 0118 & $ 2.305 \cdot\mtop$ & 5 & 23.86  \\[3pt]
      CMS   & PP8 \ttbb \hdamp down & \ttbb   & \ttbbpowheg  & NLO & \pythia 8.230 & CP5 & 4FS 3.1 NLO as 0118 & $ 0.8738 \cdot \mtop$ & 5 & 23.86 \\[3pt]
      \midrule
      ATLAS & PP8  \ttbar & \ttbar  &\powheg v2    & NLO &\pythia 8.210 & A14 & 5FS 3.0 NLO  & $1.5 \cdot \mtop$ & 5 & 451.78\footnote{cross section predicted by NNLO calculation} \\[3pt]
      CMS   & PP8  \ttbar & \ttbar  &\powheg v2    & NLO &\pythia 8.230 & CP5 & 5FS 3.1 NLO  & $1.5 \cdot \mtop$ & 5 & 451.78$^c$ \\[3pt]
      ATLAS &  PH7  \ttbar & \ttbar  & \powheg v2    & NLO &\herwig 7.13 & default & 5FS 3.0 NLO  & $1.5 \cdot \mtop$ & 5 & 451.78$^c$ \\[3pt]
      ATLAS &  aMC+P8  \ttbar & \ttbar & \amcnp      & NLO &  \pythia 8.210 & A14   &  5FS 3.0 NLO  & --- &  --- & 451.78$^c$\\[3pt] 
      CMS   &  PP8  \ttbar \hdamp up & \ttbar  &\powheg v2    & NLO &\pythia 8.230 & CP5 & 5FS 3.1 NLO  & $2.305 \cdot \mtop$ & 5 & 451.78$^c$ \\[3pt]
      CMS   &  PP8  \ttbar \hdamp down & \ttbar  &\powheg v2    & NLO &\pythia 8.230 & CP5 & 5FS 3.1 NLO  & $0.8738 \cdot \mtop$ & 5 & 451.78$^c$ \\[3pt]
      \bottomrule\\[-2pt]
    \end{tabular}
  }
\end{sidewaystable}

\begin{sidewaystable}
  \centering
  \caption{\label{tab:ttbbscalechoices}
    Scale choices used in the event generation of \ttbb and \ttbar processes  for the different generators.
  }
  \vskip6pt
  \renewcommand{\arraystretch}{1.6}
  {\footnotesize
    \begin{tabular}{lcc}
      \toprule
      ME Generator 		& \muR & \muF \\
      \midrule
     
      ATLAS \ttbbpowheg  \ttbb & $\frac{1}{2} \sqrt[4]{\mT{t}\cdot \mT{\tbar}\cdot \mT{b}\cdot \mT{\bbar}}$
                                       &  $\frac{1}{2} (\mT{t}+\mT{\tbar}+\mT{b}+\mT{\bbar}+\mT{g})$ \\
      CMS \ttbbpowheg  \ttbb &$ \frac{1}{2} \sqrt[4]{\mT{t}\cdot \mT{\tbar}\cdot \mT{b}\cdot \mT{\bbar}}$ 
       & $\frac{1}{4} (\mT{t}+\mT{\tbar}+\mT{b}+\mT{\bbar}+\mT{g})$ \\ 
      \sherpa 2.2.10                      &$ \frac{1}{2} \sqrt[4]{\mT{t}\cdot \mT{\tbar} \cdot \mT{b}\cdot \mT{\bbar}}$ 
                                         &$ \frac{1}{2} (\mT{t}+\mT{\tbar}+\mT{b}+\mT{\bbar}+\mT{g})$   \\
      \midrule
      ATLAS \powheg \ttbar	& $ \sqrt{0.5 \cdot (\mT{t}^2 + \mT{\tbar}^2)}$ & $\sqrt{0.5 \cdot (\mT{t}^2 + \mT{\tbar}^2)}$\\
       
      CMS \powheg \ttbar		&  $\sqrt{0.5 \cdot (\mT{t}^2 + \mT{\tbar}^2)}$ & $\sqrt{0.5 \cdot (\mT{t}^2 + \mT{\tbar}^2)}$\\
      ATLAS aMC \ttbar  & $ \sqrt{0.5 \cdot (\mT{t}^2 + \mT{\tbar}^2)}$ & $ \sqrt{0.5 \cdot (\mT{t}^2 + \mT{\tbar}^2)}  $ \\
      \bottomrule
    \end{tabular}
  }
  \renewcommand{\arraystretch}{1.0}
\end{sidewaystable}

\begin{table}
  \caption{Systematic variations of scales in the ME and PS codes used for all comparisons presented here.} 
  \label{tab:scalevariations}\label{tab:ttbbvariations}
  \vskip6pt
  \centering
  \renewcommand{\arraystretch}{1.2}
  \begin{tabular}{lcc}
    \toprule
    Variation &  \\
    \midrule
    Scale variation ME & \multicolumn{2}{c}{\muR $\times$  0.5, \muF $\times$ 0.5;  \muR $\times$ 2, \muF $\times$ 2} \\
    ISR variation (PS) &  \multicolumn{2}{c}{$\alpha_{\text{s}}^{\text{ISR}} \times 0.5; {\alpha_{\text{s}}}^{\text{ISR}} \times 2.0$} \\
    FSR variation (PS) & \multicolumn{2}{c}{$\alpha_{\text{s}}^{\text{FSR}} \times 0.5; {\alpha_{\text{s}}}^{\text{FSR}} \times 2.0$} \\
    \bottomrule
  \end{tabular}
  \renewcommand{\arraystretch}{1.0}
\end{table}

\newpage
\subsection{Object reconstruction, fiducial volume and observables}\label{sec:ttbb:rivet}
The object definition and event selection applied in this comparison study is defined at particle level and is the same for ATLAS and CMS.
All objects are defined using stable final-state particles with a mean lifetime of $\tau > \SI{3e-11}{\second}$.
Jets are reconstructed from all stable final-state particles (but excluding leptons and neutrinos from the top quark decay chain) using  the anti-$k_{\mathrm{t}}$ jet algorithm~\cite{Cacciari:2008gp,Cacciari:2011ma} with a radius parameter of $R = 0.4$.
Jets which contain at least one ghost-associated~\cite{Cacciari:2008gn} $B$-hadron with $\pt>\SI{5}{\GeV}$ are defined as \bjets, all other jets are considered ``light'' jets.
The four-momentum of the bare leptons from top quark decay are modified (``dressed'') by adding the four-momenta of all radiated photons within a cone of size $\Delta R=0.1$.
All objects are considered within pseudo-rapidity $|\eta|<2.5$ and with $\pt>\SI{27}{\GeV}$ for leptons and  $\pt>\SI{25}{\GeV}$ for jets and \bjets .

Leptons are removed if they are separated from a jet by less than $\Delta R=0.4$, where $\Delta R = \sqrt{(\Delta \eta )^2 + (\Delta \phi)^2}$.
Events are selected with at least four \bjets, and further separated into two analysis regions: events with exactly one lepton and at least six jets (single lepton channel) and events with exactly two leptons and at least four jets (dilepton channel).

A set of observables relevant for the \ttHbb analysis is studied within this fiducial phase space.
All observables are studied for both the single lepton and the dilepton channel, however only the variables listed in Table\,\ref{tab:variables} are shown in the following figures, as no significant qualitative difference is observed between the different top quark decay channels.

\begin{table}[hb]
  \centering
  \caption{
    The list of observables used for the comparison of the generators for the \ttbb process.
  }\label{tab:variables}
  \vspace{0.25cm}
  {\small
    \setlength\tabcolsep{1.5pt}
    \renewcommand{\arraystretch}{1.1}
    \begin{tabular}{lll}
      \toprule
      Variable & Description  & Channel \\
      \midrule
      $\Delta R_{bb}^{\text{min} \Delta R}$ & $\Delta R$ of the two \bjets in the event which are closest in $\Delta R$ \,\,\,& dilepton \\ 
      $m_{bb}^{\text{min} \Delta R}$  & Invariant mass of the two \bjets closest in $\Delta R$                & dilepton \\ 
      $N_{\text{jets}}$   & Number of  jets in the event (all jet  flavours)      & dilepton    \\
      Light jet \pt  & Transverse momentum of the light jets in the event & dilepton \\
      \midrule
      $N_{\text{\bjets}}$   & Number of \bjets in the event      & single lepton  \\
      $\HT^{\text{jets}}$ & Scalar sum of \pt of jets in the event (all jet flavours) & single lepton \\
      Leading \bjet \pt \,\,\,& \pt of $b$-jet with largest \pt in the event & single lepton \\
      Fourth \bjet \pt & \pt of \bjet with fourth largest \pt in the event & single lepton\\
      
      
      
      
      \bottomrule
    \end{tabular}
    \renewcommand{\arraystretch}{1.0}
  }
\end{table}

\subsection{Results}\label{sec:ttbb:results}
Three sets of generator predictions are compared for the observables given in Table~\ref{tab:variables} as follows. All comparisons  are performed with respect to the  \ttbb PP8 sample.
 The PP8 \ttbb sample and the alternative predictions are normalised to an integral of one, after all selections and in each histogram individually for a shape-only comparison.
The scale uncertainty variations on PP8 \ttbb are derived as  listed in  Table\,\ref{tab:scalevariations}  and the differences are added in quadrature to the statistical uncertainties to form the  shaded area displayed in the figures.

\begin{figure}[!htb]
  \centering
  \includegraphics[width=0.4\textwidth]{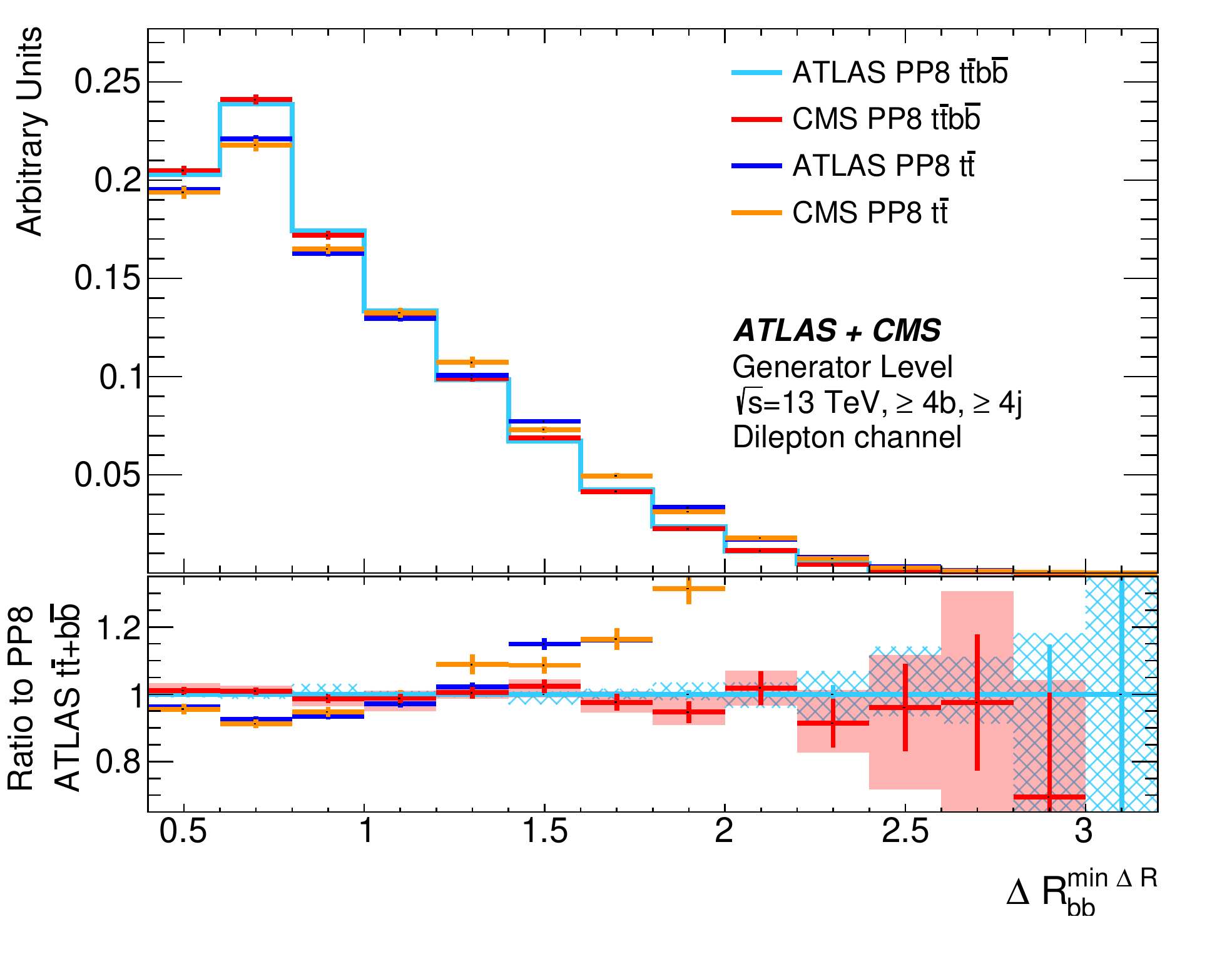}
  \includegraphics[width=0.4\textwidth]{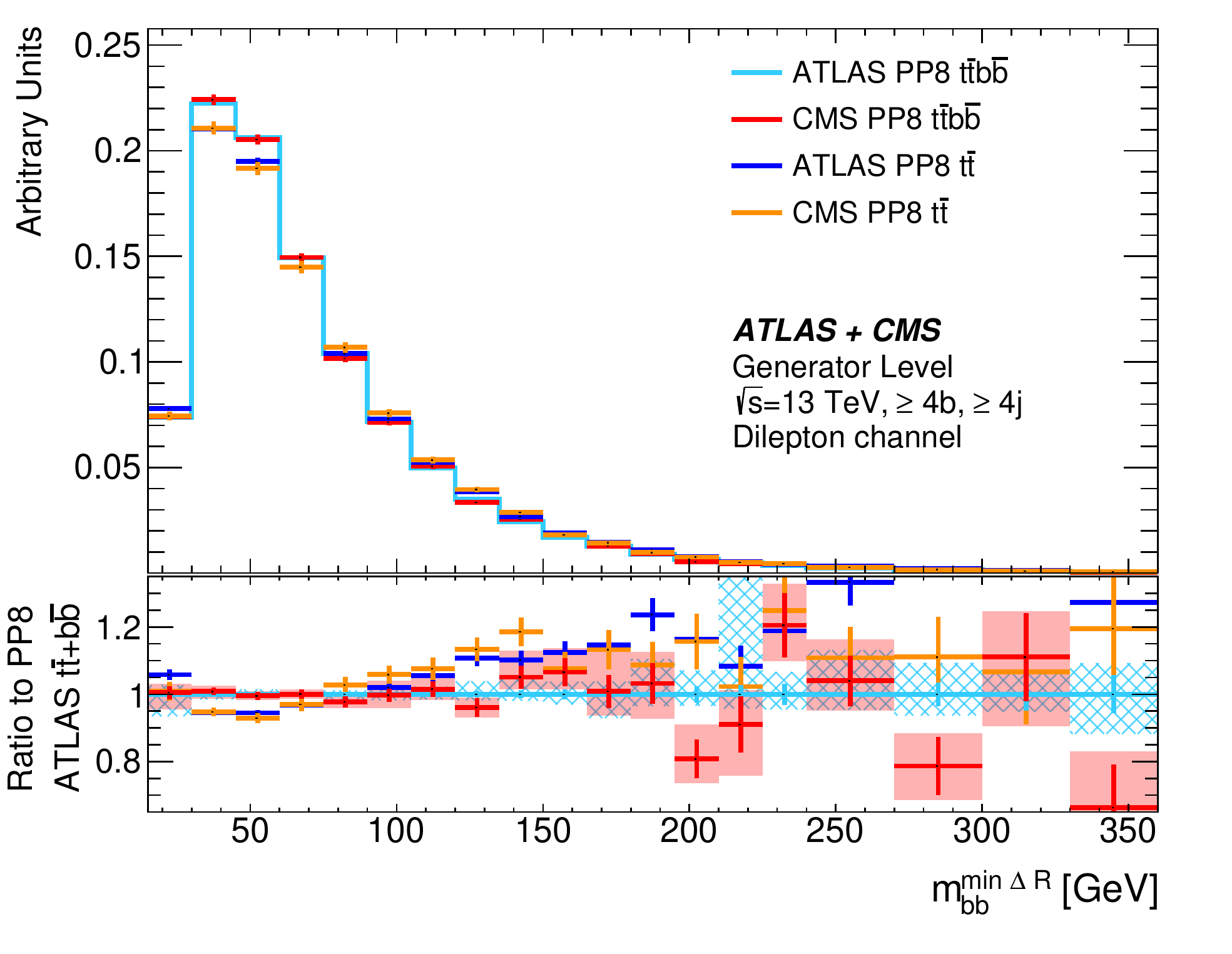}
  \includegraphics[width=0.4\textwidth]{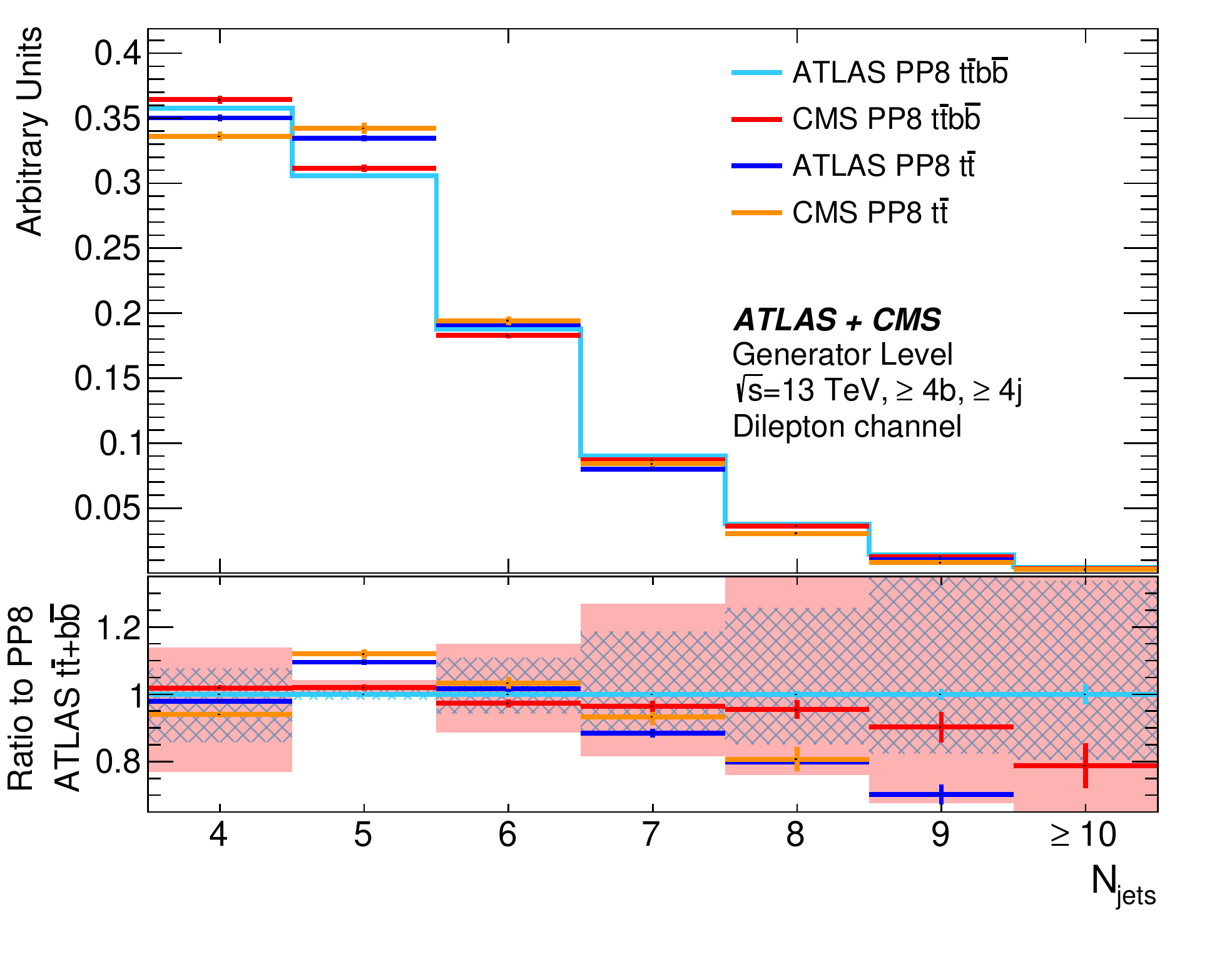}
  \includegraphics[width=0.4\textwidth]{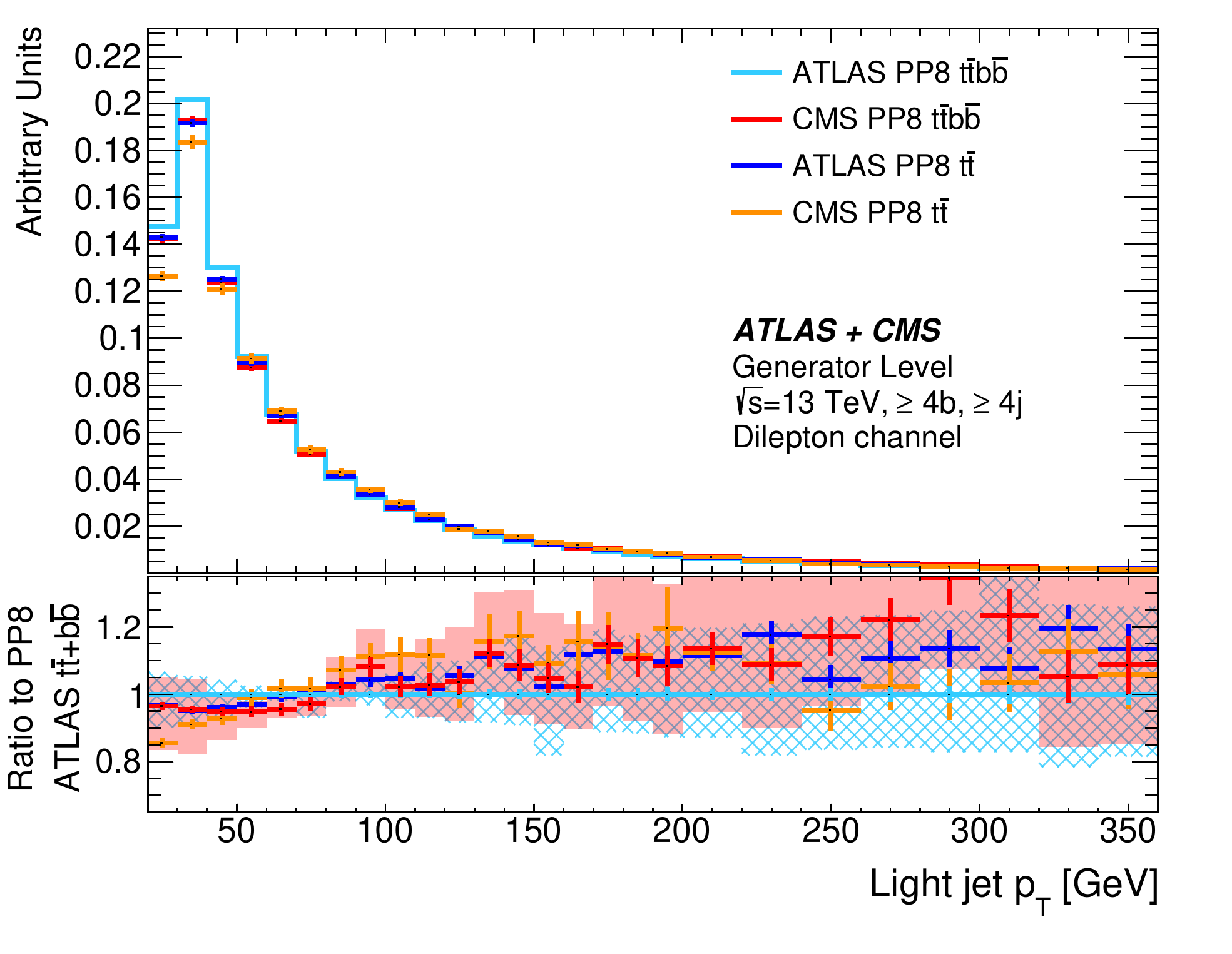}
  
  \includegraphics[width=0.4\textwidth]{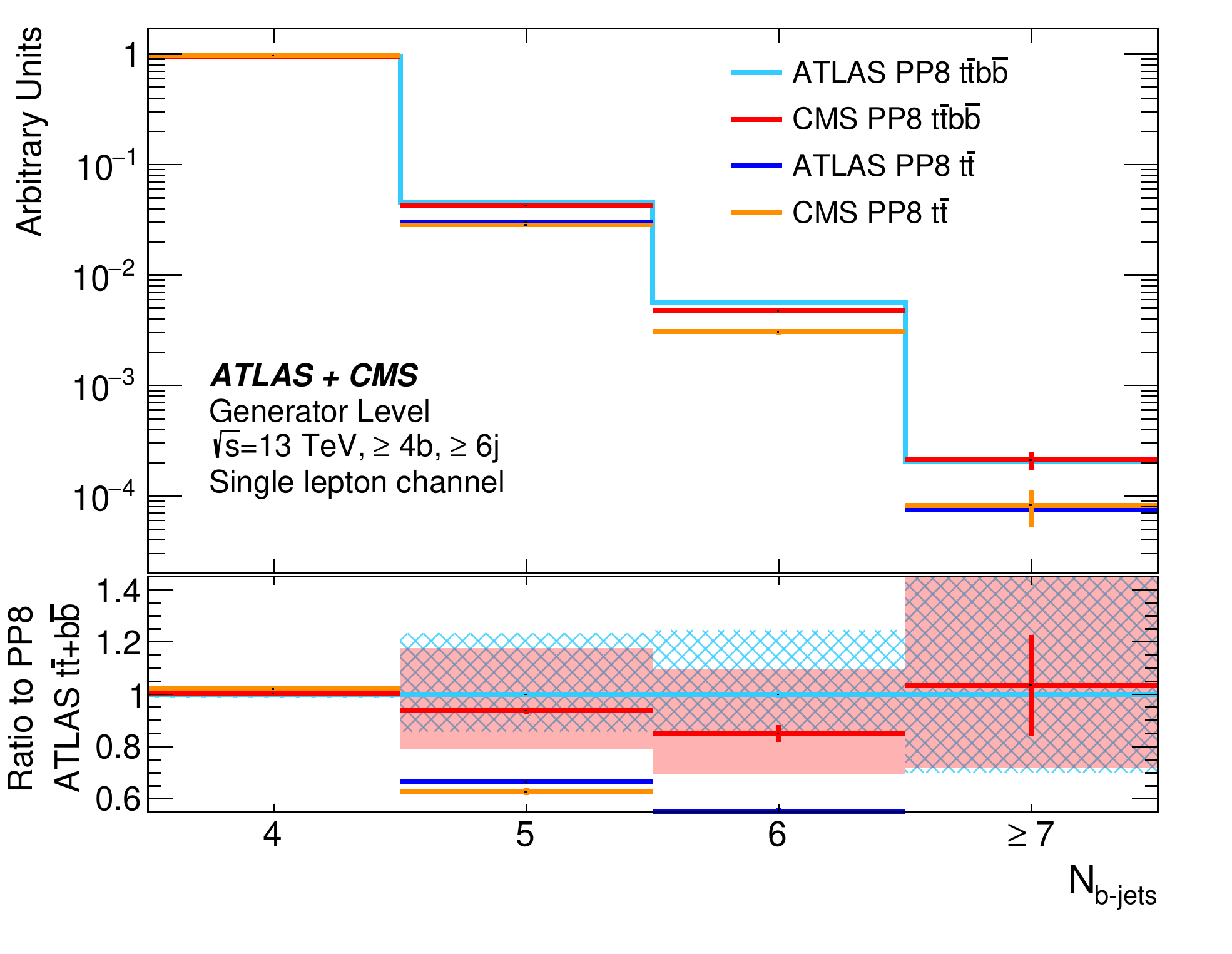}
  \includegraphics[width=0.4\textwidth]{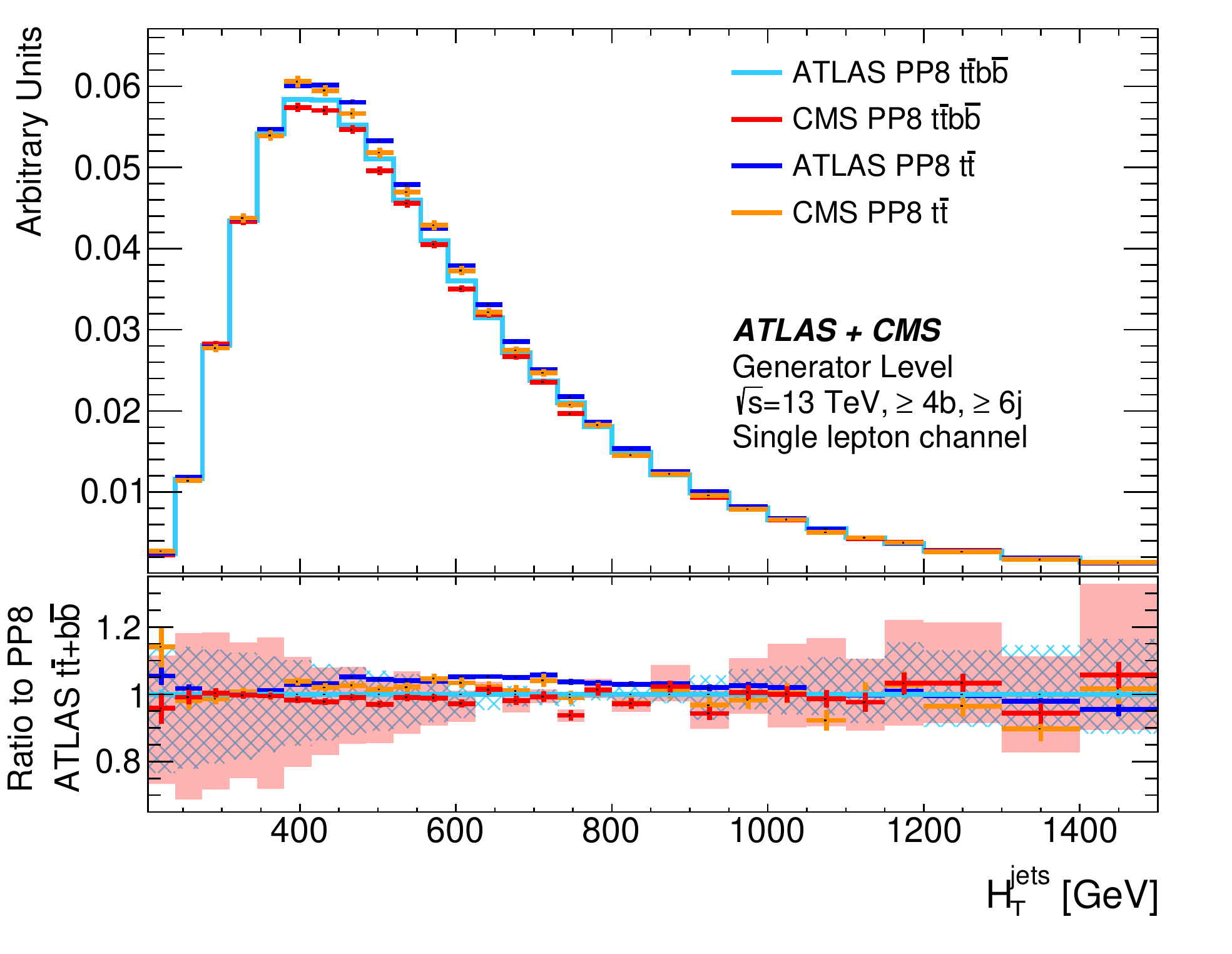}
  \includegraphics[width=0.4\textwidth]{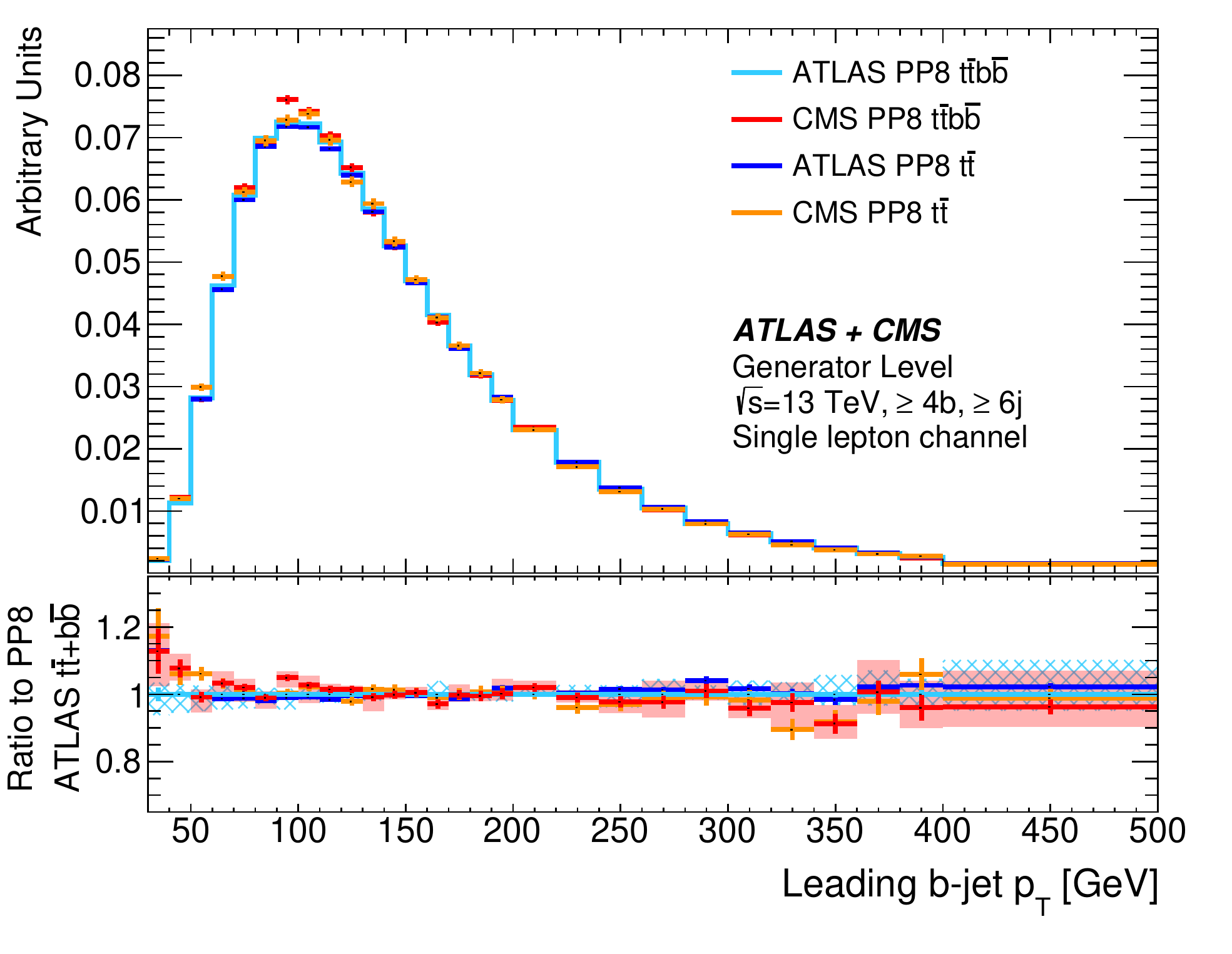}
  \includegraphics[width=0.4\textwidth]{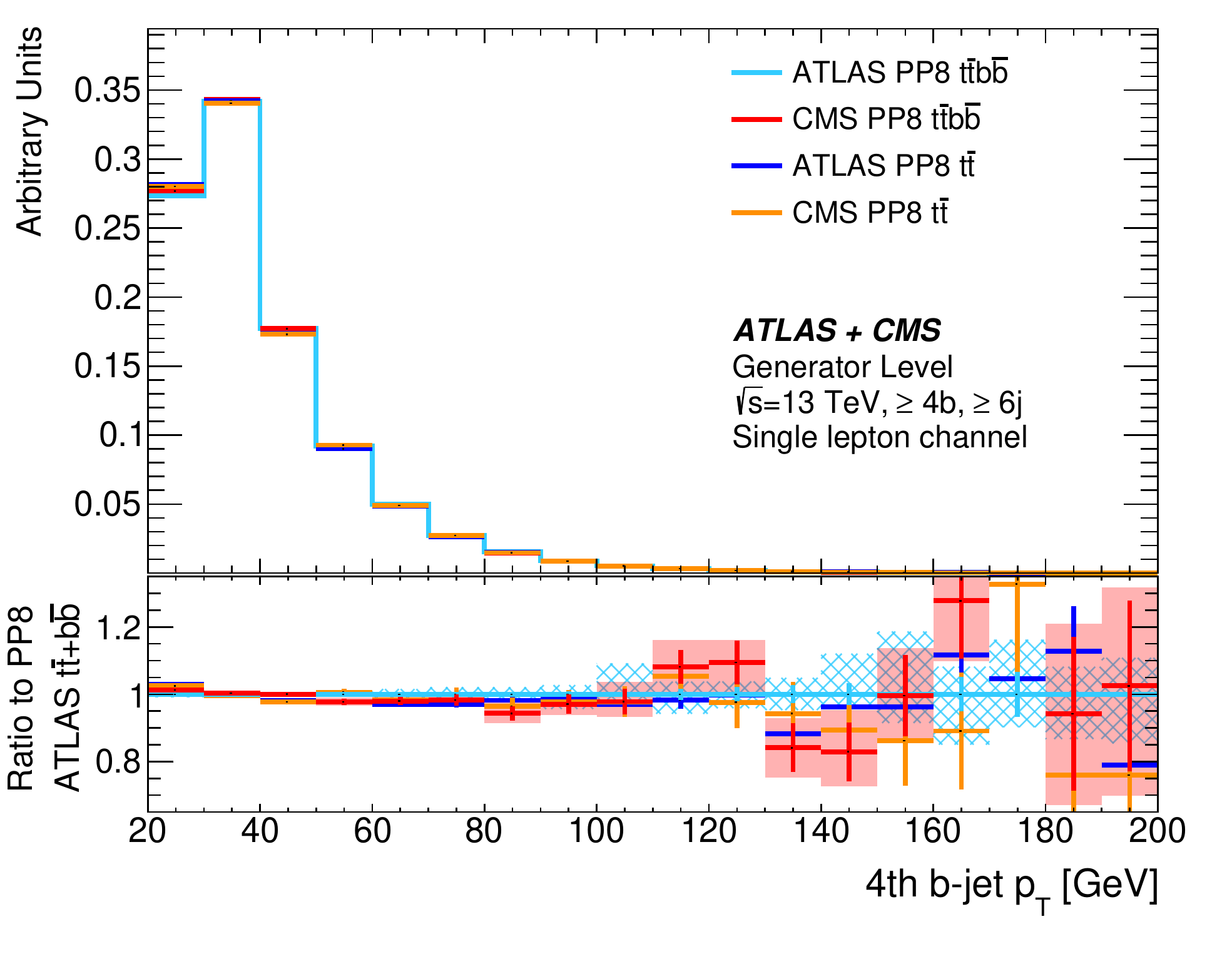}
  \caption{
    Comparison of  PP8 predictions for \ttbb and \ttbar with the described settings using the observables defined in Table~\ref{tab:variables} in the fiducial analysis phase space.
    All predictions are normalised to one.
    The error bands are constructed from the statistical uncertainties  and the scale variations (ME and PS) for the ATLAS PP8 \ttbb (blue) and the CMS PP8 \ttbb (red) samples.
    Statistical uncertainties are indicated by vertical lines. 
    The ratio shows the different curves divided by the ATLAS PP8 \ttbb prediction.
  }\label{fig:comp1}
\end{figure}

Figure~\ref{fig:comp1} shows the nominal \ttbb predictions from ATLAS and CMS to be used in future analyses compared to the nominal predictions used in the early Run-2 analyses.
The differences between ATLAS and CMS set-ups  cause only minor differences between the predictions.
However, significant differences between the PP8 \ttbb predictions and the PP8 \ttbar predictions  are observed in $\Delta R_{\bbbar}^{\text{min} \Delta R}$, the jet multiplicity and in the number of events with more than four \bjets. 
Furthermore, the  uncertainty band is  slightly larger in the CMS \ttbb predictions, potentially caused by the lower factorisation scale. 

\begin{figure}[!htb]
  \centering
  \includegraphics[width=0.4\textwidth]{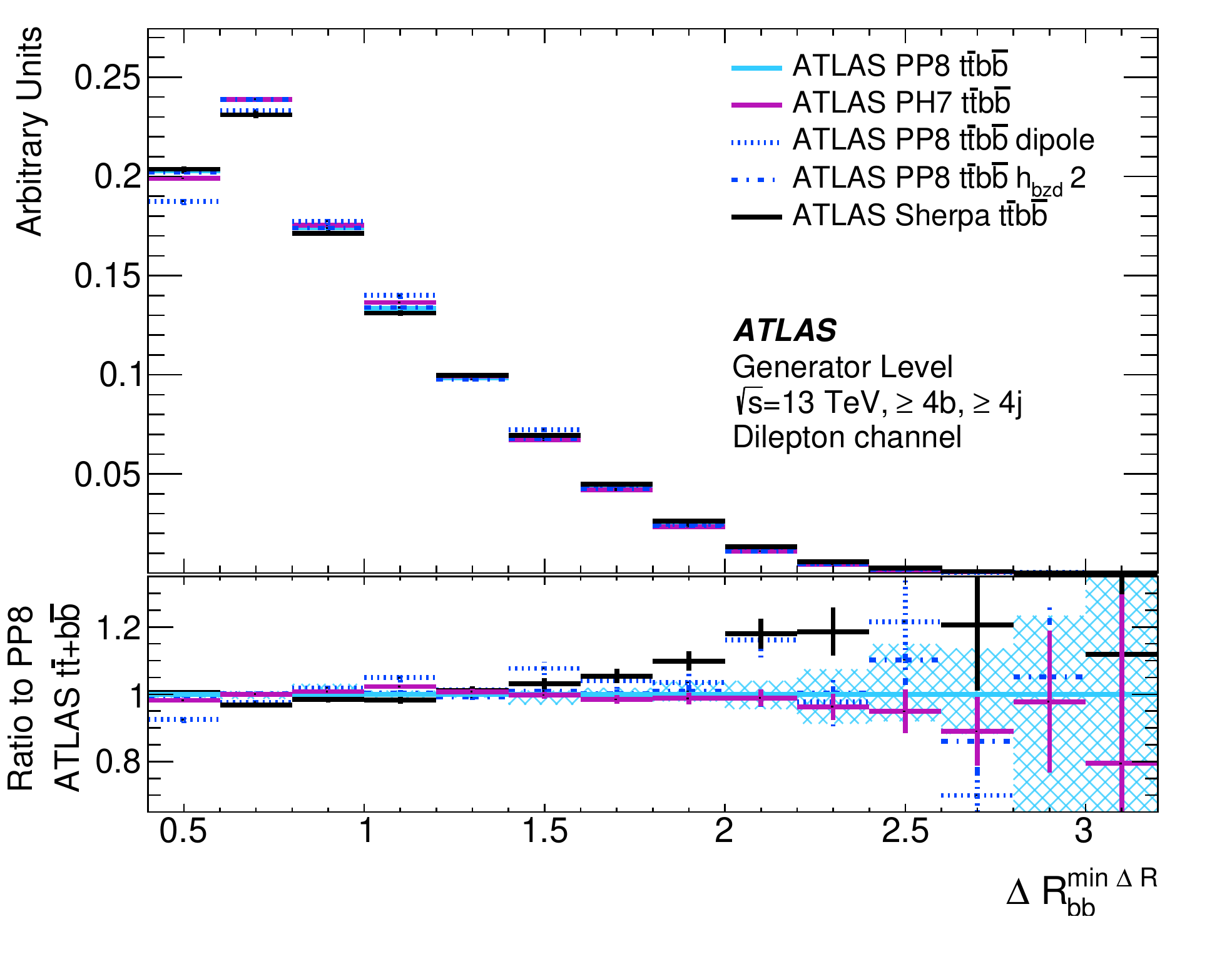}
  \includegraphics[width=0.4\textwidth]{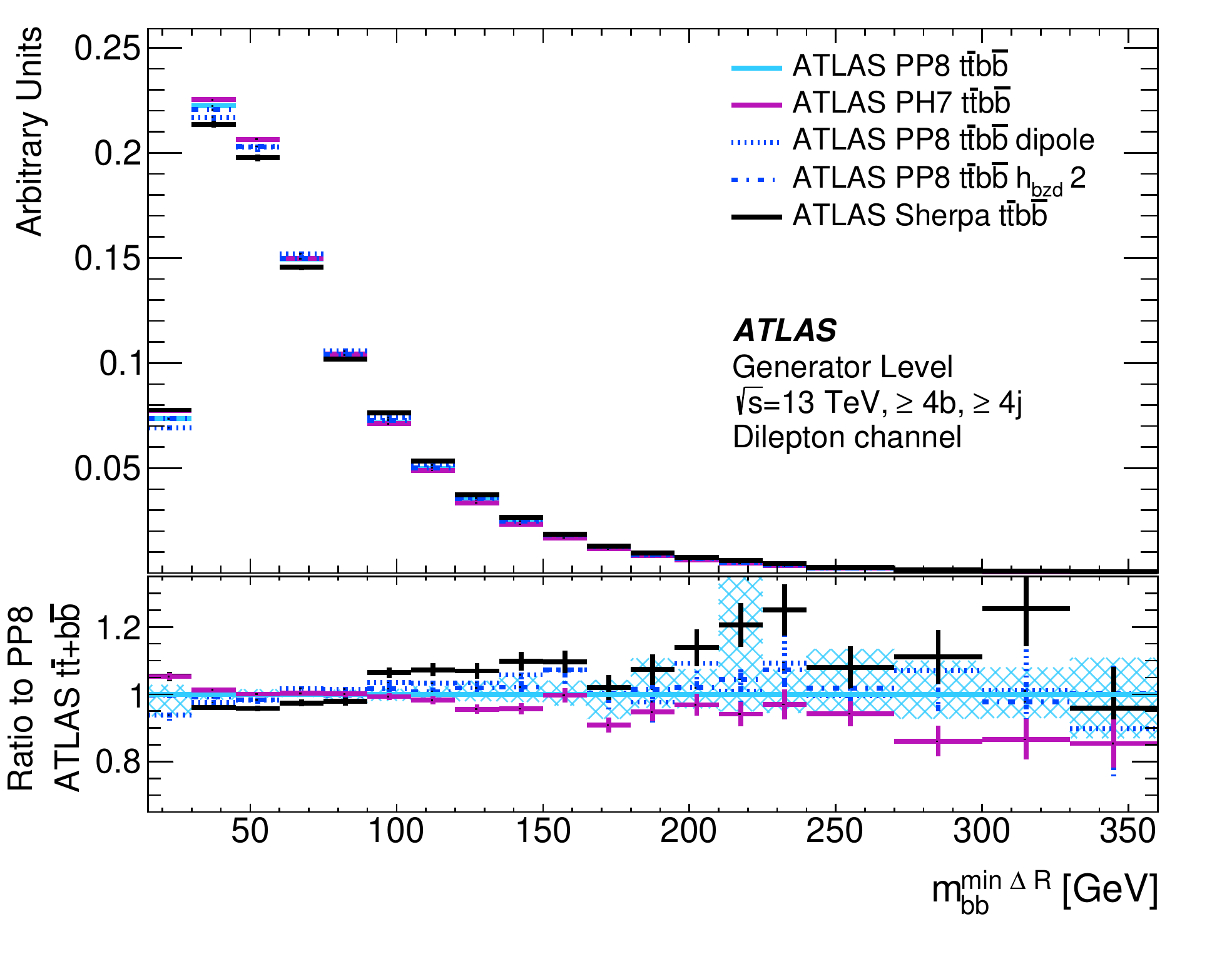}
  \includegraphics[width=0.4\textwidth]{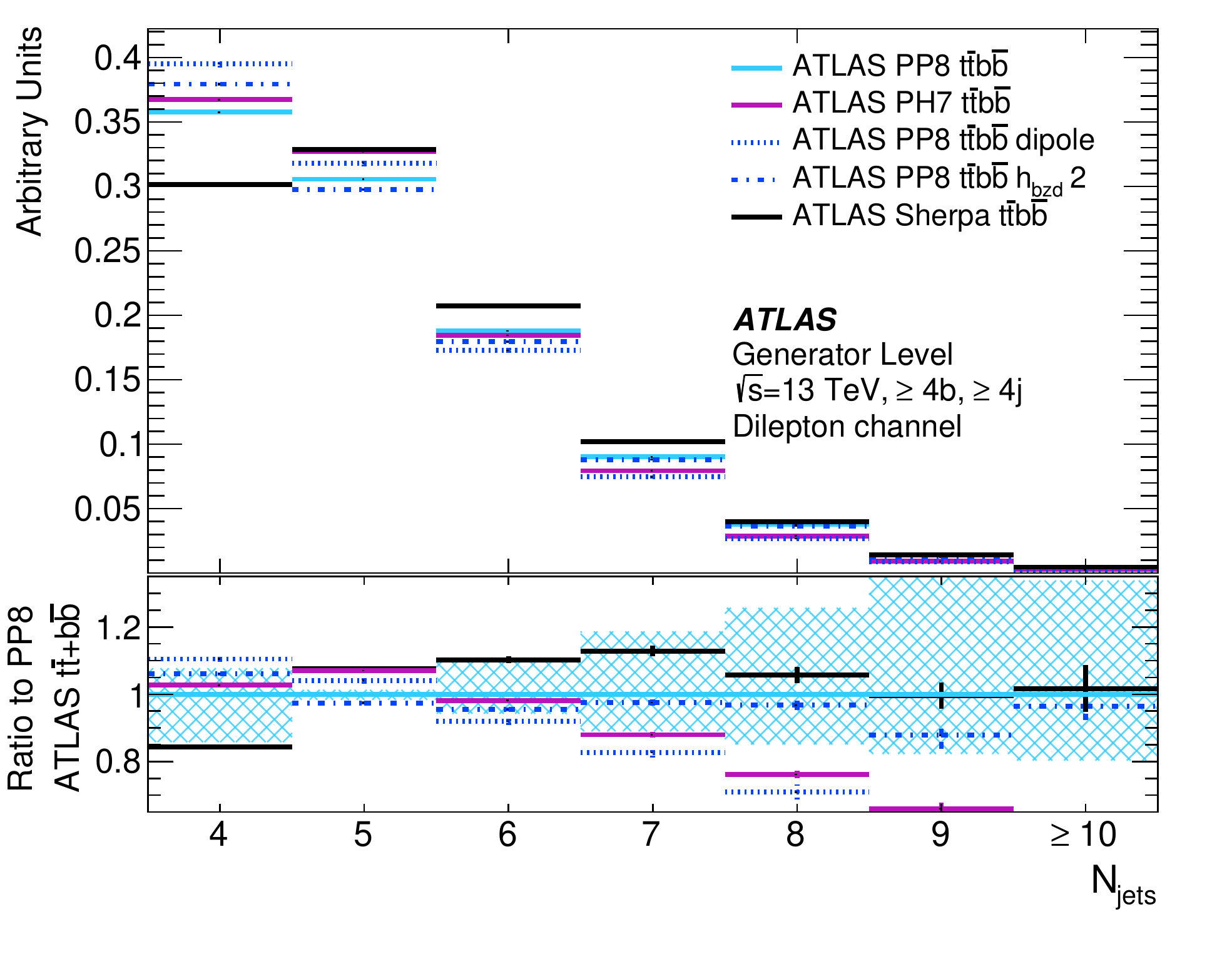}
  \includegraphics[width=0.4\textwidth]{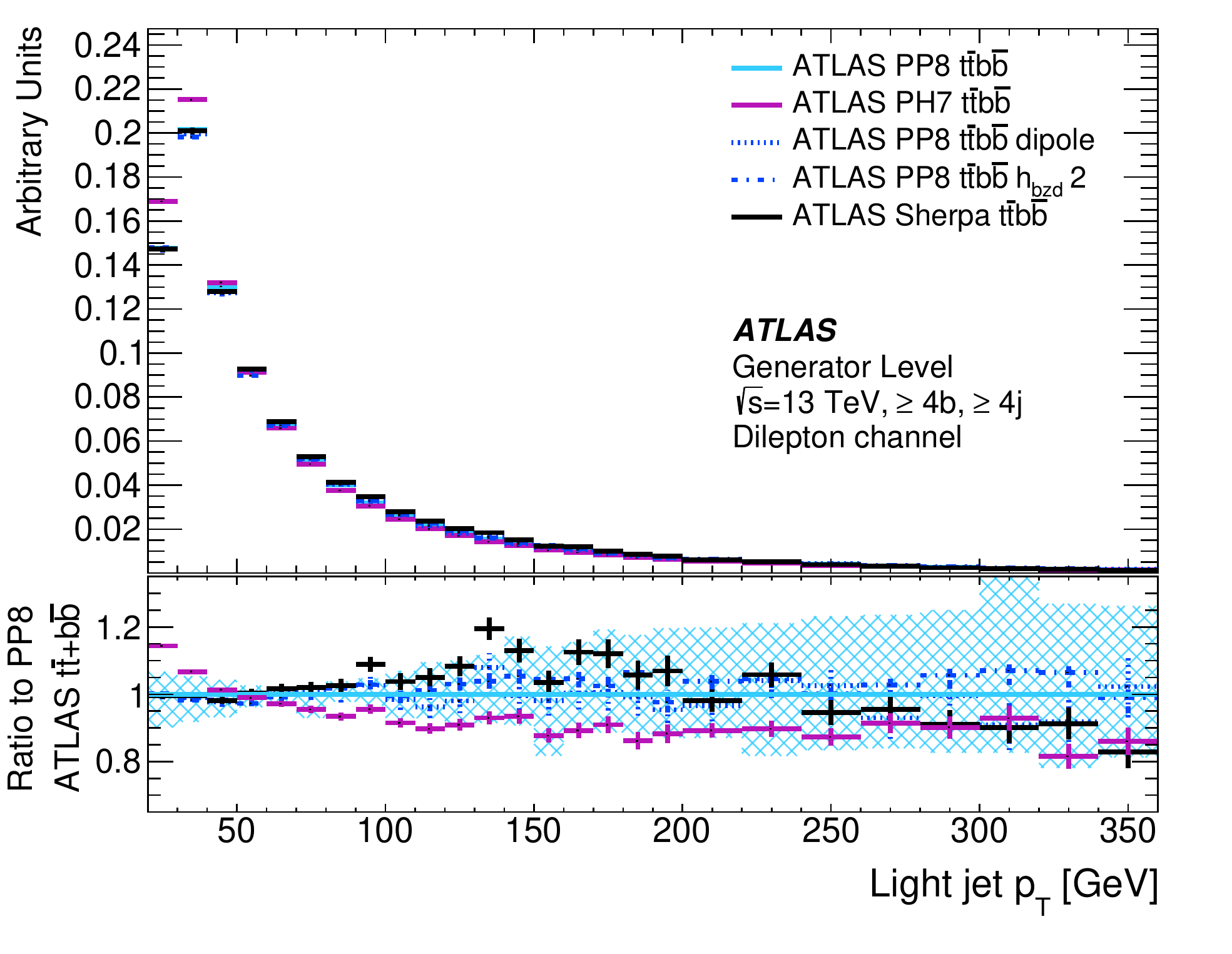}
  
  \includegraphics[width=0.4\textwidth]{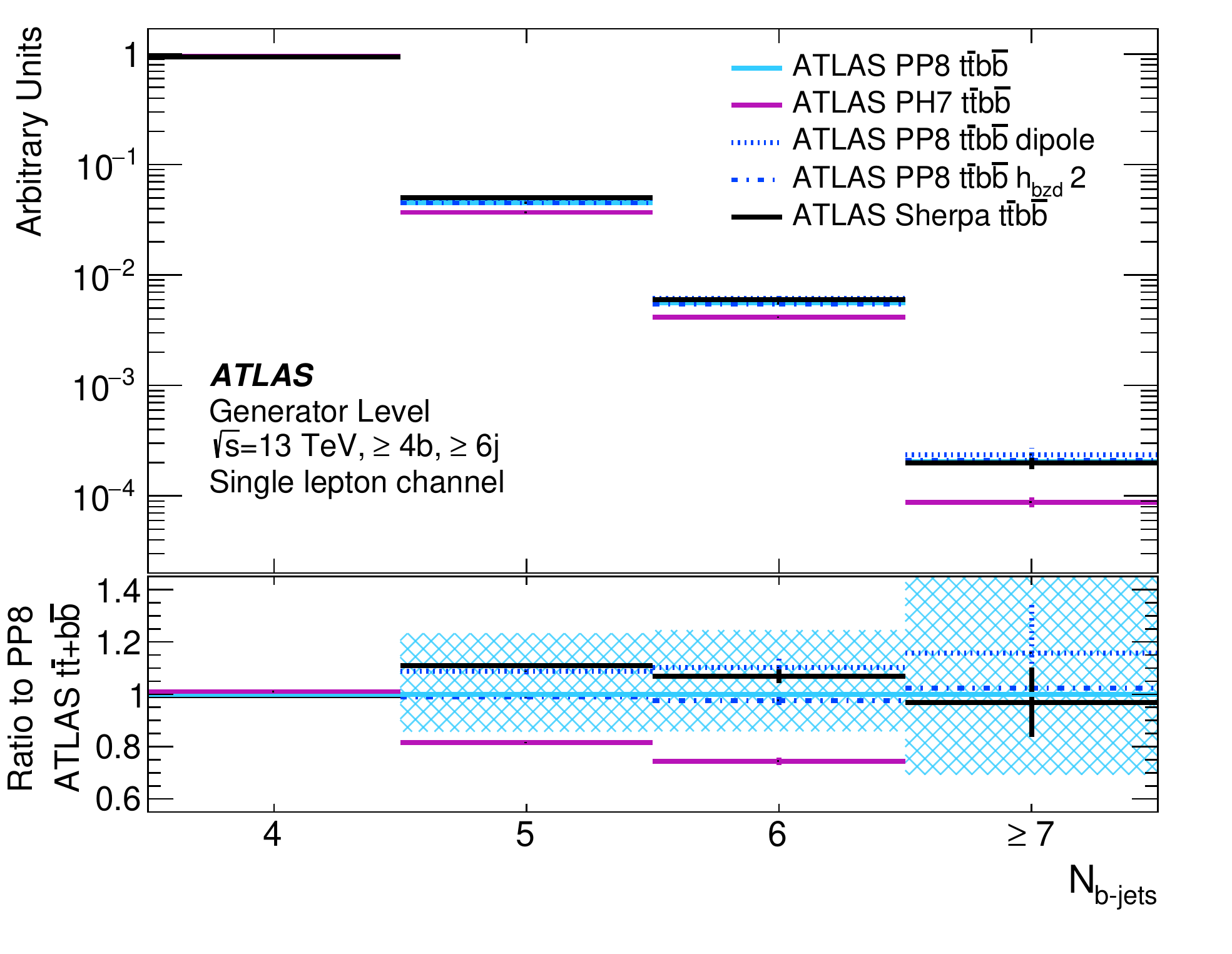}
  \includegraphics[width=0.4\textwidth]{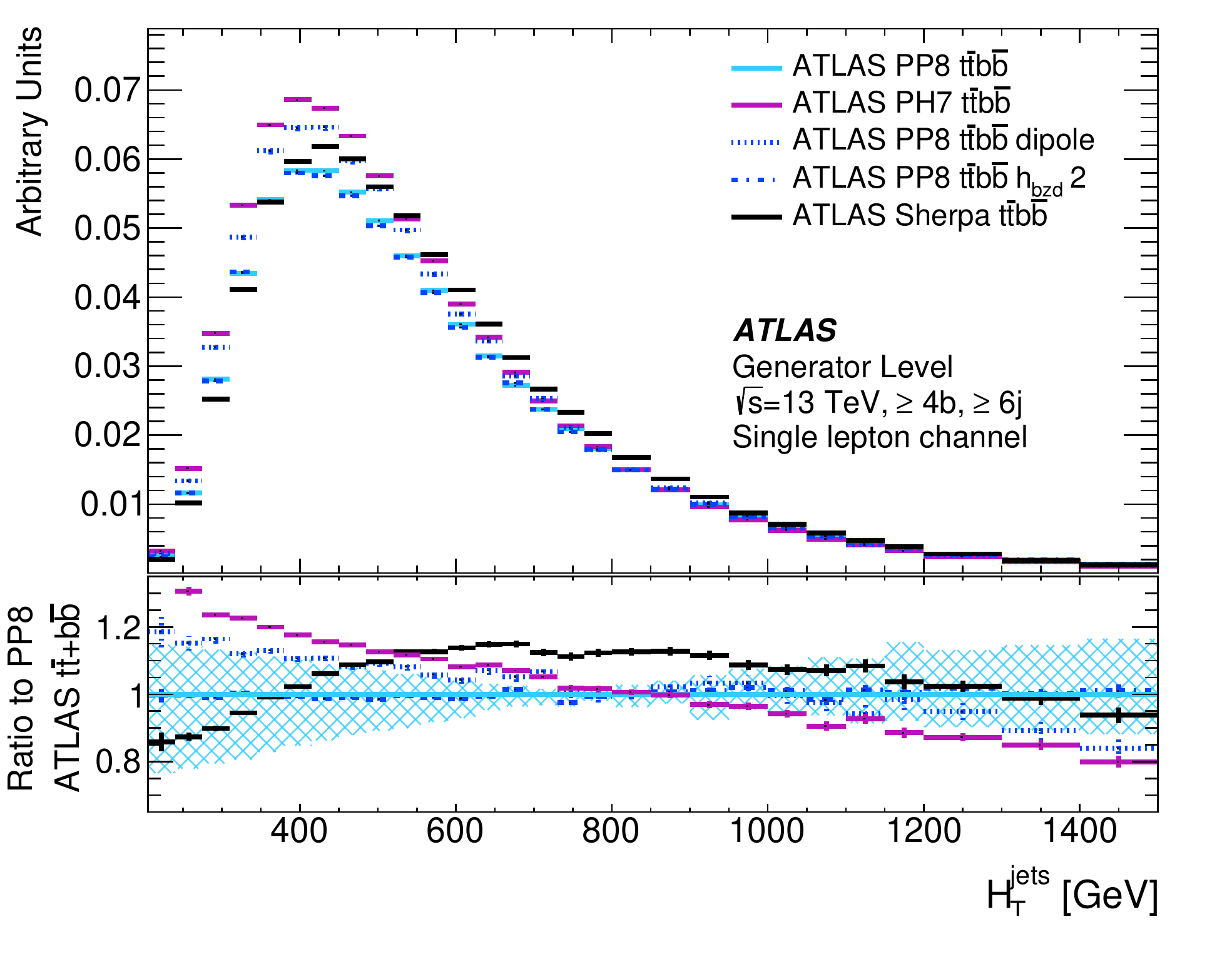}
  \includegraphics[width=0.4\textwidth]{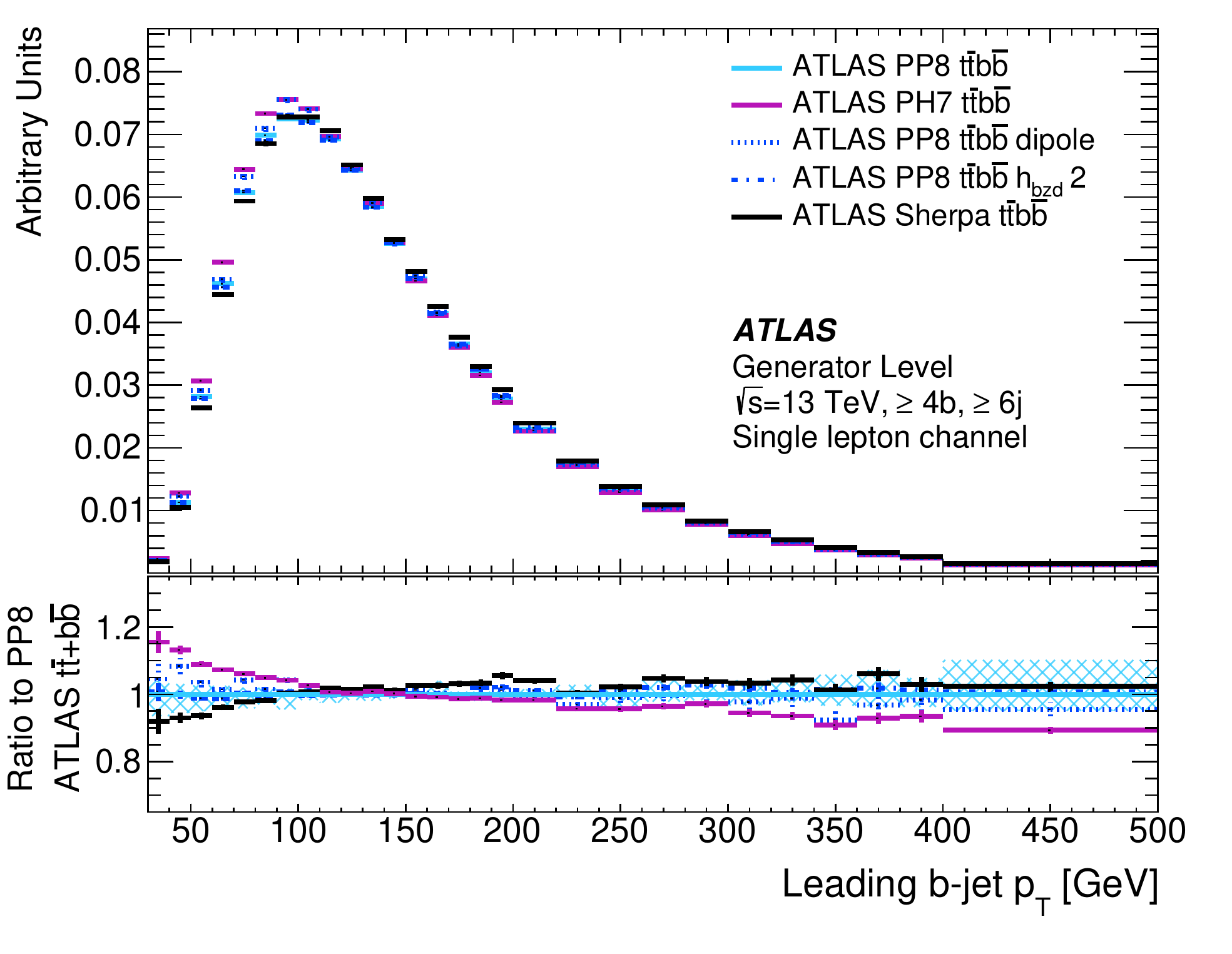}
  \includegraphics[width=0.4\textwidth]{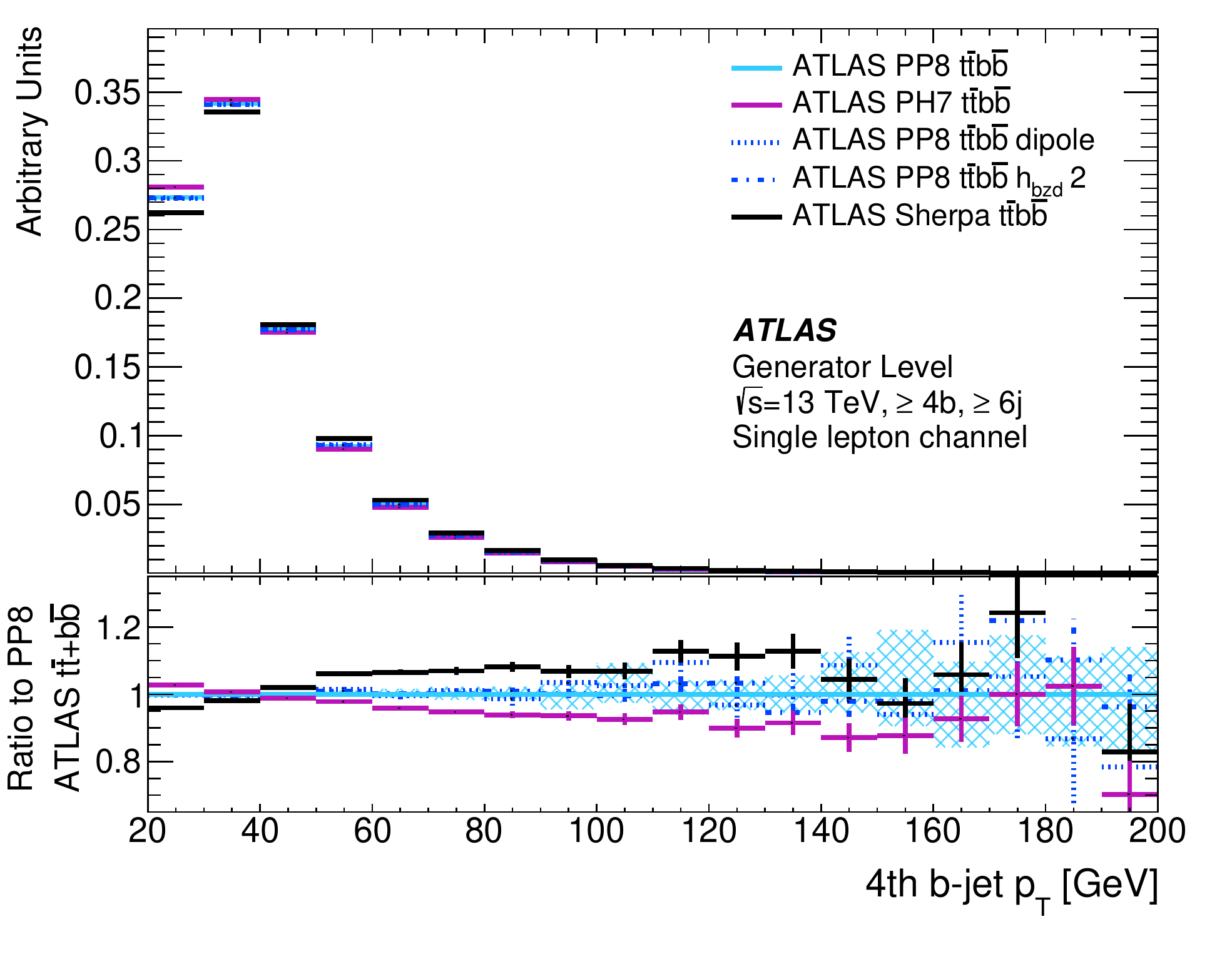}
  \caption{
    Comparison of ATLAS PP8 predictions for \ttbb with different matching and PS settings and \sherpa. 
    All distributions  are normalised to one.
    The ratio shows the different curves divided by PP8 \ttbb. The error band contains the statistical uncertainty  and the scale variations (ME and PS) for the ATLAS PP8 \ttbb sample. 
        Statistical uncertainties are indicated by vertical lines.
  }\label{fig:comp2}
\end{figure}

In Fig.~\ref{fig:comp2}, the ATLAS nominal PP8 \ttbb prediction is compared to all generator variations potentially considered as modelling uncertainties for future ATLAS \ttHbb analyses, i.e.\ variations in \ttbbpowheg  and \pythiaa parameter settings as well as \sherpa as alternative generator.
As already discussed in Ref.~\cite{ttbbpubnote}, the parameter $\bornzero$ has only a minor influence on the observables.
Interestingly, predictions of \ttbbpowheg  matched to \pythiaa using the dipole shower and  matched to  the \herwiga PS  both show a significant decrease with respect to the nominal PP8 \ttbb  in the jet multiplicity and \HT.
\sherpa differs up to  10--20\,\% in all distributions  with significant differences in shape. 

\begin{figure}[!htb]
  \centering
  \includegraphics[width=0.4\textwidth]{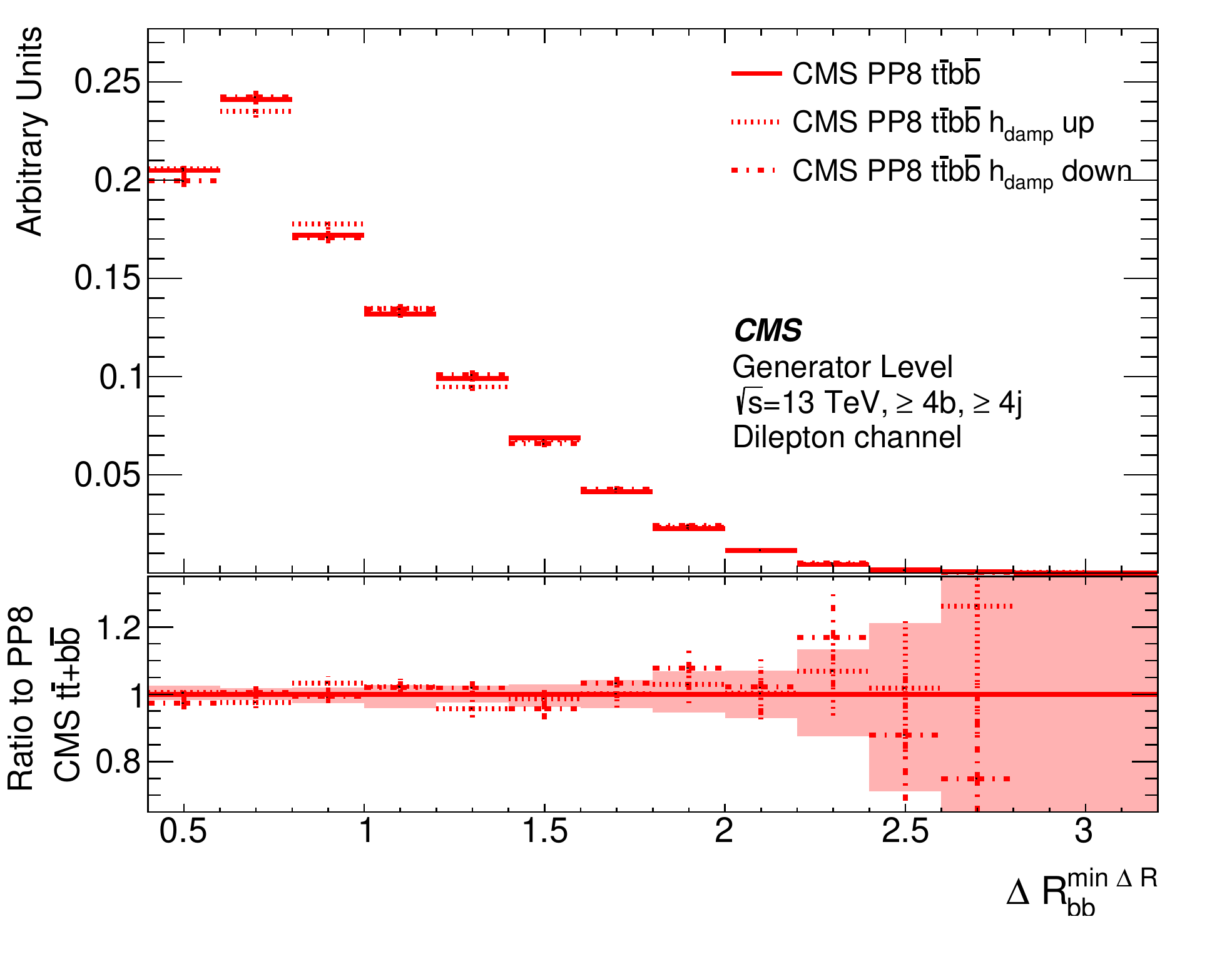}
  \includegraphics[width=0.4\textwidth]{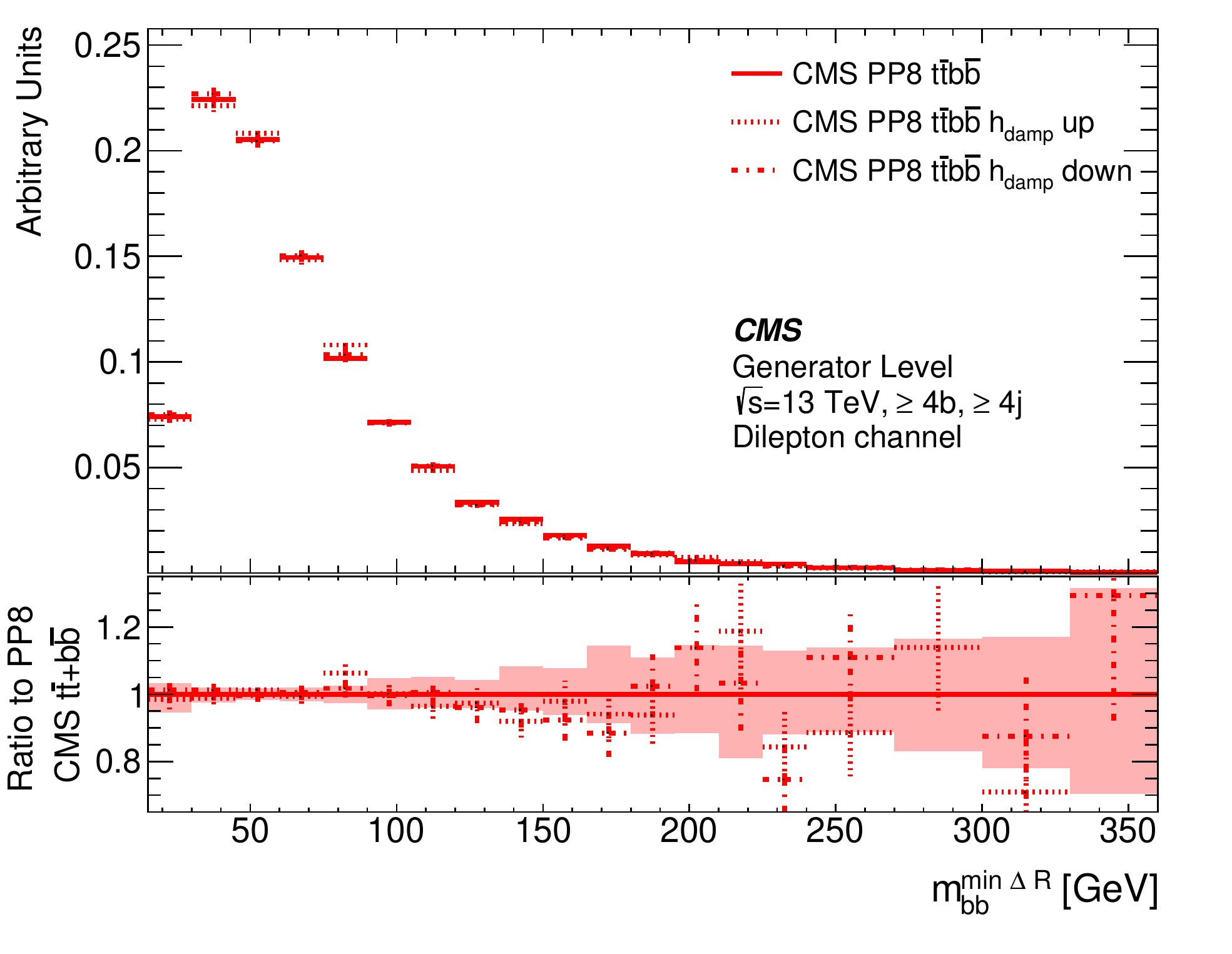}
  \includegraphics[width=0.4\textwidth]{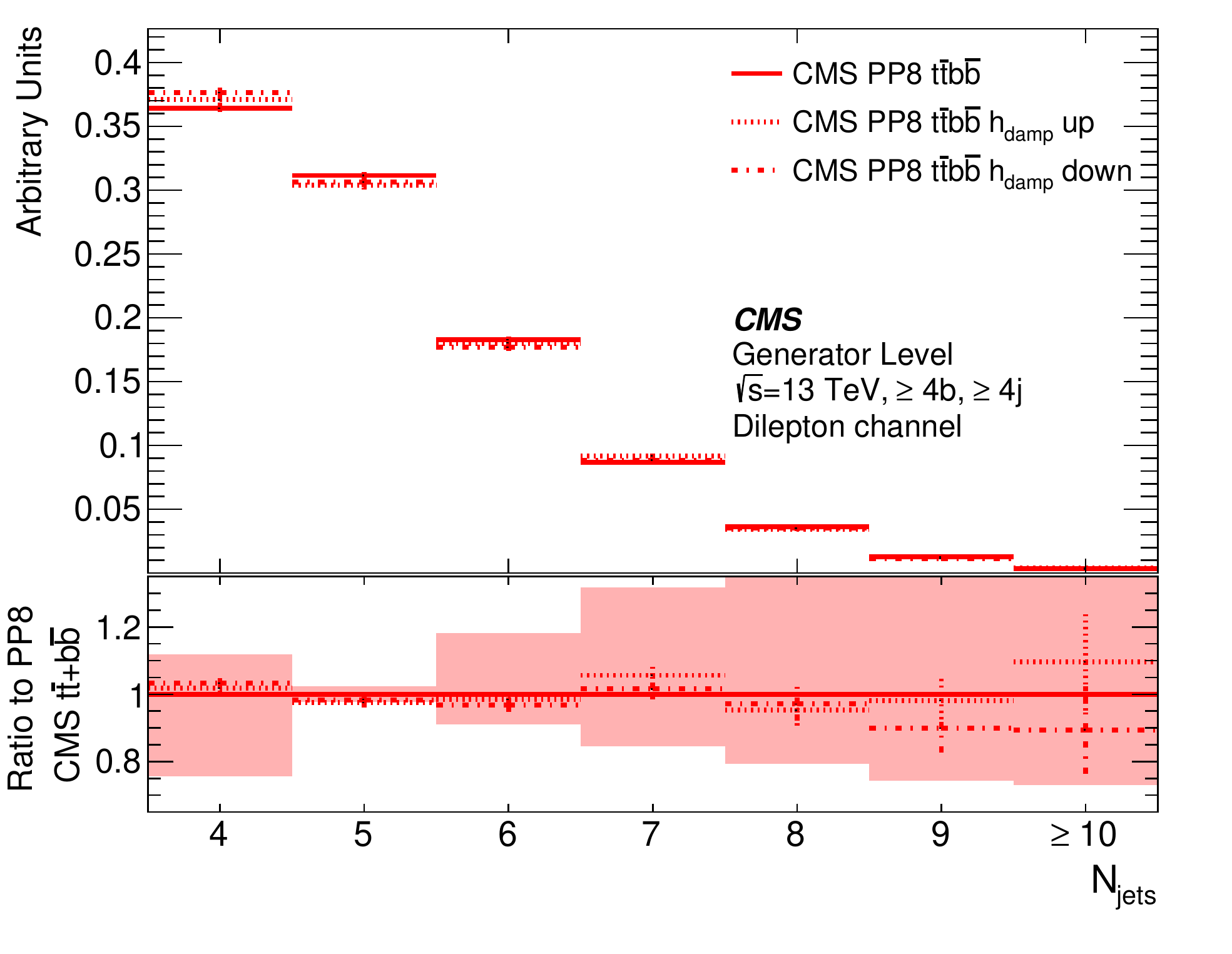}
  \includegraphics[width=0.4\textwidth]{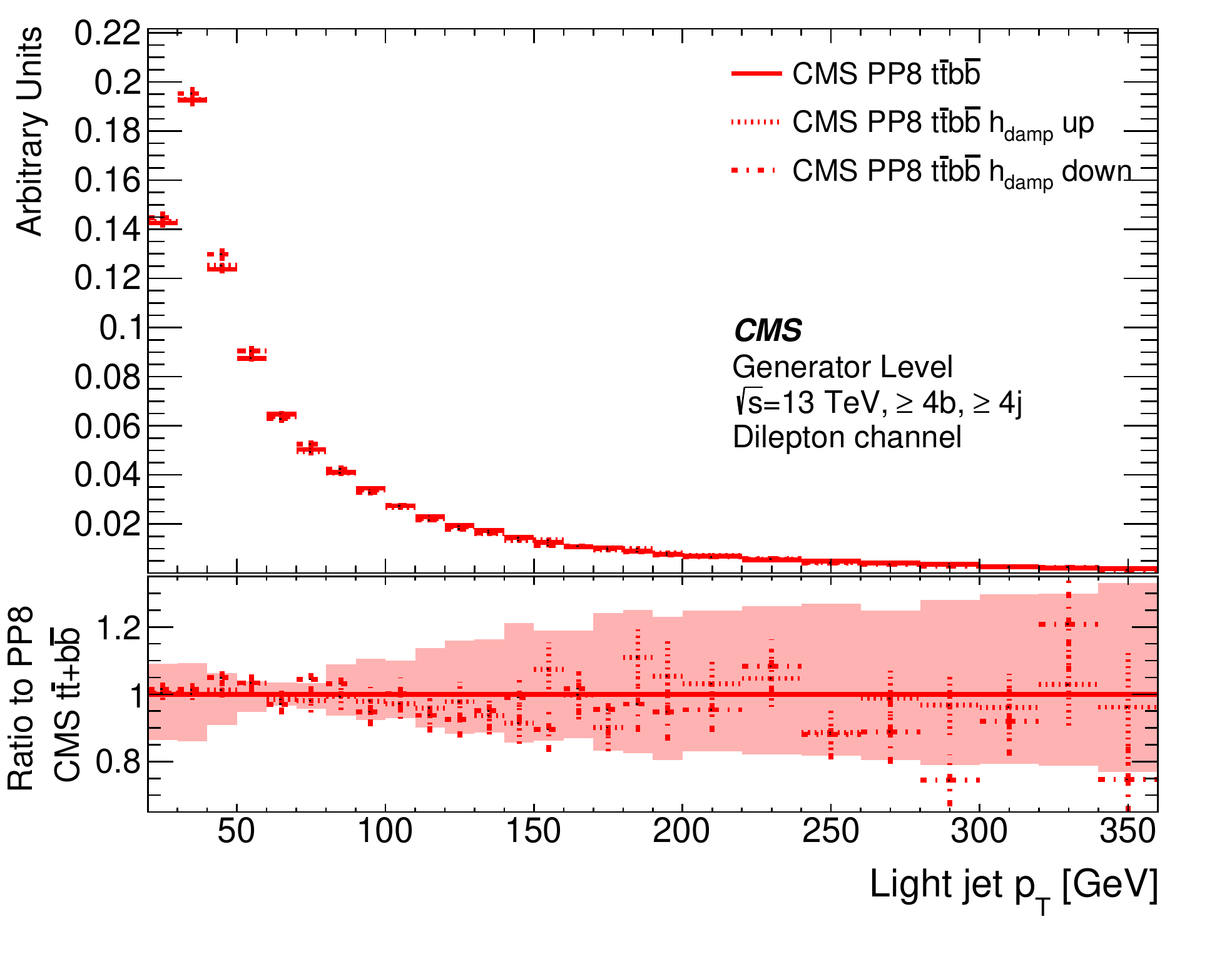}
  
  \includegraphics[width=0.4\textwidth]{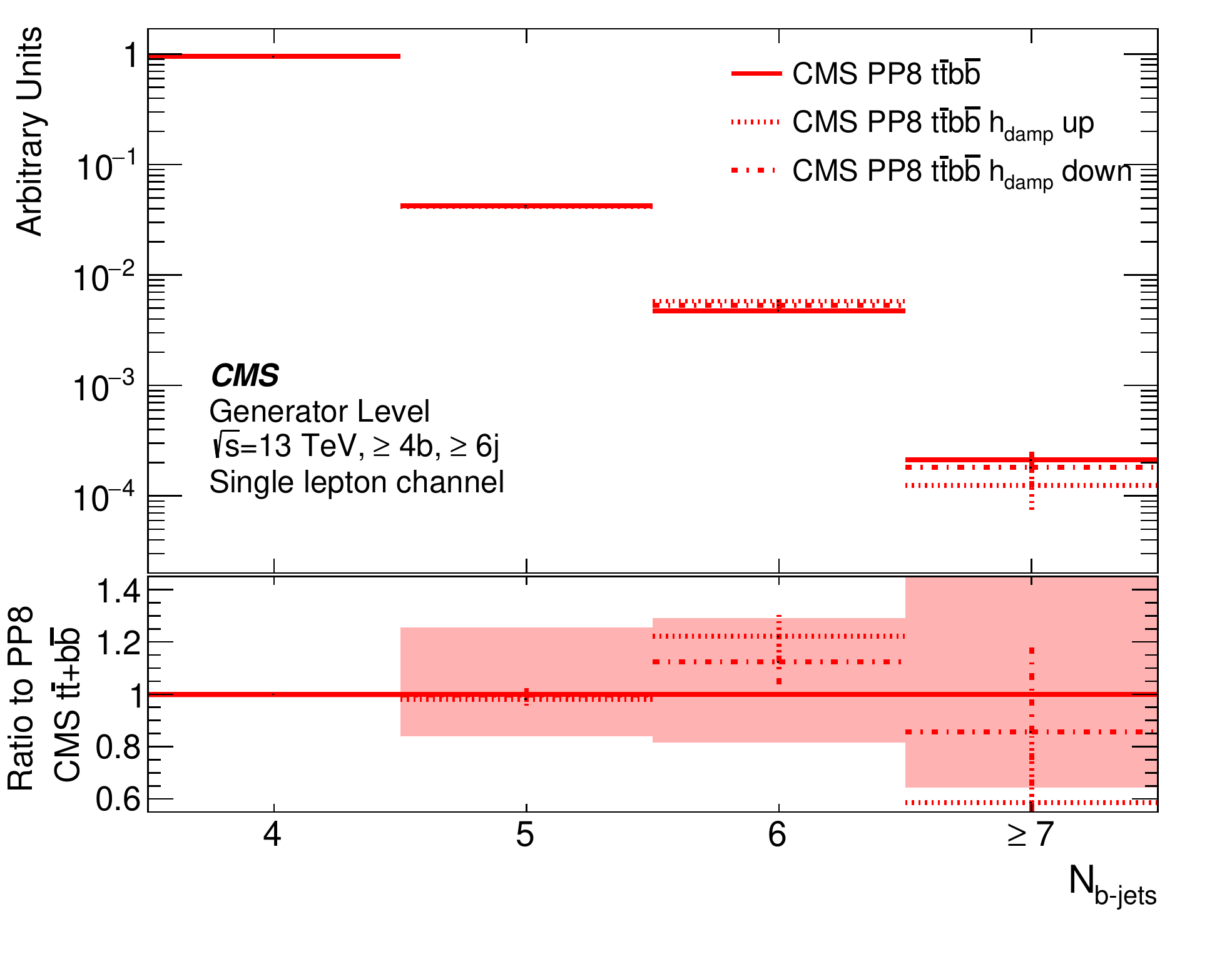}
  \includegraphics[width=0.4\textwidth]{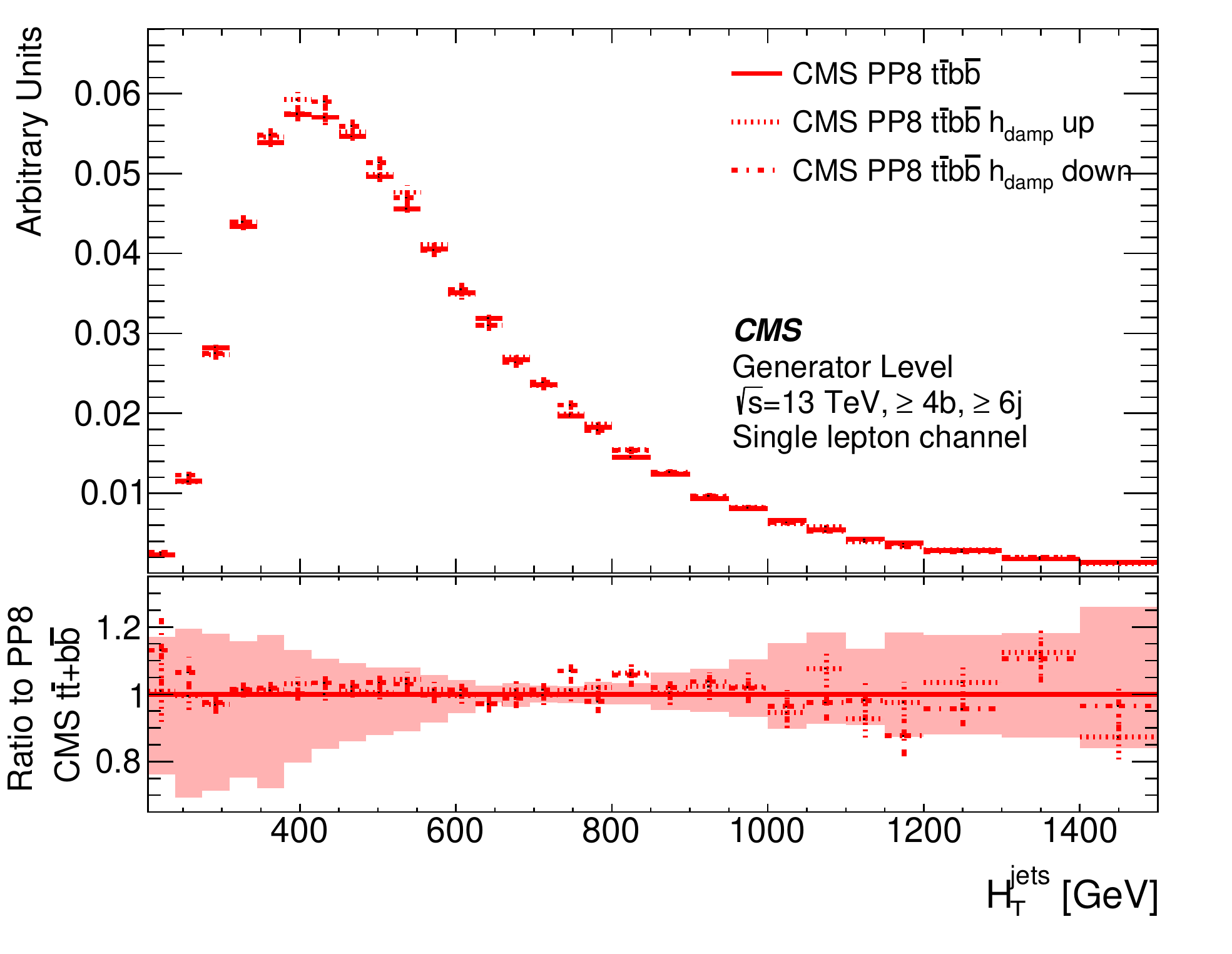}
  \includegraphics[width=0.4\textwidth]{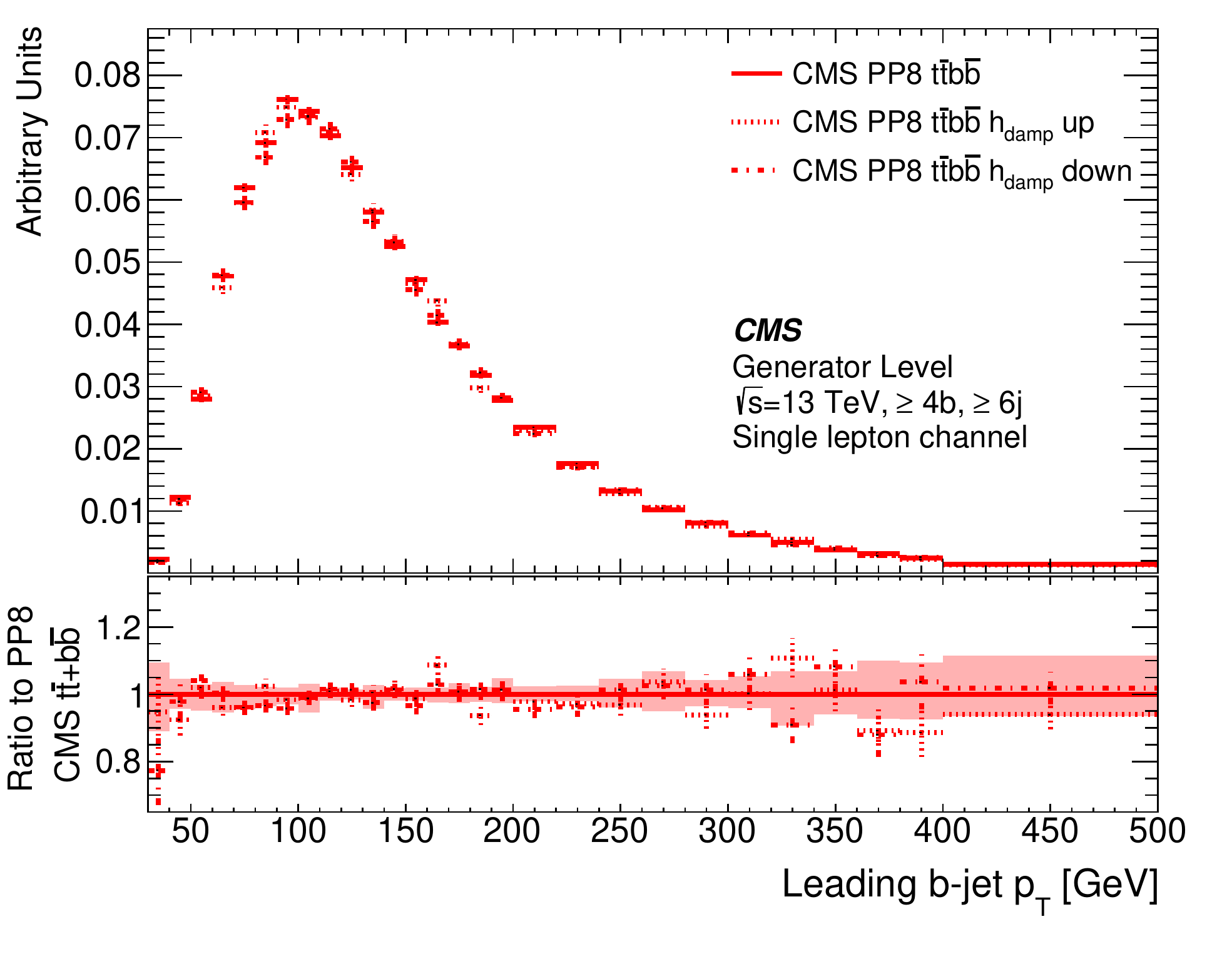}
  \includegraphics[width=0.4\textwidth]{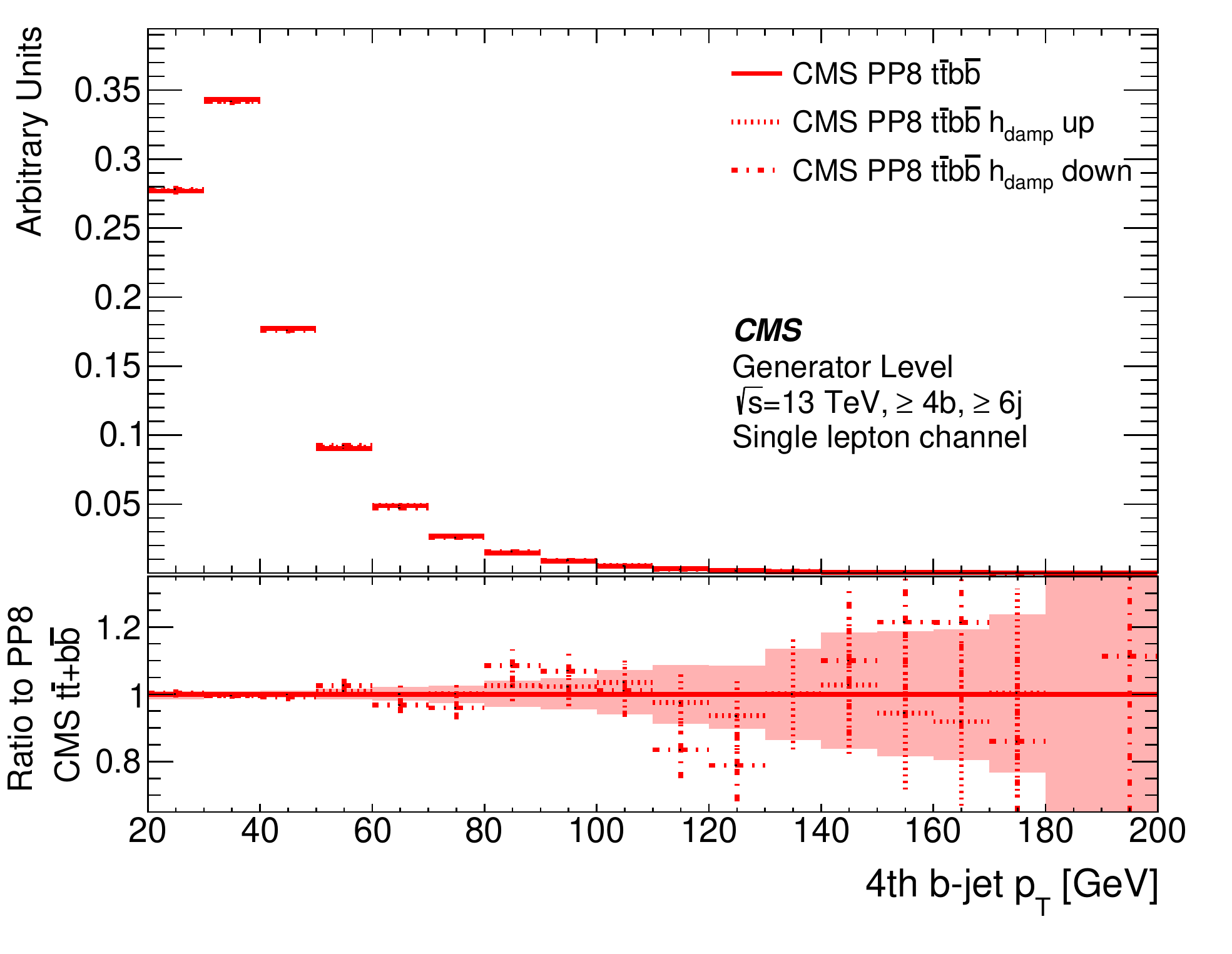}                                                                          
  \caption{
    Comparison of CMS PP8 predictions for \ttbb with different settings of the parameter \hdamp.  
    All predictions are normalised to one.
    The ratio shows the different curves divided by PP8 \ttbb.
    The error band contains the statistical uncertainty  and the scale variations (ME and PS) for the CMS PP8 \ttbb sample. 
    Statistical uncertainties are indicated by vertical lines.
  }\label{fig:comp3}
\end{figure}

In Fig.~\ref{fig:comp3}, the CMS nominal PP8 \ttbb prediction is compared to generator variations potentially considered for the CMS \ttHbb analysis.
The scale uncertainties, which include the scale variations in the ME and the PS, are significantly larger than the differences observed for the different \hdamp variations, except at very low \HT and low leading \bjet \pt where the \hdamp down variations shows up to 20\,\% differences.
Significant statistical fluctuations are observed at regions of low event yields, which are, however, not expected to be relevant for the analysis.

\begin{figure}[!htb]
  \centering
  \includegraphics[width=0.4\textwidth]{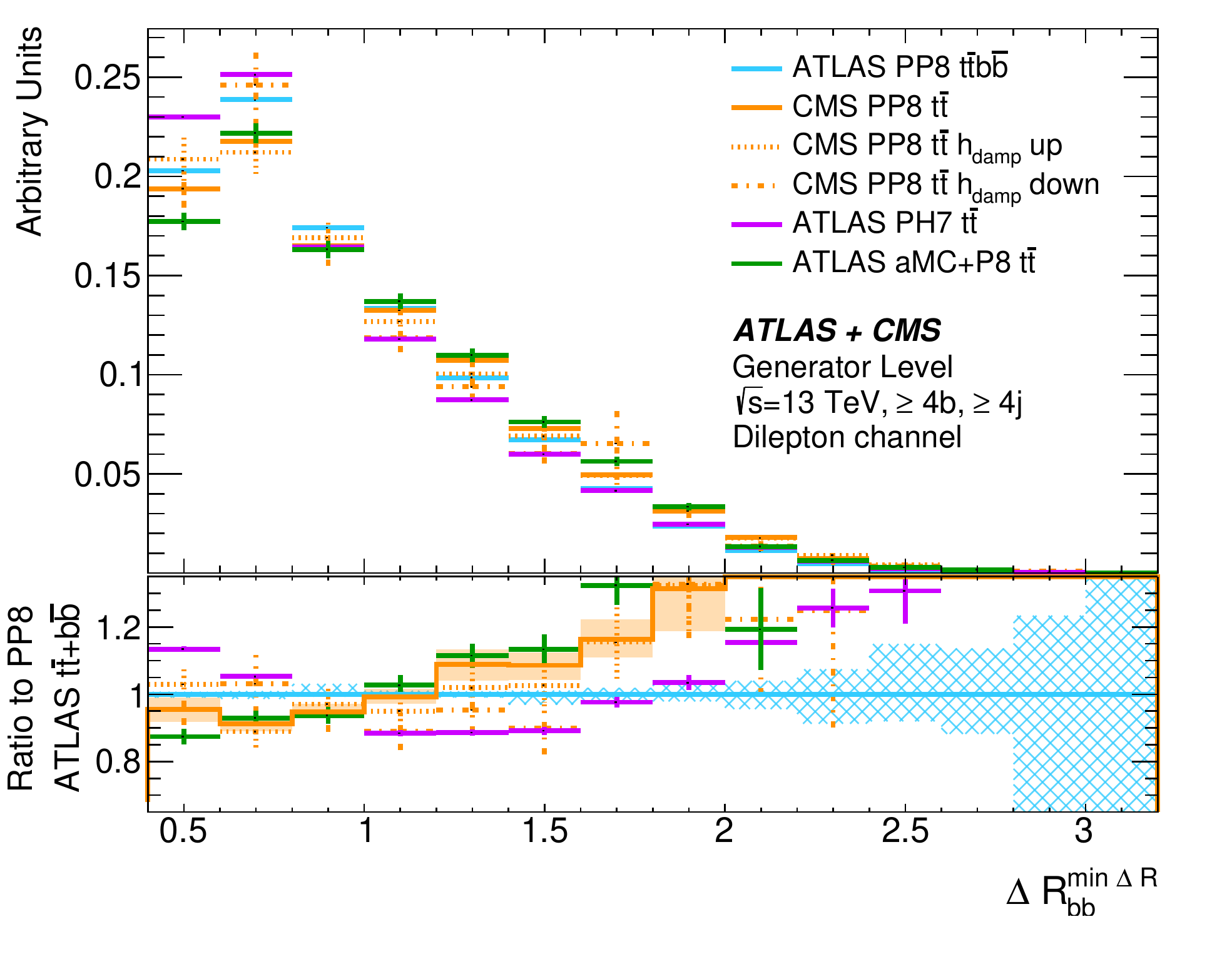}
  \includegraphics[width=0.4\textwidth]{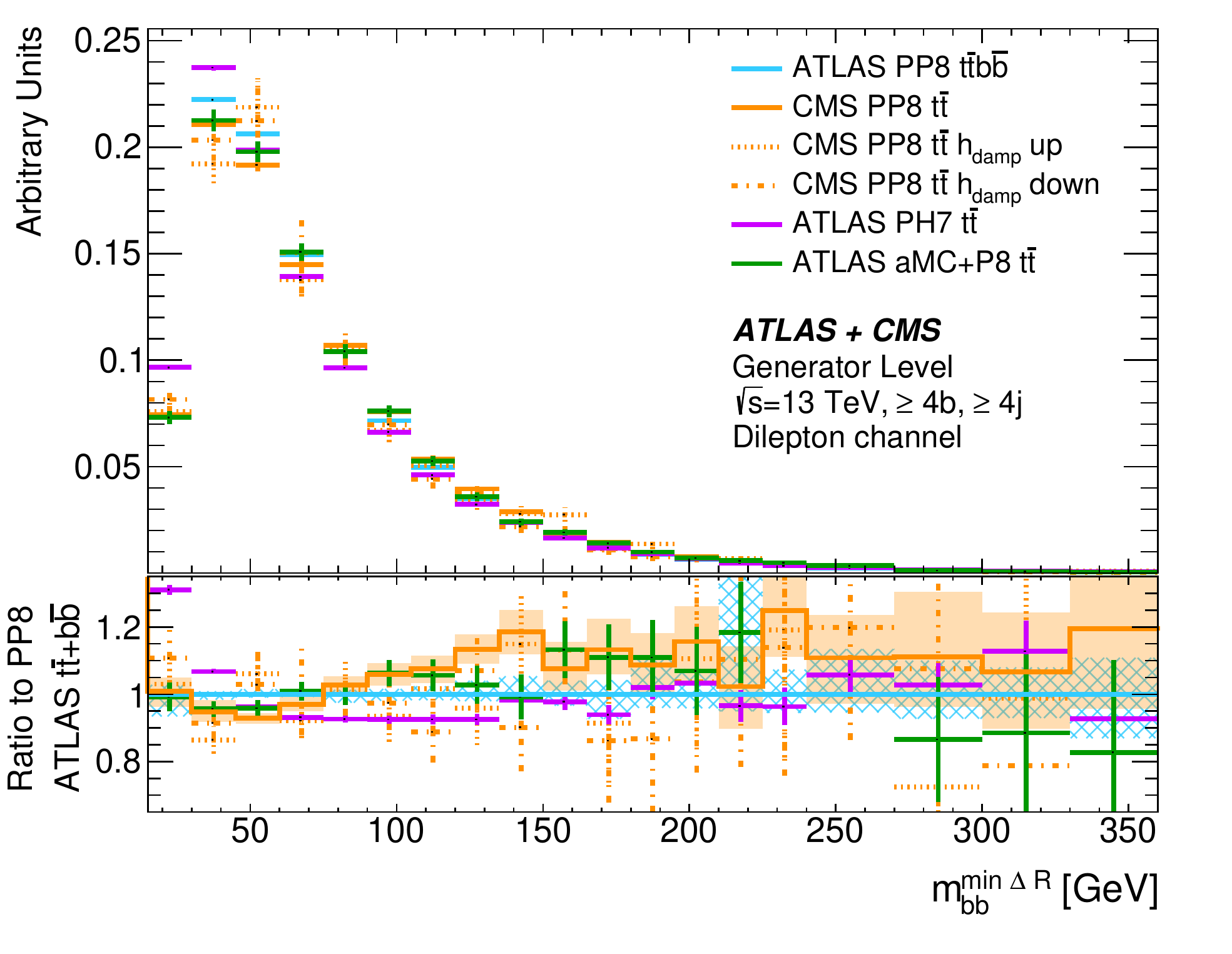}
  \includegraphics[width=0.4\textwidth]{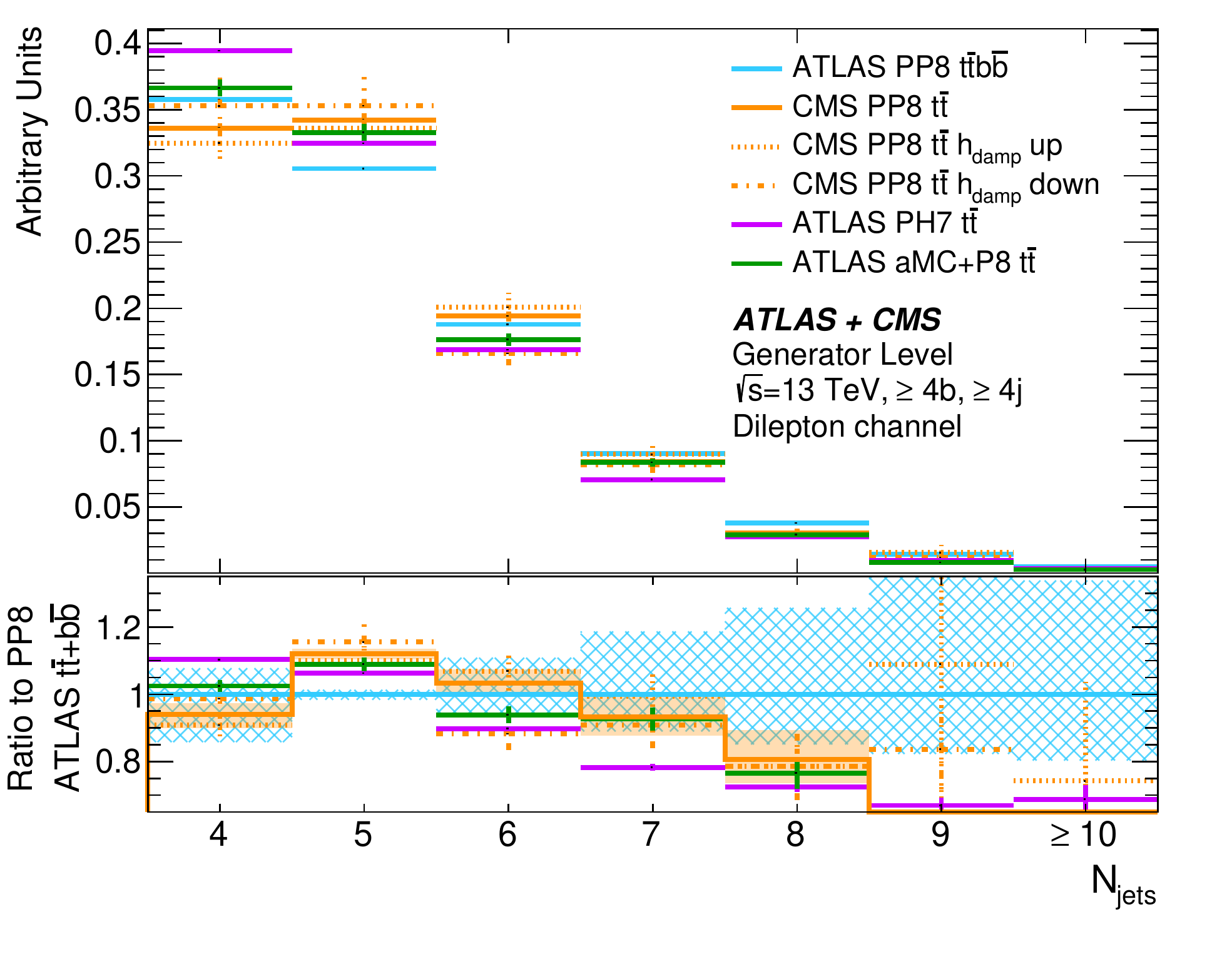}
  \includegraphics[width=0.4\textwidth]{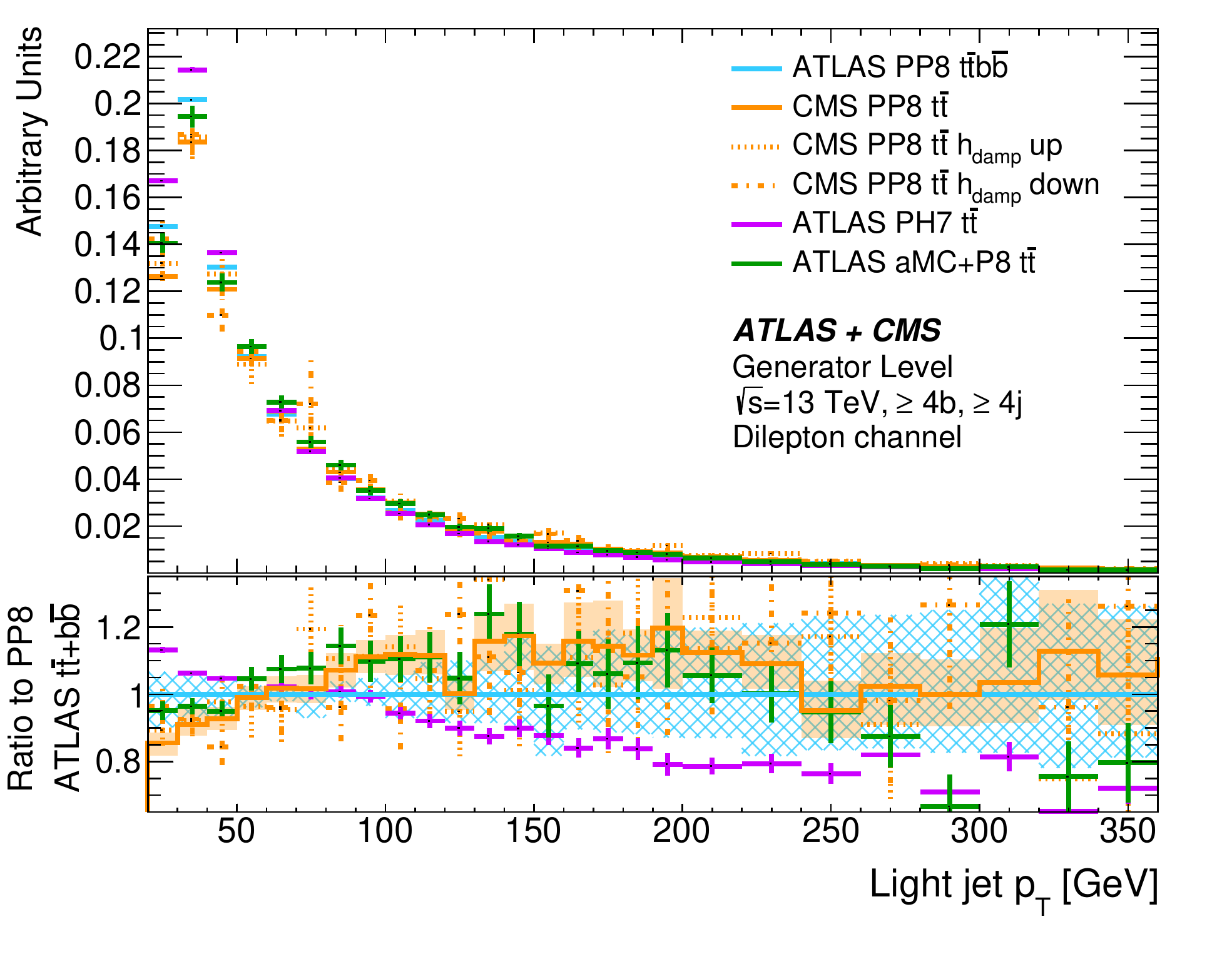}
  
  \includegraphics[width=0.4\textwidth]{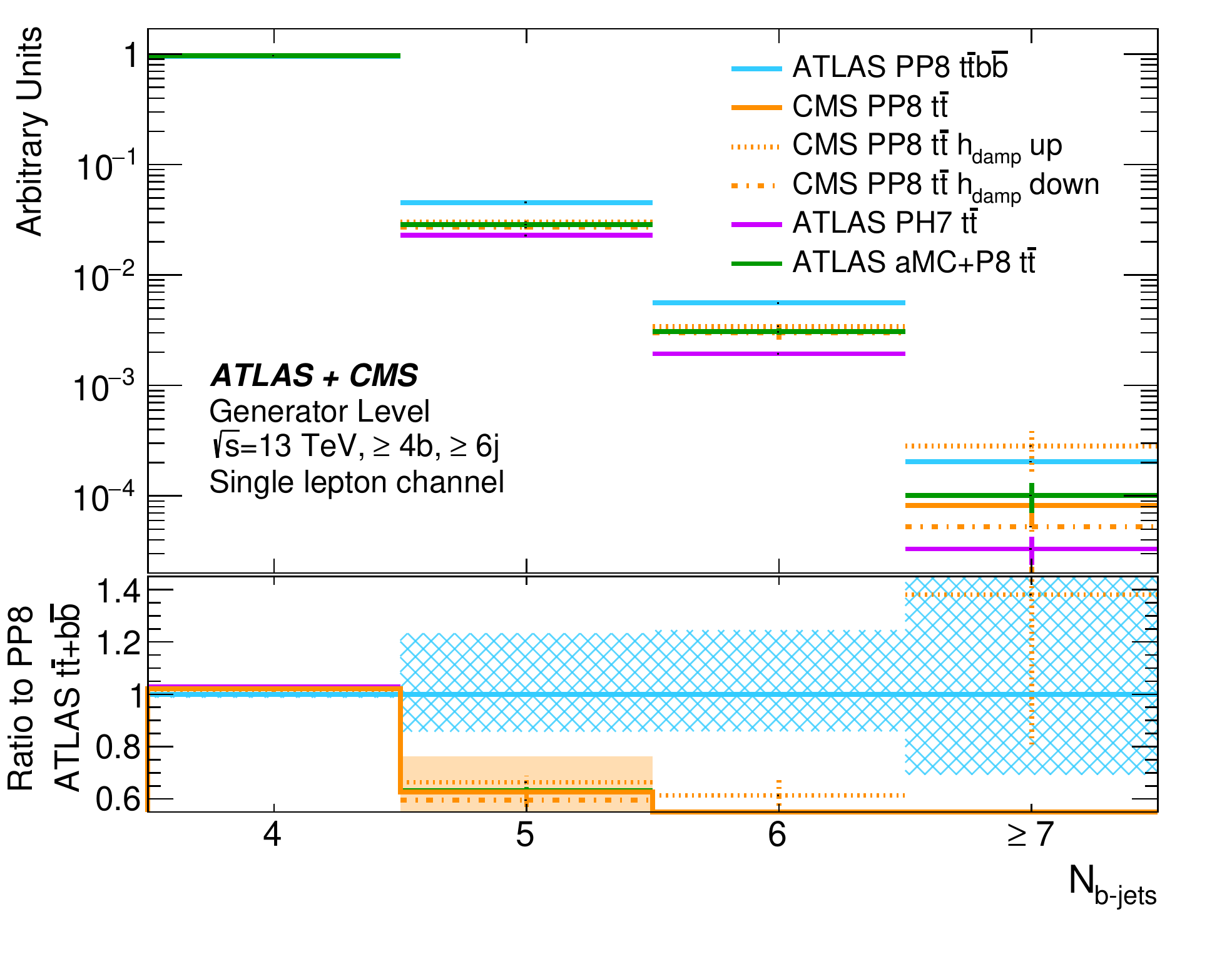}
  \includegraphics[width=0.4\textwidth]{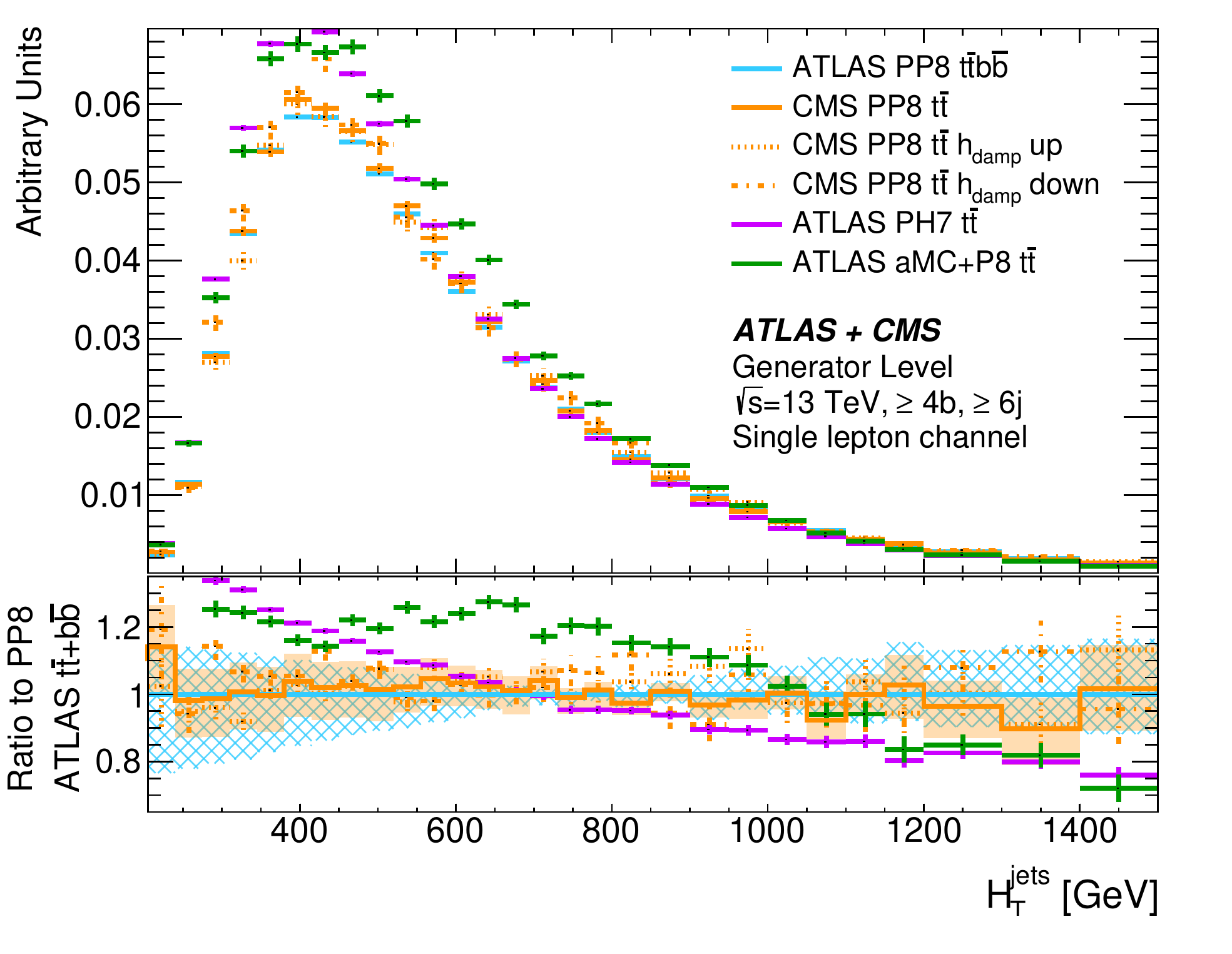}
  \includegraphics[width=0.4\textwidth]{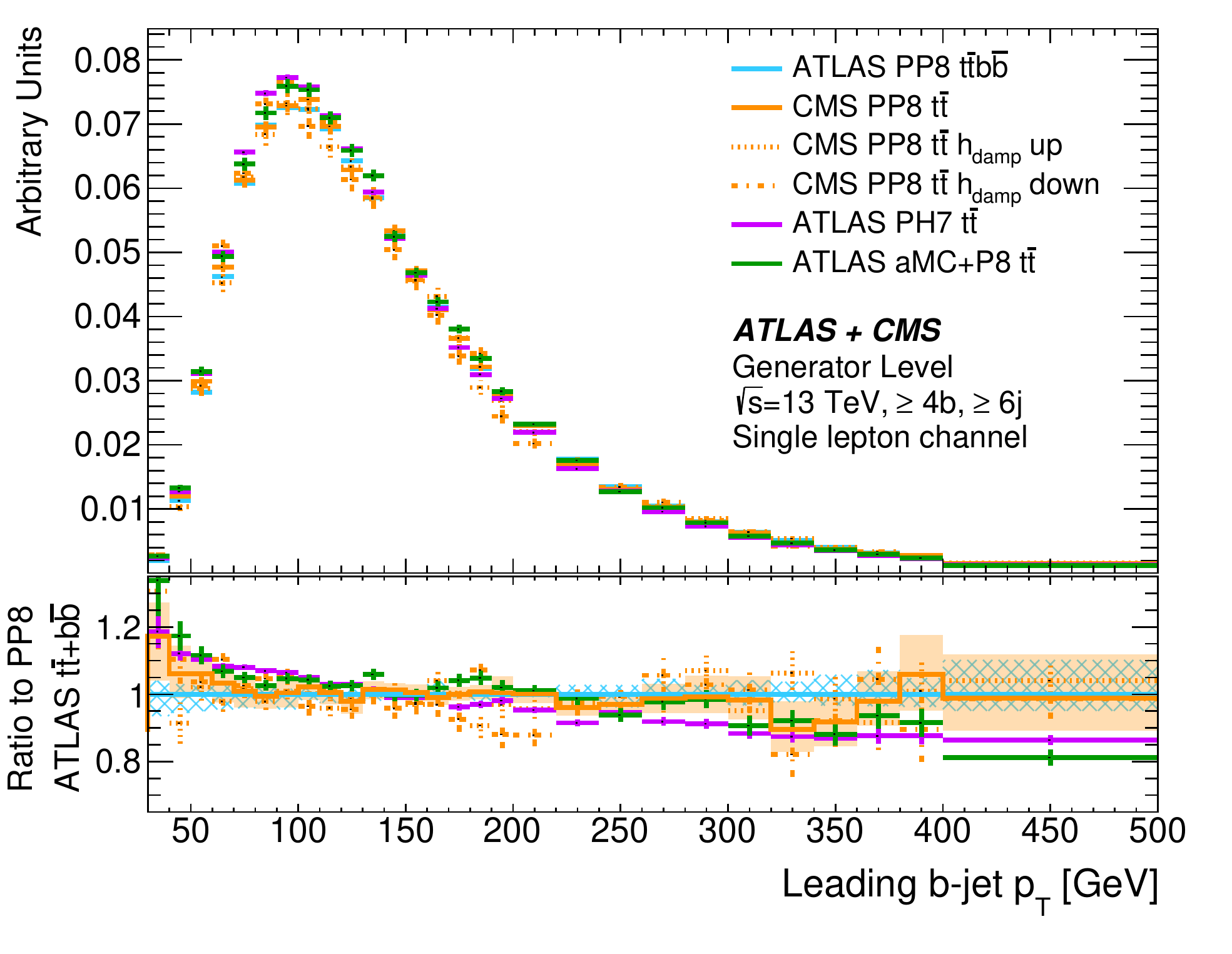}
  \includegraphics[width=0.4\textwidth]{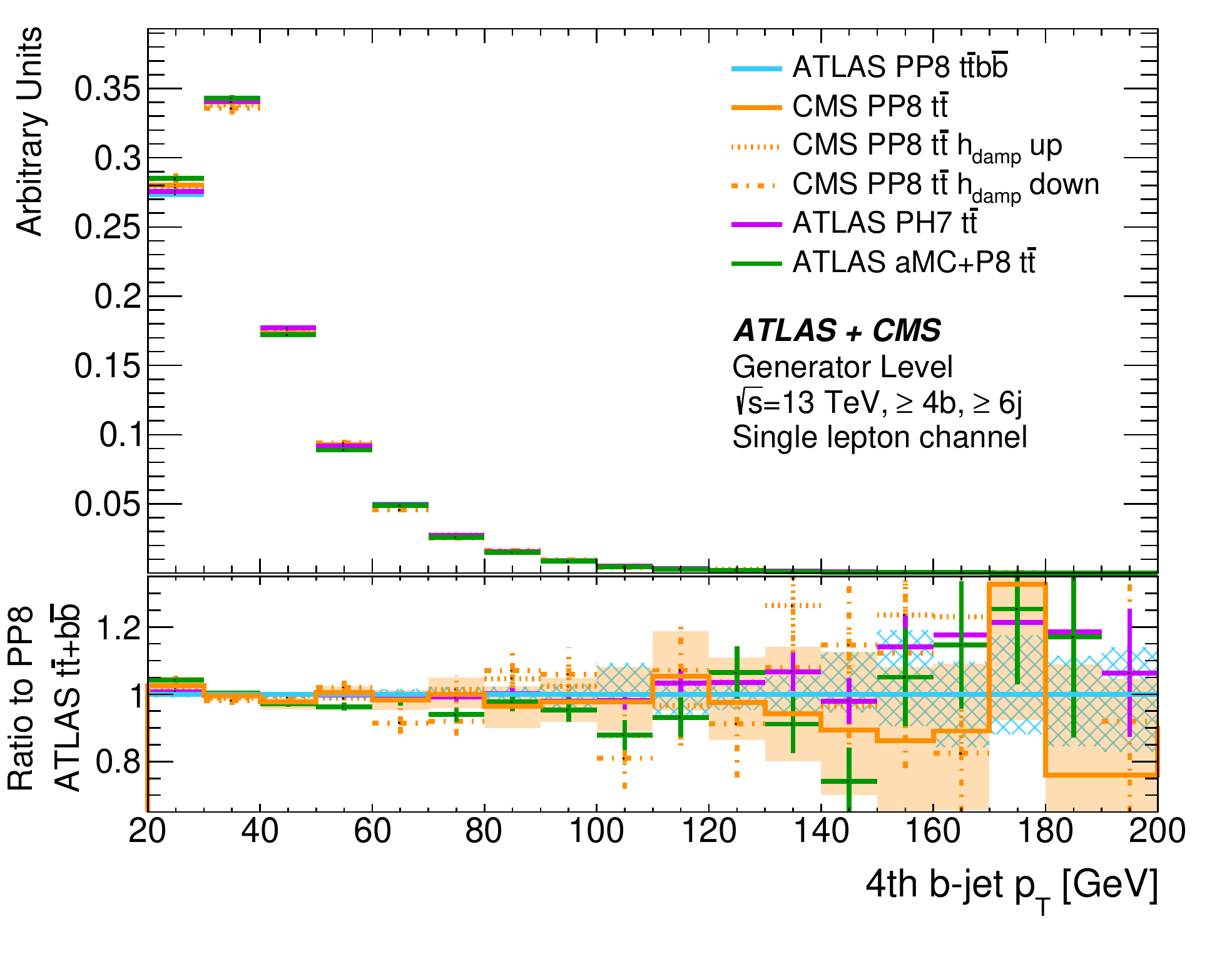}
  \caption{
    Comparison of  predictions used for the systematic uncertainties of the first Run-2 analysis by CMS~\cite{CMS-HIG-17-026} and of the first full Run-2 analysis by ATLAS~\cite{HIGG-2020-23}. 
    All distributions are normalised to one.
    The ratio shows the different curves divided by ATLAS PP8 \ttbb. The error bands are constructed from the statistical uncertainties  and the scale variations (ME and PS) for the ATLAS PP8 \ttbb (blue) and the CMS PP8 \ttbb (red) samples.
    Statistical uncertainties are indicated by vertical lines.
  }\label{ttbb:comp4}
\end{figure}

Finally, Fig.~\ref{ttbb:comp4} shows the distributions used to estimate the systematic modelling uncertainties of the first Run-2 analysis by CMS~\cite{CMS-HIG-17-026} and of the first full Run-2 analysis by ATLAS~\cite{HIGG-2020-23}.
In addition to the scale and PS \alphas variations, the uncertainty on the \ttbb PP8 prediction is estimated in case of ATLAS by assigning the relative difference between PP8 \ttbar and alternative \ttbar predictions listed in Table\,\ref{tab:ttbbsamples} to the \ttbb prediction, and in case of CMS by the \hdamp variations, where also a cross check with \sherpa \ttbar has been made but was not included in the fit.
Due to displaying purposes, the ATLAS PP8 \ttbar prediction, which is very similar to the CMS \ttbar prediction as demonstrated in Fig.~\ref{fig:comp1}, is not shown.

\subsection{Conclusions}
Comparisons of generator predictions used by ATLAS and CMS in a typical phase space of the \ttHbb measurement were presented.
Two sets are used for comparison: the generators used in the most recent published analyses involving \ttbar inclusive predictions based on 5FS PDFs to estimate uncertainties and the set of generators in the future effort using \ttbb calculations at NLO based on 4FS PDFs.

The difference between the predictions exceeds the uncertainties from the scale variations both for the uncertainties considered in the published \ttHbb analysis and for the future analyses. 
The uncertainties due to the choice of PS and NLO generator are  reduced when estimating them based on \ttbb ME  predictions compared to the previously used \ttbar ME matched predictions.

Despite differences in the set-ups between the experiments  for the nominal PP8 \ttbb generator, only small differences are observed in the predictions. 
However, the considerations of the modelling uncertainties differs significantly:  CMS considers  inherent variations of the chosen model as uncertainty, while ATLAS studied  inherent variations and  differences obtained with alternative generator choices and the latter dominates the uncertainties. Scale variations are applied by both experiments, however the details of the estimates differ between ATLAS and CMS in the published analysis but the effect of the different treatment are not yet studied for the future analyses. 

The presented studies shall  be used as input to discussions between the experiments and theorists to define theory uncertainties for future combinations of  ATLAS and CMS.
\clearpage
\section{Comparisons of Monte Carlo predictions for the $t\bar{t}W$ process}
\label{sec:ttW}


The  ATLAS\,\cite{ATLAS-CONF-2019-045} and CMS\,\cite{CMS-HIG-19-008} experiments  measured the \ttH production cross section in multi-lepton final states,
 which are primarily sensitive to the decays of $H\rightarrow WW^*$, $H\rightarrow \tau \tau$ and $H\rightarrow ZZ^*$. The dominant background in these measurements stems from \ttW production. These measurements along with the recent CMS measurement of \ttW production~\cite{CMS-PAS-TOP-21-011} show some tension with the SM \ttW predictions which were used to calculate the inclusive cross section  and the acceptance  in the analysis phase space.

Different nominal MC predictions were used by the experiments for these measurements, ATLAS used \sherpa~2.2.1~\cite{Sherpa} and CMS used \amcnp~2.4.2 matched to \pythiaa using the FxFx merging scheme\,\cite{Frederix:2012ps} and including sub-leading electroweak (EW) corrections of the order $\alpha_s \alpha^3$  where $\alpha$ ($\alphas$) refers to the EW (QCD) coupling constant. The experiments applied different corrections  to predict the theoretical inclusive \ttW cross section that entered the calculation of the scale factor  to data, resulting in a value of   \SI{727}{\femto\barn}  for ATLAS\,\cite{ATLAS-CONF-2019-045}  and \SI{650}{\femto\barn} for CMS\,\cite{CMS-HIG-19-008}. Both experiments estimate the uncertainty of the MC prediction related to missing higher order corrections by varying the renormalisation and factorisation scales in the ME. However, ATLAS considers additionally  uncertainties associated with the modelling of additional QCD radiation  by comparing the nominal \ttW prediction with that of \amc as alternative MC generator differing in particular in  the number of additional partons in the ME calculation, the  parton shower and merging algorithm.


In recent times there have been significant theoretical developments
in \ttW modelling despite the challenges associated with calculations
of \ttW with higher order corrections in the QCD, \alphas, and EWK,
$\alpha$, couplings. Even at LO in \alphas, complications
arise because \ttW is a $q\overline{q}$-initiated process in which the radiation of the
$W$-boson from one of the initial state quarks polarises the incoming
quark, making spin correlations all the more important~\cite{Maltoni:2014zpa}.
Initial calculations of \ttW production at next-to-leading order (NLO)
in QCD at fixed order~\cite{Campbell:2012dh} and later matched to a parton
shower~\cite{Garzelli:2012bn,Maltoni:2015ena} were later augmented with NLO
EWK corrections (of order $\alpha^2\alpha_s^2$)~\cite{Frixione:2015zaa} to provide the higher
order cross sections used across the LHC programme for a number of
years~\cite{YR4}. Furthermore, full NLO calculations including fixed-order corrections matched to parton shower in the POWHEG-BOX framework and accounting for LO spin-correlation of decay product have 
recently been provided in \cite{FebresCordero:2021kcc}.

Since then there  has been significant theoretical progress in
calculating more complex and precise predictions. Higher order QCD
corrections including \ttW production with additional partons open
gluon-initiated production modes with significant contributions to the
total cross section. Recent studies show that these contributions also
have large next-to-leading order (NLO)
corrections~\cite{Alwall:2014hca} and that \ttW$jj$ can be
large~\cite{vonBuddenbrock:2020ter}, both of which require NLO-merged
calculations~\cite{Frederix:2021agh} for such effects to be properly
included. Furthermore, beyond the
traditionally ``leading'' NLO EWK corrections (of order
$\alpha^2\alpha_s^2$) there are even larger contributions from
traditionally ``sub-leading'' NLO corrections (of order
$\alpha^3\alpha_s$)~\cite{Dror:2015nkp,Frederix:2017wme,FebresCordero:2021kcc}
due to the existence of $tW$ scattering contributions embedded in to
the \ttW{}$j$ process.
Calculations at NLO in QCD accounting for next-to-next-to-leading
logarithmic effects (NNLL) are also available~\cite{Kulesza:2018tqz} as well
as recent predictions at NLO+NNLL in QCD also with NLO EWK
corrections~\cite{Broggio:2019ewu,Kulesza:2020nfh}.
Full off-shell calculations at NLO in
QCD~\cite{Bevilacqua:2020pzy,Denner:2020hgg,Bevilacqua:2020srb} are
also now available and more recently the NLO EWK corrections have also
been incorporated~\cite{Denner:2021hqi} into these calculations, along
with the development of procedures to apply the off-shell corrections
to NLO+PS setups~\cite{Bevilacqua:2021tzp}.

A first attempt to formulate an uncertainty estimate in view of these  theoretical predictions has been made in \cite{FebresCordero:2021kcc} where different generator codes at NLO QCD are compared with fixed order calculations to demonstrate that a robust theoretical prediction of  hadronic \ttW production cannot be expressed as a simple recipe covering the specifics of all experimental observables. Therefore the value of comparing several well tested tools is emphasised.

For future analyses, updated MC models will be used and the estimate of systematic uncertainty is under development. In particular, ATLAS  is considering  \sherpa predictions including several higher order EW corrections in addition to the predictions at NLO in the strong coupling, namely of the order $\alpha^3$, $\alpha^{2}\alphas^{2}$  and  $\alpha^{3}\alphas$. Furthermore, calculations of \amc employing the FxFx merging scheme will be considered. For inclusive predictions,  \powheg predictions \cite{FebresCordero:2021kcc} are also considered. CMS will continue to use \amc with the FxFx merging scheme including subleading EW corrections however the EW corrections are not included in the present document in order to facilitate the comparison between the setups used by each experiment. The samples will be described in the following and an overview with detailed information on the samples is given in Table~\ref{tab:mcconfig}. The use of other theoretical developments, already outlined, will also be considered in future but are beyond the scope of this document.

 Comparisons are performed  using stable final-state particles in a fiducial phase space similar to the experimental measurements in the two same-sign leptons (2lSS) channel as implemented in a dedicated routine in  the \rivet analysis toolkit~\cite{rivet}. Two sets of distributions are presented, one
where the histograms are normalised to unit area to asses shape differences in the differential distributions and another set where the generator cross sections are set to  \SI{600.8}{\femto\barn} the value reported in Ref.\,\cite{YR4}. This allows to study differences in acceptance for the different generator predictions.

  The chapter is organised as follows: Section\,\ref{sec:ttW:mcsetup} gives the detailed set-up for the generator samples, Section\,\ref{sec:ttW:ttw_fid} describes the object reconstruction and event selection, Section\,\ref{sec:ttW:results} gives the two sets of results and finally conclusions are drawn in Section\,\ref{sec:ttW:conclusion}.

\subsection{MC generator set-ups}\label{sec:ttW:mcsetup}
This chapter describes in detail the set-up of the MC generator set-ups used  for the ATLAS and CMS samples.

\subsubsection*{ATLAS setup}

The nominal sample for the comparison of this note  was generated using the \sherpa~2.2.10~\cite{Sherpa,Gleisberg:2008ta} generator with the NNPDF3.0 NLO PDF set.
The \ttW matrix element was calculated for up to one additional parton at NLO and up to two partons at leading order (LO) accuracy using
\textsc{Comix}~\cite{Gleisberg:2008fv} and \textsc{OpenLoops}~\cite{Cascioli:2011va}, and merged with the \sherpa parton shower~\cite{Schumann:2007mg} using the MEPs@NLO prescription~\cite{Hoeche:2012yf} with a merging scale of \SI{30}{\GeV}.
The choice of renormalisation and factorisation scales of the core process  is $\muR = \muF = \HT/2$, where $\HT$ is defined as the scalar sum of the transverse masses $\sqrt{\pt^2+m^2}$ of all final state particles.
Systematic uncertainties due to missing higher-order QCD corrections are estimated in the nominal sample by varying the factorisation and renormalisation scales together with  \alphas in the parton shower  by a factor of 0.5 (2.0) with respect to the central value.

In addition to this nominal prediction at NLO in the strong coupling, a separate sample is produced which contains also higher order corrections relating to EW contributions. These are added in two ways. First, event-by-event correction factors
are applied that provide virtual NLO EW corrections of the order $\alpha^{2}\alphas^{2}$ derived using the formalism described in Ref.~\cite{Kallweit:2015dum} along with LO corrections of order $\alpha^3$, both are implemented using the prescription outlined in Refs.~\cite{Sherpa,Gutschow:2018tuk}.
  Second, sub-leading EW corrections at order $\alpha^{3}\alphas$\,\cite{Frederix:2017wme}
are partially accounted for (only the real emission contribution) via the addition of an independent \sherpa~2.2.10 sample produced at LO in QCD for this final state. This sample is marked as ``QCD+EW'' in the following.


Alternative \ttW predictions are produced  using the \amcnp~2.3.3 program to generate \ttW production with up to one additional parton  in the final state at NLO accuracy in the strong coupling. The renormalisation and factorisation scales are the same as in the nominal sample. Another sample is generated using \amcnp~2.9.3  for up to one additional parton at NLO accuracy and up to two additional partons at LO accuracy in the ME and merging the different jet multiplicities  using the FxFx NLO matrix-element and parton-shower merging prescription~\cite{Frederix:2012ps}, see detailed description in \cite{atlasFXFXPub}. As part of the FxFx merging algorithm,  scales are dynamically chosen  and set to the characteristic scale of the hard process. In both samples, spin correlation effects between the ME decay products are accounted for by Madspin~\cite{Artoisenet:2012st} and the showering and subsequent hadronization is performed using \pythia~8.210 and \pythia~8.245~\cite{Sjostrand:2014zea}, respectively, with the A14 tune~\cite{ATL-PHYS-PUB-2014-021}. These samples are referred to as ``ATLAS MG5\_aMC+Py8'' and ``ATLAS MG5\_aMC+Py8 FxFx'' in the following.

\subsubsection*{CMS setup}
CMS simulates proton-proton to $\ttbar\ell\nu$ processes at NLO accuracy in the matrix element calculation using \amcnp~2.4.2. Spin correlation effects between the ME decay products are accounted for by Madspin~\cite{Artoisenet:2012st}. The ME calculation includes diagrams with up to one additional parton at NLO and any further partons are generated by the parton shower. The renormalisation and factorisation scales are set to the characteristic scale of the hard process. They are chosen dynamically and are dependent kinematics of the event after the FxFx merging prescription\footnote{see in particular section 2.2.3
  of Ref.~\cite{Frederix:2012ps} where elements of Refs.~\cite{Catani:2001cc,Hamilton:2012np} are taken into account}.

Theoretical uncertainties associated with missing higher-order QCD corrections from the ME calculation are estimated by varying the renormalisation and factorisation scale by a factor of 0.5 and 2.0. All possible combinations of these variations, implemented using a dedicated set of per-event weights, are then used to construct the uncertainty envelope.

The parton shower, hadronization processes and decays of $\tau$\, leptons (including polarisation effects) are modelled using \pythia~8.226 with the CP5 tune. The samples is called ``CMS MG5\_aMC+Py8 FxFx'' in the following.


\newpage

\begin{sidewaystable}
\footnotesize{
\begin{center}
\caption{\label{tab:mcconfig} The configurations used for the event generation of the \ttW processes. Scale settings given in terms of $\HT = \sum_{i=0}^{N}\sqrt{p_{T,i}^2+m_i^2}$, where N corresponds to the number of final state particles.}
\vspace{0.25cm}{
\setlength\tabcolsep{6.pt}
\renewcommand{\arraystretch}{1.5}
\begin{tabular}{|l|l|l|l|l|l|}
\hline\hline
\bf{Label} & ATLAS Sherpa 2.2.10  & ATLAS Sherpa 2.2.10  & ATLAS MG5\_aMC+Py8 FxFx & ATLAS MG5\_aMC+Py8 & CMS MG5\_aMC+Py8 FxFx \\
 & & QCD+EW & & & \\ \hline
\bf{Process} & \ttW inclusive & \ttW inclusive & \ttW inclusive & \ttW inclusive & $\ttbar\ell\nu$   (\ttW inclusive) \\ \hline
\bf{Generator} & \sherpa~2.2.10\,\cite{Sherpa} & \sherpa~2.2.10\,\cite{Sherpa} & \amcnp~2.9.3\,\cite{Alwall:2006yp} & \amcnp~2.3.3\,\cite{Alwall_2014}  & \amcnp~2.4.2 \\ \hline
\bf{order of QCD ME } & 0,1 $j$@NLO\footnote{\label{note1}In addition to the implicit 2$j$@LO contribution from the real emission part of the 1$j$@NLO calculation, Sherpa adds the 2$j$@LO as an explicit separate process within the merging such that the ME is supplemented with higher-order improvements such as the CKKW scale choice and Sudakov factors."} & 0,1 $j$@NLO\footnoteref{note1} & 0,1 $j$@NLO  & NLO & 0,1  $j$@NLO \\ \hline
\bf{ME or core scale} & $\muR=\muF=\HT/2$
 & $\muR=\muF=\HT/2$ & dynamic scale  choice \cite{Frederix:2012ps,Catani:2001cc,Hamilton:2012np} & $\muR=\muF=\HT/2$ &  dynamic scale  choice \cite{Frederix:2012ps,Catani:2001cc,Hamilton:2012np}  \\ \hline
\bf{order of EW corr.} & - & $\alpha^3$,  $\alpha^{2}\alphas^{2}$, $\alpha^3\alphas$ & - & - & - \\ \hline
\bf{Parton Shower} & \sherpa~2.2.10 & \sherpa~2.2.10 & \pythia~8.245\,\cite{Sjostrand:2014zea} & \pythia~8.210\,\cite{Sjostrand:2014zea} & \pythia~8.226  \\ \hline
\bf{Merging Scheme} & MEPs@NLO\,\cite{Hoeche:2012yf} & MEPs@NLO\,\cite{Hoeche:2012yf} & FxFx\,\cite{Frederix:2012ps} & -  & FxFx \\ \hline
\bf{Merging Scale} & \SI{30}{\GeV} & \SI{30}{\GeV} & \SI{30}{\GeV} & - & \SI{42}{\GeV} \\ \hline
\bf{PDF} & NNPDF3.0 NNLO\,\cite{NNPDF:2014otw} & NNPDF3.0 NNLO & NNPDF3.0 NLO & NNPDF3.0 NLO & NNPDF3.1 NLO\,\cite{NNPDF:2017mvq}  \\ \hline
\bf{Tune} & \sherpa default & \sherpa default & A14\,\cite{ATL-PHYS-PUB-2014-021} & A14  & CP5\,\cite{Sirunyan:2019dfx}  \\ \hline
\bf{Cross section}\footnote{ $\sigma_{\text{tot}}$=600.8~fb from YR4 is used for all samples in the generator comparisons in section\,\ref{sec:ttW:ttw_gen} except for \sherpa QCD+EW } & \SI{597}{\femto\barn} & \SI{615}{\femto\barn} & \SI{613}{\femto\barn} & \SI{548}{\femto\barn} & \SI{220}{\femto\barn} (666\,fb\footnote{calculated from $\ttbar\ell\nu$ as 0.2198 x (1/ (3 x 0.11) )}) \\ \hline
\hline
\end{tabular}
}
\end{center}
}
\end{sidewaystable}


\subsection{Object reconstruction, fiducial volume and observables}
\label{sec:ttW:ttw_fid}
Object and event selection is defined at stable particle-level that closely matches the detector-level described in reference~\cite{ATLAS-CONF-2019-045} (ATLAS) and~\cite{CMS-HIG-19-008} (CMS). 
Jets are reconstructed from all stable final state particles with a mean lifetime of $\tau\gr\SI{3e-11}{\second}$ (but excluding leptons and neutrinos from the top quark decay chain), using the anti-$k_t$ algorithm with a radius parameter of $R=0.4$.
Jets are required to satisfy $\pt\gr\SI{25}{\GeV}$ and $|\eta|\less2.5$.
Jets that are matched to a $b$-hadron\footnote{no \pt cut is applied} by ghost matching~\cite{Cacciari:2008gn} are referred to as $b$-jets.
Electrons and muons, referred to as light leptons $\ell$, are required to be separated from selected jets by $\Delta R\gr0.4$ and are otherwise removed.
Hadronically decaying $\tau$ leptons are required to satisfy $\pt\gr\SI{25}{\GeV}$ and $|\eta|\less2.5$.
Events are selected with exactly two light leptons. The four-momentum of the bare leptons from top quark decay are modified (”dressed”) by adding the four-momenta of all radiated photons within a cone of size $\Delta R = 0.1$.
Leptons are required to have $|\eta|\less2.5$ and $\pt\gr25(20)\,\si{\GeV}$ for leading $\ell_0$ (subleading $\ell_1$) lepton (\pt ordered).
Leptons are required to have same charge, targeting the semi-leptonic \ttbar decay and leptonic $W$ decay.

Events with at least 3 jets and at least one of them being a $b$-jet are considered in the fiducial volume. The object definition and event selection is summarised in Tables\,\ref{tab:ttWobjects} and \ref{tab:ttWeventselection}.
These are then split into five regions, categorized by the number of jets of any flavour (three or  $\geq$4), $N_{b-\text{jets}}$ (one or $\geq$2) as well as the presence of hadronically decaying $\tau$ lepton, as summarised in Table~\ref{tab:ttWregions}.

\begin{table}
\begin{center}
\caption{\label{tab:ttWobjects}
The object reconstruction used in the \rivet analysis of the \ttW processes. Leptons are ordered in \pt.}
\vspace{0.25cm}
{\small
\setlength\tabcolsep{1.5pt}
\begin{tabular}{|l|l|}
\hline
\textbf{Object  } & \textbf{reconstruction and selection}\\
\hline
\hline
jets &  stable final state particles with anti-k$_t$ algorithm,  radius R = 0.4 \\
 & prompt "dressed" leptons and neutrinos are vetoed from jet \\
 & $\pt\gr\SI{25}{\GeV}$ and $|\eta|\less2.5$ \\
 $b$-jets &  jets ghost matched to $B$-hadrons  \\
&  $\pt\gr\SI{25}{\GeV}$ and $|\eta|\less2.5$ \\
\hline
light leptons (electrons and muons) & dressed with photons within $\Delta R < 0.1$ \\
& $|\eta|\less2.5$ and $\pt\gr25(20)\,\si{\GeV}$ for leading  (subleading) lepton\\
overlap removal & remove light lepton if $ \Delta R(\text{jet}, \text{lepton}) < 0.4 $\\
hadronicaly decaying $\tau$ leptons (before decay) &  $\pt\gr\SI{25}{\GeV}$ and $|\eta|\less2.5$ \\

\hline
\end{tabular}
}
\end{center}
\end{table}

\begin{table}
\begin{center}
\caption{\label{tab:ttWeventselection}
The event selection used in the \rivet analysis for the \ttW processes. $N_{\text{jets}}$ refers to all jets independent of jet flavour, i.e.\ $b$-jets are included.}
\vspace{0.25cm}
{\small
\begin{tabular}{l}
\hline
\hline
\textbf{Event selection for 2$\ell$SS}\\
\hline
exactly 2 leptons with same charge  \\
$N_{\text{jets}}\geq$3 \\
 $N_{b-\text{jets}}\geq$1 \\
\hline
\hline
\end{tabular}
}
\end{center}
\end{table}

\begin{table}
\begin{center}
\caption{\label{tab:ttWregions}
The  region definitions used in the \rivet analysis for the \ttW processes.}
\vspace{0.25cm}
{\small
\begin{tabular}{l|l}
\hline
\hline
\textbf{Region} & \textbf{Selection} \\ \hline
1  & $N_{b-\text{jets}}=$1, $N_{\text{jets}}\geq$4, 0-$\tau_{\text{had}}$ \\
2 & $N_{b-\text{jets}}\geq$2,   $N_{\text{jets}}\geq$4, 0-$\tau_{\text{had}}$ \\
3 & $N_{b-\text{jets}}=$1,  $N_{\text{jets}}$=3, 0-$\tau_{\text{had}}$ \\
4 & $N_{b-\text{jets}}\geq$2, $N_{\text{jets}}$=3, 0-$\tau_{\text{had}}$ \\
5 & $N_{b-\text{jets}}\geq$1, $N_{\text{jets}}\geq$3, 1-$\tau_{\text{had}}$\\
\hline
\hline
\end{tabular}
}
\end{center}
\end{table}

The definitions of the regions are motivated by the $t\bar{t}H$ multi-lepton analysis strategy.
Regions 1 and 2 corresponds to the signal regions\footnote{slightly different then in Ref.~\cite{ATLAS-CONF-2019-045}, in order to define a common selection with the CMS Collaboration.} and Regions 3 and 4 are used as control regions in the 2$\ell$ same-sign  0-$\tau_{\text{had}}$ \ttH channel.
Definition of Region 5 is closely followed\footnote{requirement on jet multiplicity is relaxed.} by the selections in the 2$\ell$ same-sign 1-$\tau_{\text{had}}$ \ttH channel.



\begin{table}[]
\begin{center}
\caption{\label{tab:ttw_varlist}
List of the observables for the comparison of  \ttW predictions. Leptons  and $b$-jets are ordered in \pt. }
\vspace{0.25cm}
{\small
\setlength\tabcolsep{1.5pt}
\begin{tabular}{l|l|c}
\hline\hline
Variable & Description & Regions \\ \hline
$N_{\text{jets}}$   &     Jet multiplicity        &      1,2,5   \\ 
 $N_{b-\text{jets}}$       &     Number of  $b$-jets       &   1,2,5      \\ 
$\HT^{\text{jets}}$      & Scalar sum of transverse momentum of all jets in the event         &  1,2,3,4       \\ 
$\pt^{b0}$       &      Leading $b$-jet transverse momentum       &   1,2      \\ 
 $p_T^{\ell 0}$      &   Leading lepton transverse momentum           &     1,2,5    \\ 
$\Delta R _{ \ell 0\text{jets} }$      &     Minimum angular separation between the leading lepton $\ell 0$ and the nearest jet         &  1,2       \\ 
$\Delta R _{\ell 0\ell1 }$      &       Angular distance between the two leptons      &    1,2,5    \\ 
$max |\eta _{\ell}|$      &    Value of the highest lepton's pseudorapidity in the event        &     1,2    \\ 
    \hline\hline
\end{tabular}
}
\end{center}
\end{table}
The list of variables for the comparison of the \ttW generators presented in this note are summarised in Table~\ref{tab:ttw_varlist}.
\clearpage
\subsection{Results}\label{sec:ttW:results}

The samples described in Table~\ref{tab:mcconfig} are compared in the following.
The ratio plots show the ratios of the all MC samples with respect to ATLAS Sherpa~2.2.10, the shaded band represents scale variations.  
The same set of distributions are presented twice with different focus: in Sect.\,\ref{sec:ttW:ttw_shape}   shapes  are compared and in Sect.\,\ref{sec:ttW:ttw_gen}   acceptance effects are studied.  

\subsubsection{Shape comparison}
\label{sec:ttW:ttw_shape}
\begin{figure}[!htb]
\centering
\includegraphics[width=0.45\textwidth]{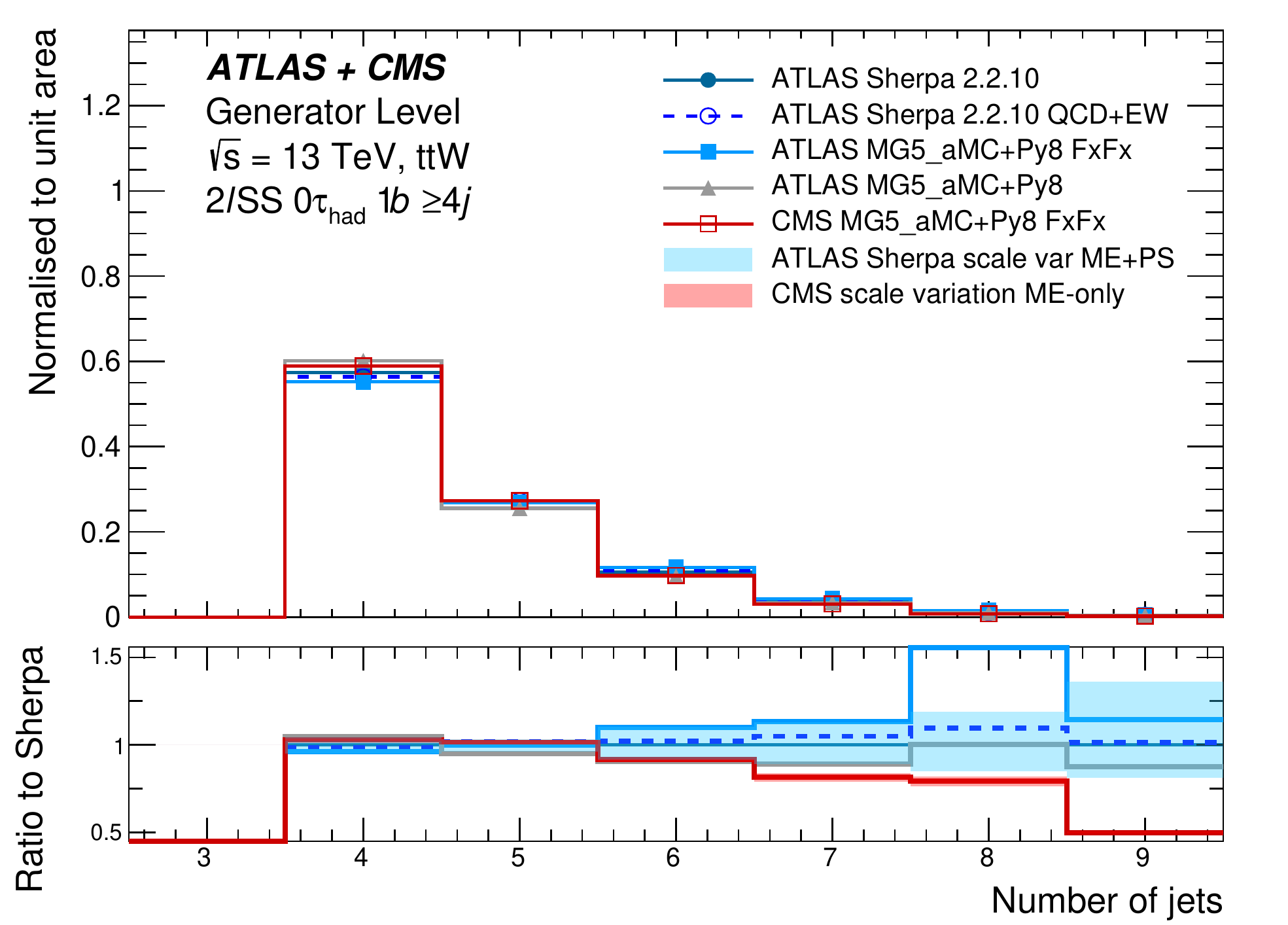}
\includegraphics[width=0.45\textwidth]{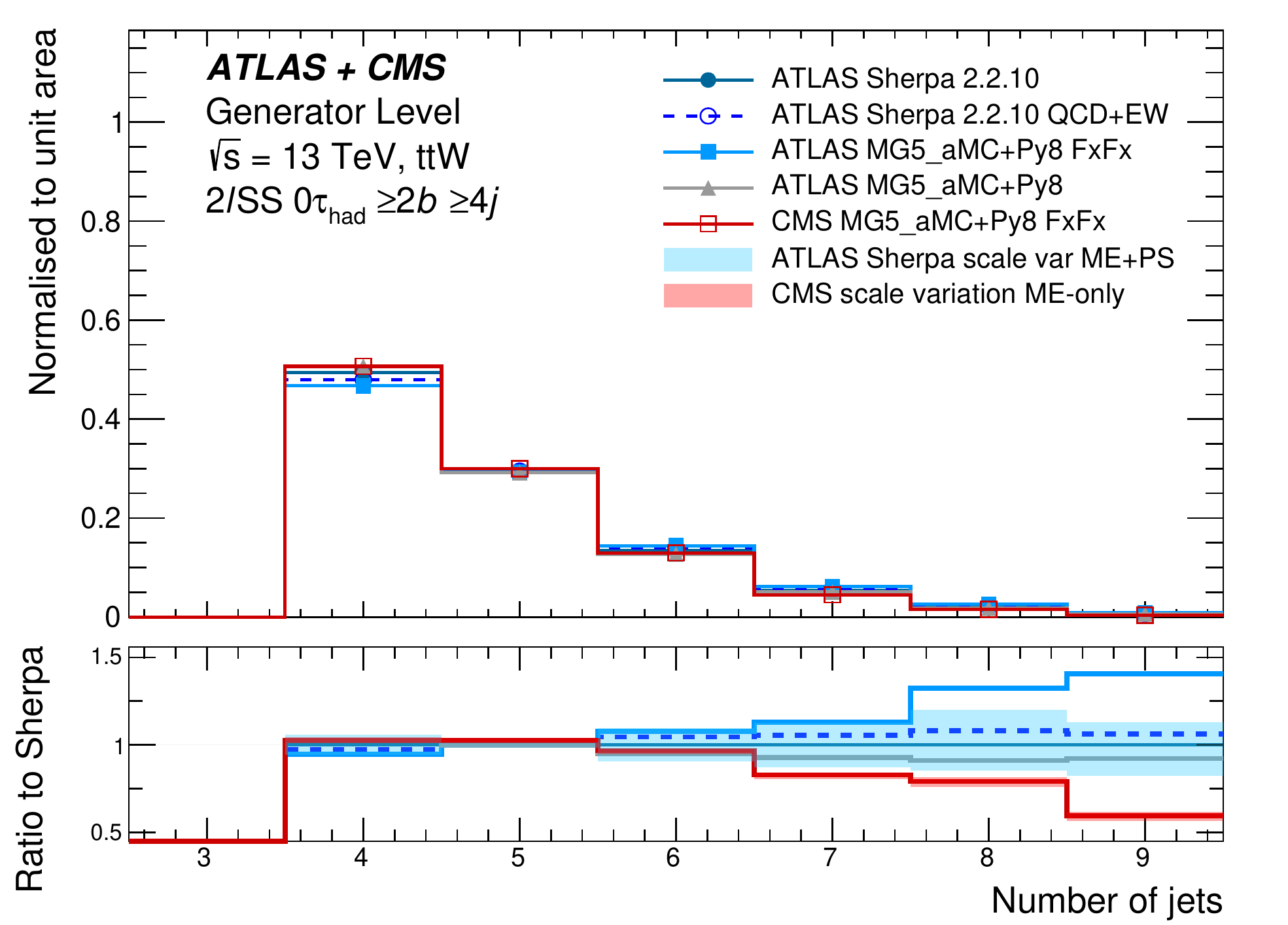}\\
\includegraphics[width=0.45\textwidth]{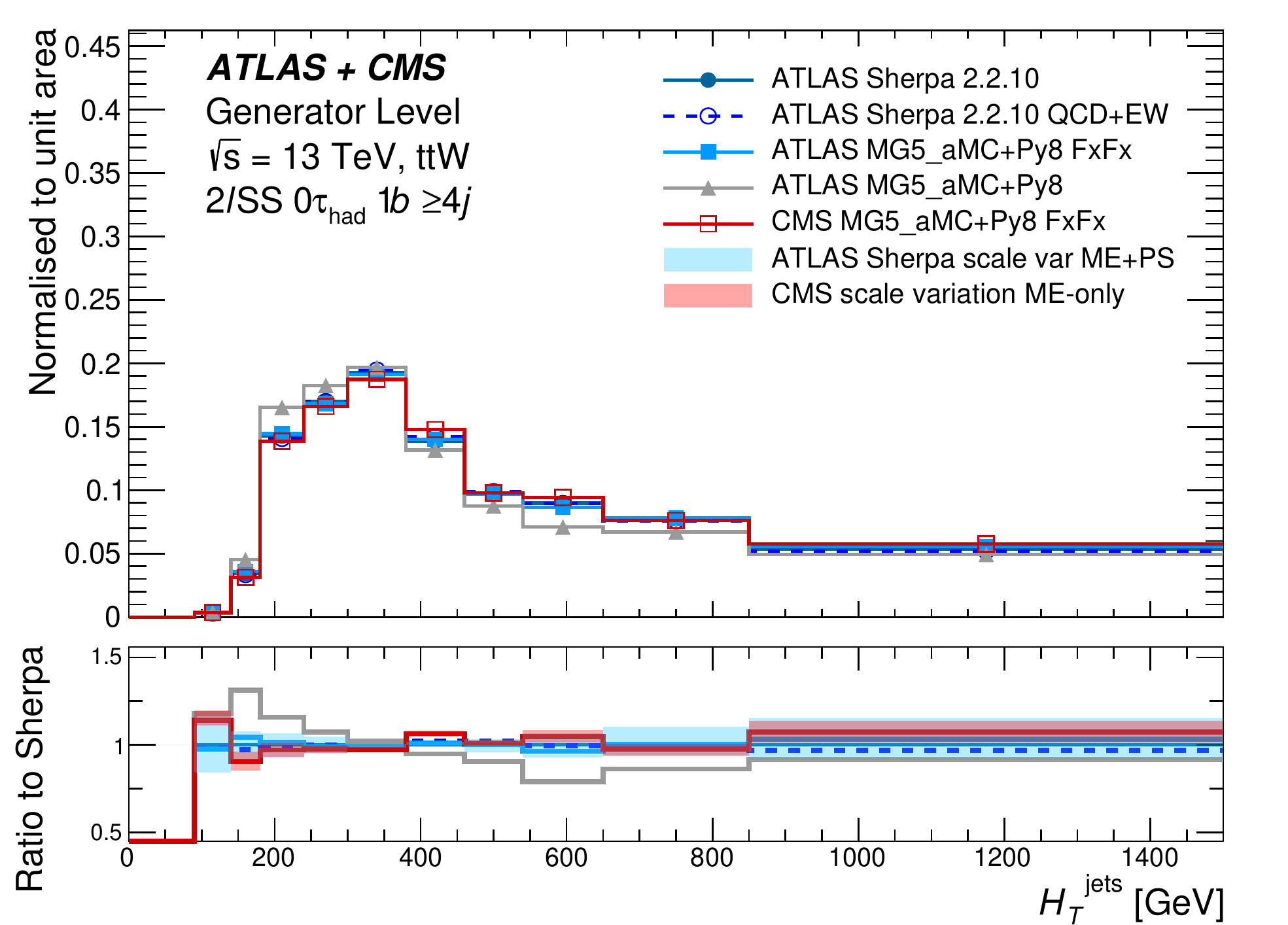}
\includegraphics[width=0.45\textwidth]{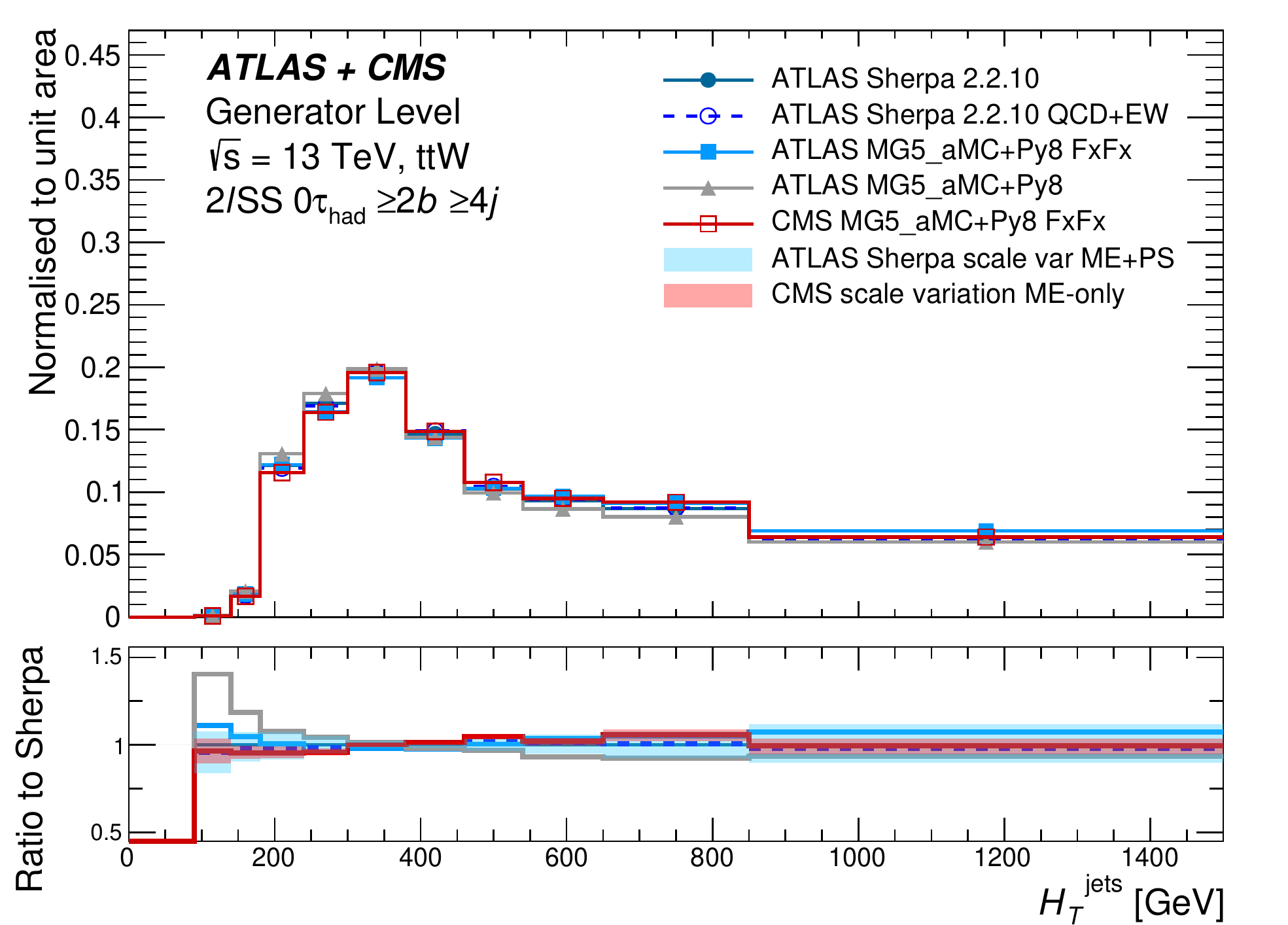}\\
\includegraphics[width=0.45\textwidth]{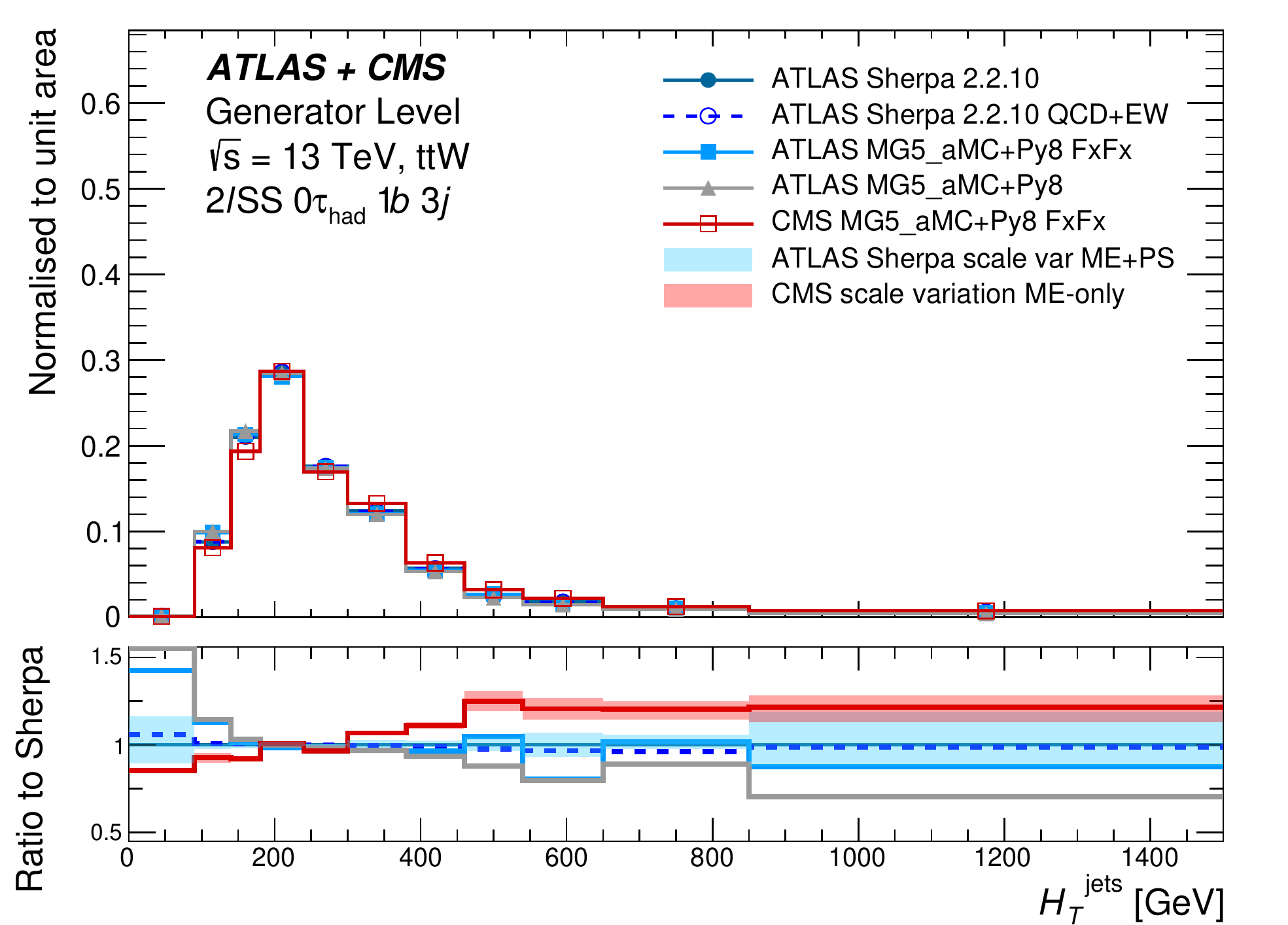}
\includegraphics[width=0.45\textwidth]{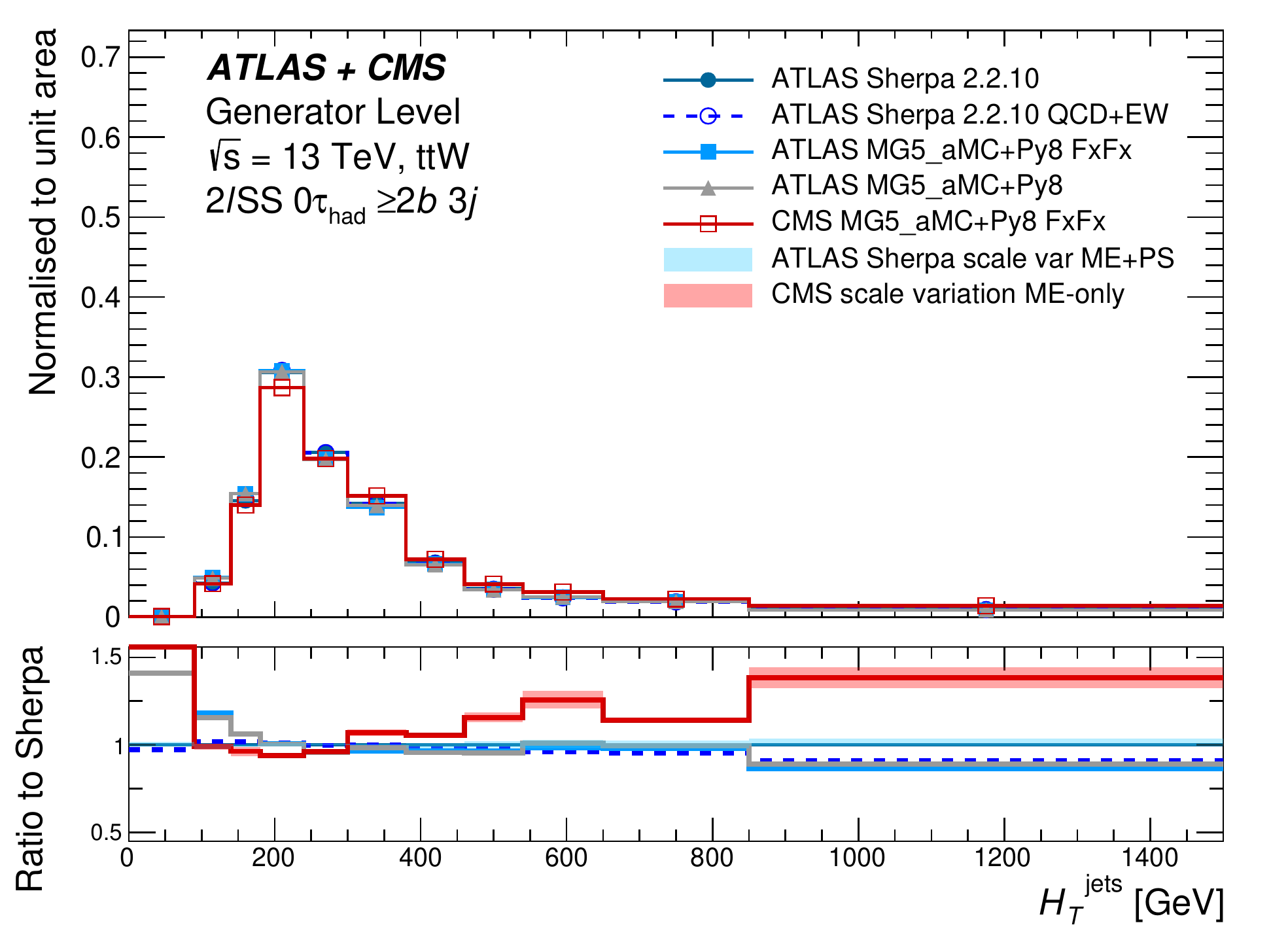}\\
  \caption{Distribution of the jet multiplicities (top) and the scalar sum of jets transverse momentum, $H_{\text{T}}^{\text{jets}}$ (middle), for the Region 1 with $N_{b-\text{jets}}=$1 (left) and Region 2 with  $N_{b-\text{jets}}\geq$2 (right) selection requiring four and more jets, and  for the Region 3 $N_{b-\text{jets}}$ = 1 (bottom, left) and Region 4 with $N_{b-\text{jets}}\geq$2 (bottom, right) selection requiring exactly three jets.  \label{ttW:4j12b_shape}\label{ttW:3j12b_shape}}
\end{figure}

\begin{figure}[!htb]
\centering
\includegraphics[width=0.45\textwidth]{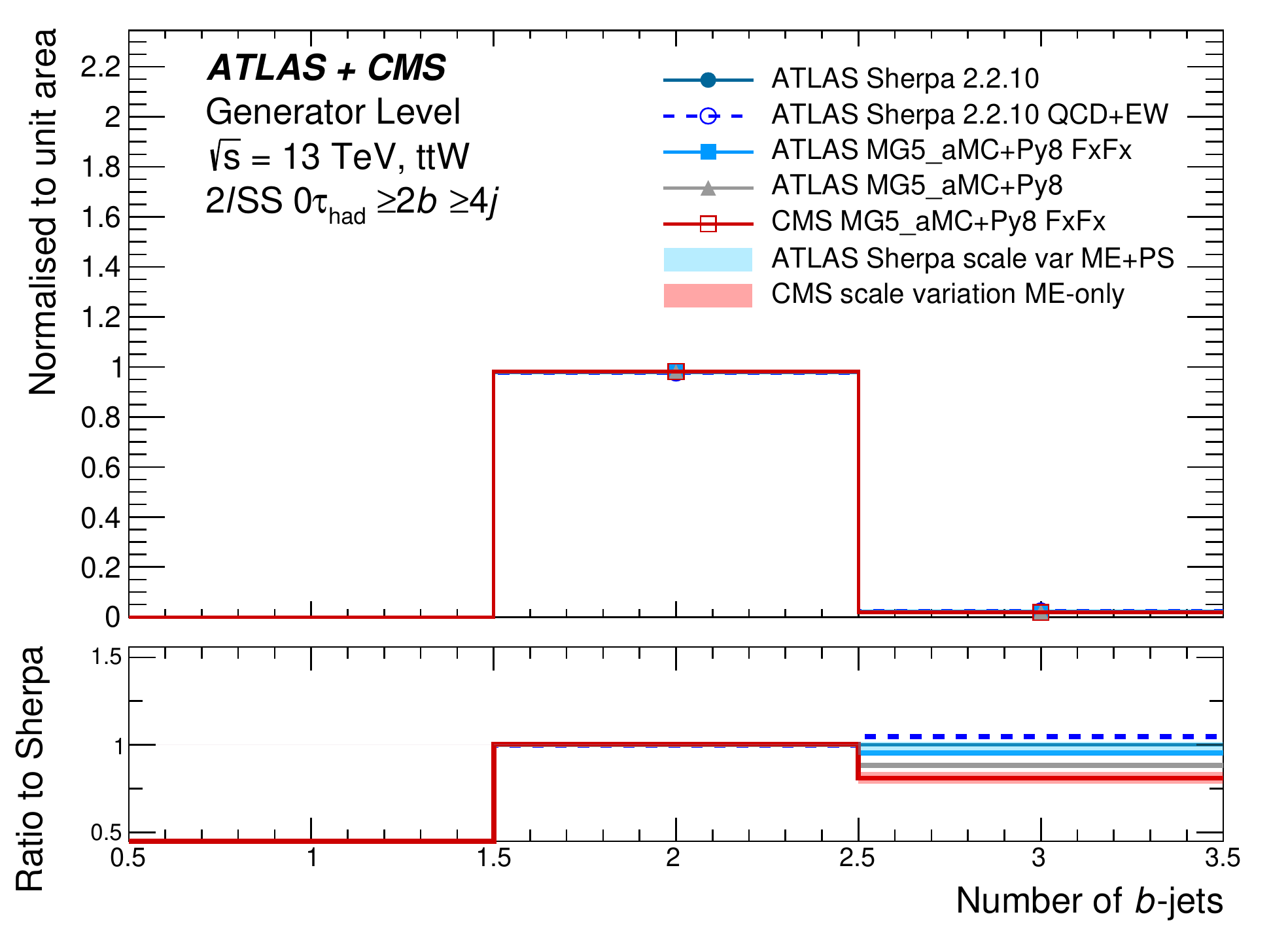}\\
\includegraphics[width=0.45\textwidth]{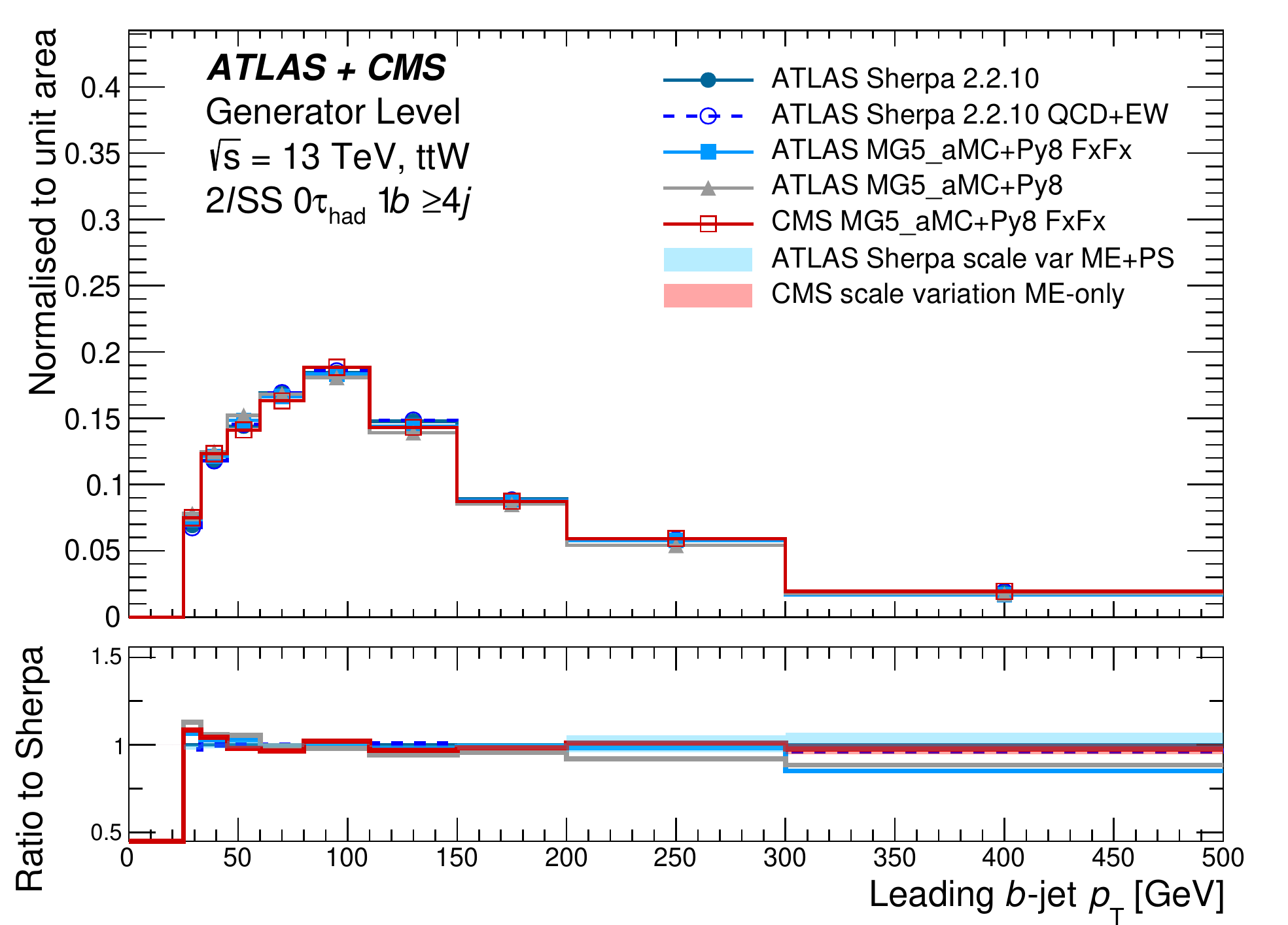}
\includegraphics[width=0.45\textwidth]{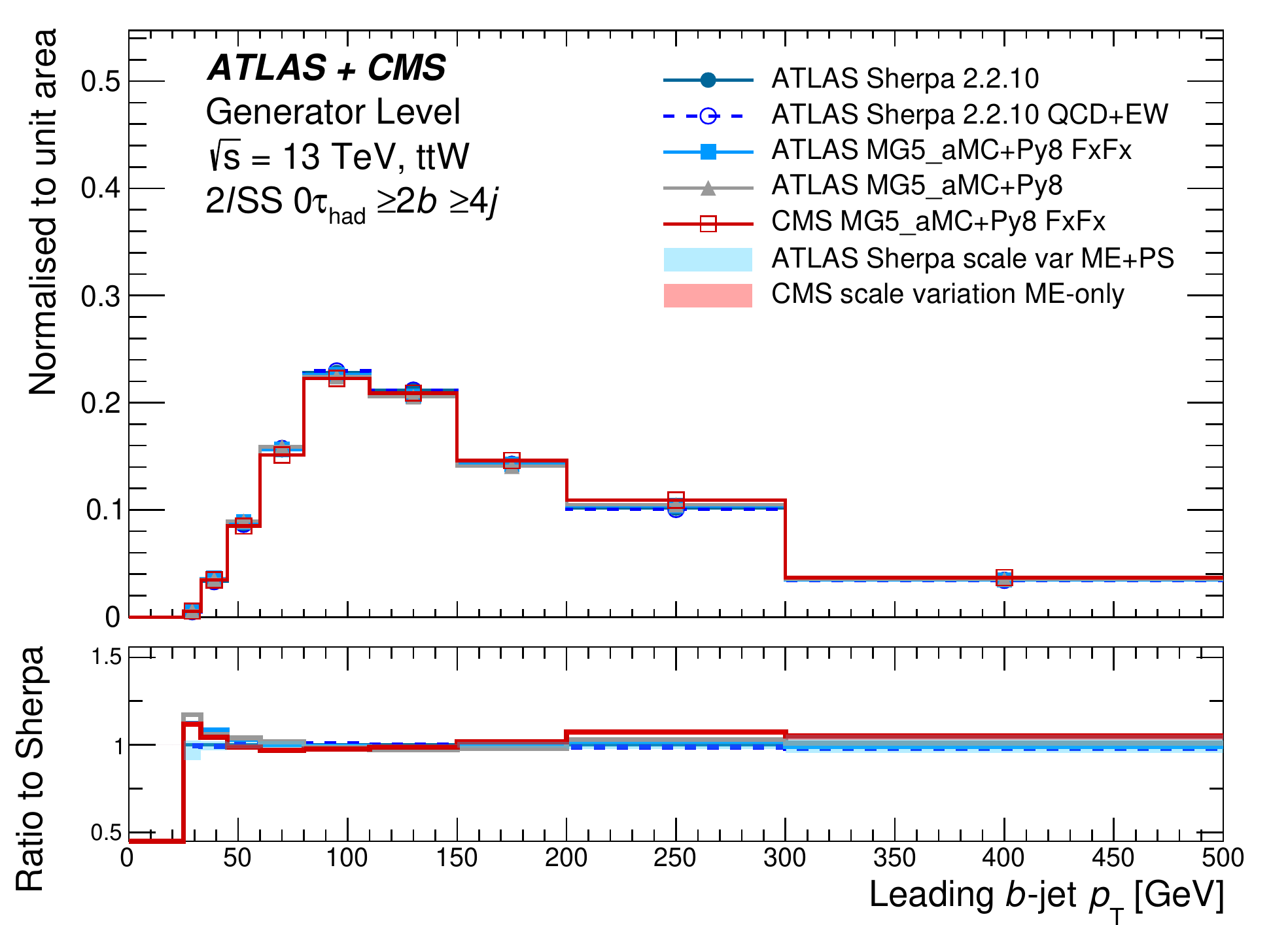}\\
  \caption{Distribution of the $b$-jet multiplicities (top) and the leading $b$-jet transverse momentum (bottom), for the Region 1 with $N_{b-\text{jets}}$=1 (left) and Region 2 with $N_{b-\text{jets}}\geq$2 (right) selection requiring four and more jets.  \label{ttW:4jbinfo_shape}}
\end{figure}

In the following, shape comparisons between nominal and alternative generators will be presented, i.e.\ the distributions are normalised to unit area. The modelling of jet based distributions are presented in Fig.~\ref{ttW:4j12b_shape}  for the regions without hadronic $\tau$ leptons.
Sizeable discrepancies in the modelling of high jet multiplicities can be observed between the ATLAS and CMS \amcnp FxFx predictions  which are in opposite direction compared to \sherpa \ttW. All predictions except ATLAS \amc agree well on \HT in regions with at least four jets, but larger discrepancies are observed for the three jet regions. The distributions of \bjet \pt differ  more in the regions with one \bjet, as shown in Fig.\,\ref{ttW:4jbinfo_shape}.

\begin{figure}[!htb]
\centering
\includegraphics[width=0.45\textwidth]{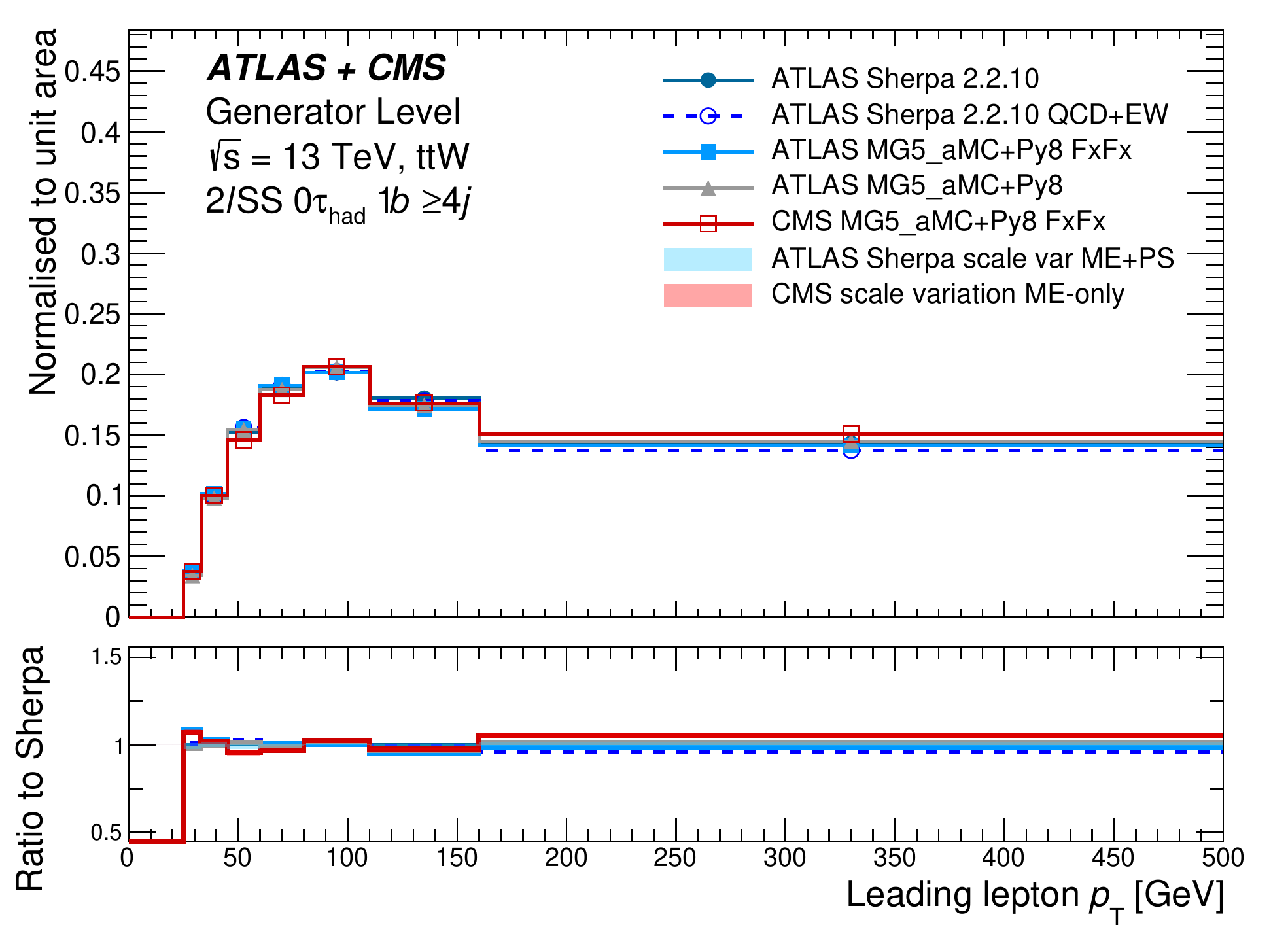}
\includegraphics[width=0.45\textwidth]{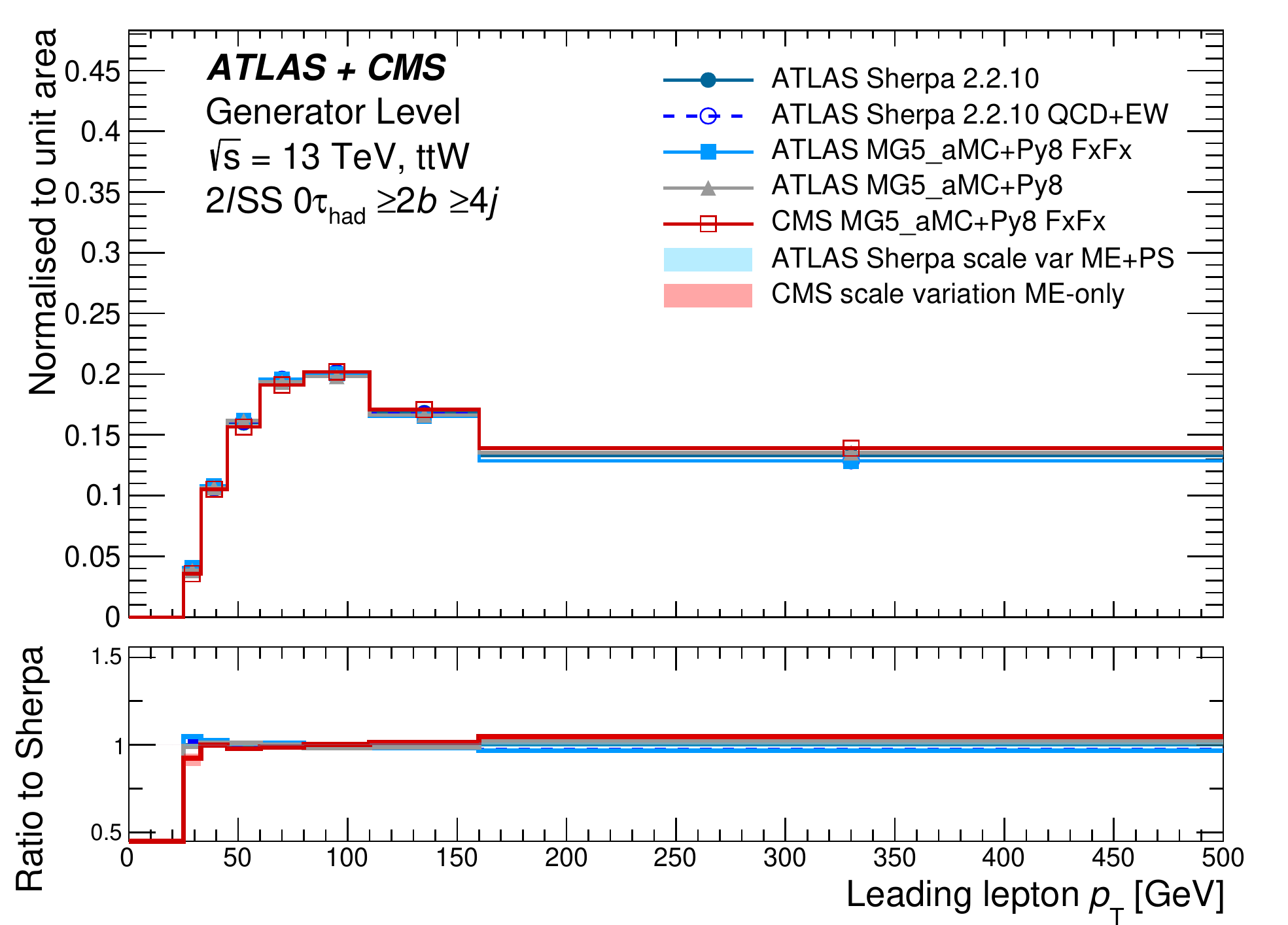}\\
\includegraphics[width=0.45\textwidth]{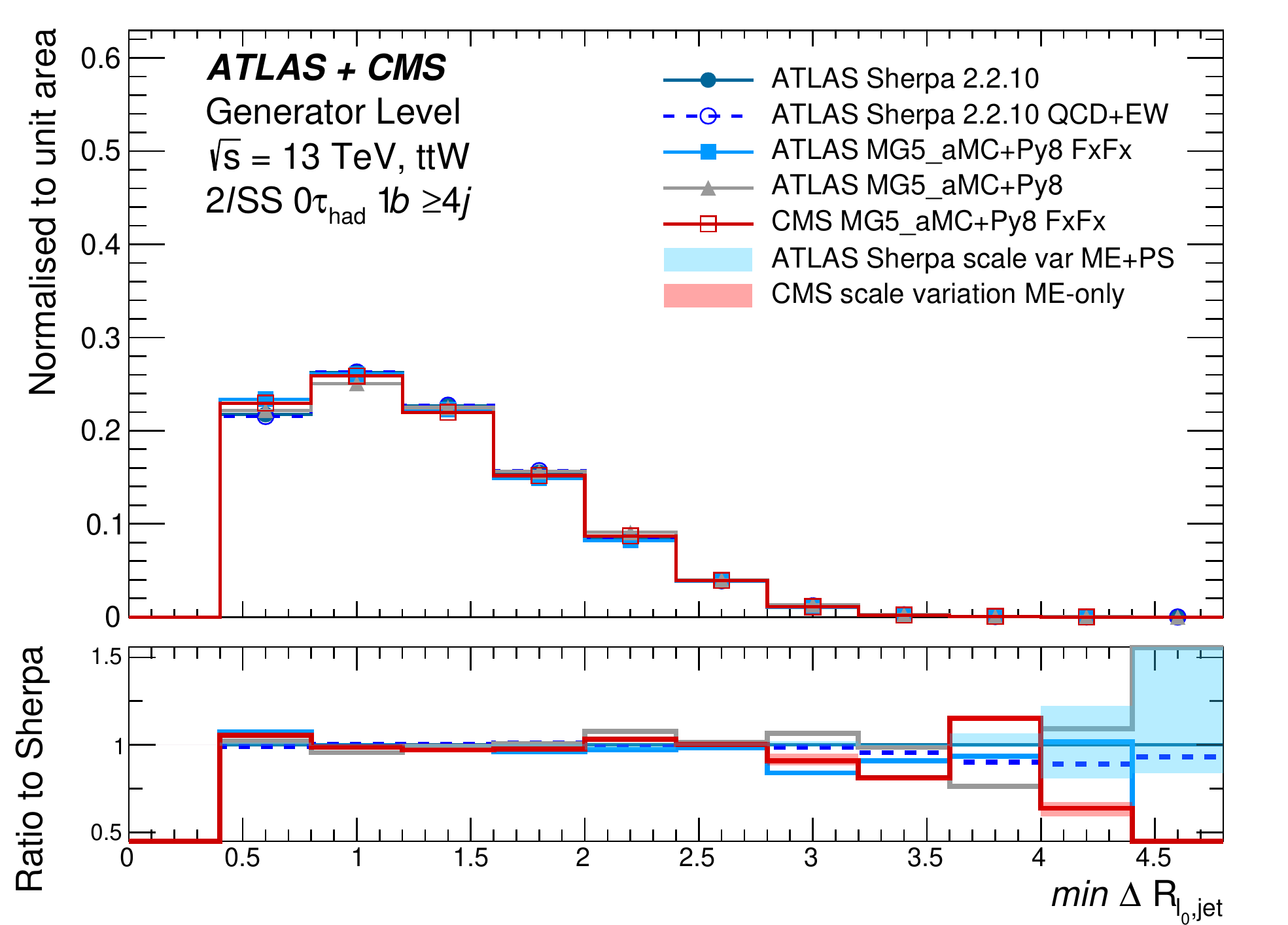}
\includegraphics[width=0.45\textwidth]{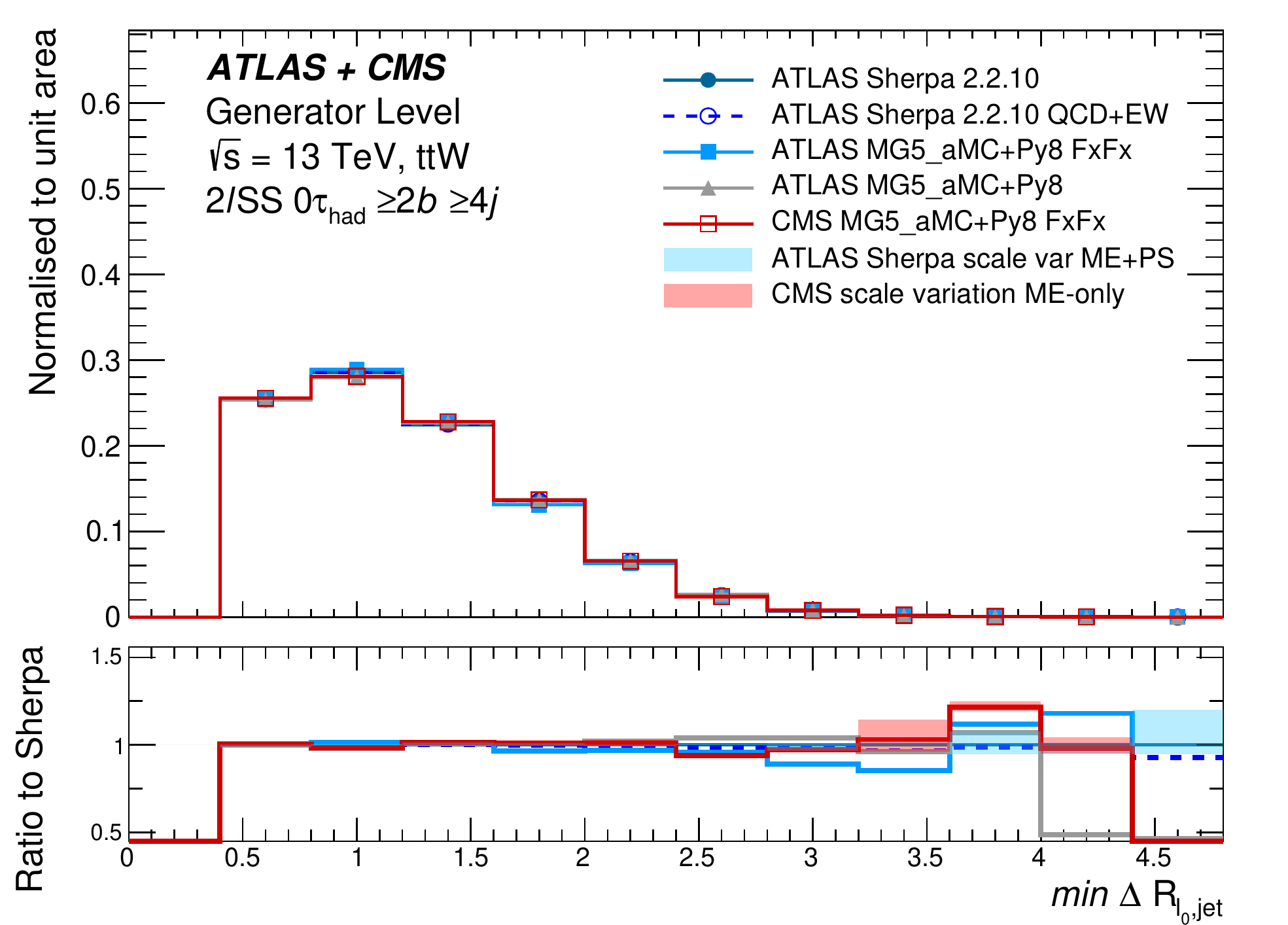}\\
  \caption{Distribution of the leading lepton transverse momentum (top) and the minimum angular separation between the leading lepton and the nearest jet (bottom), for the Region 1 with $N_{b-\text{jets}}$=1 (left) and Region 2 with $N_{b-\text{jets}}\geq$2 (right) selection requiring four and more jets.
  \label{ttW:lep_kin_shape}}
\end{figure}

\begin{figure}[!htb]
\centering
\includegraphics[width=0.45\textwidth]{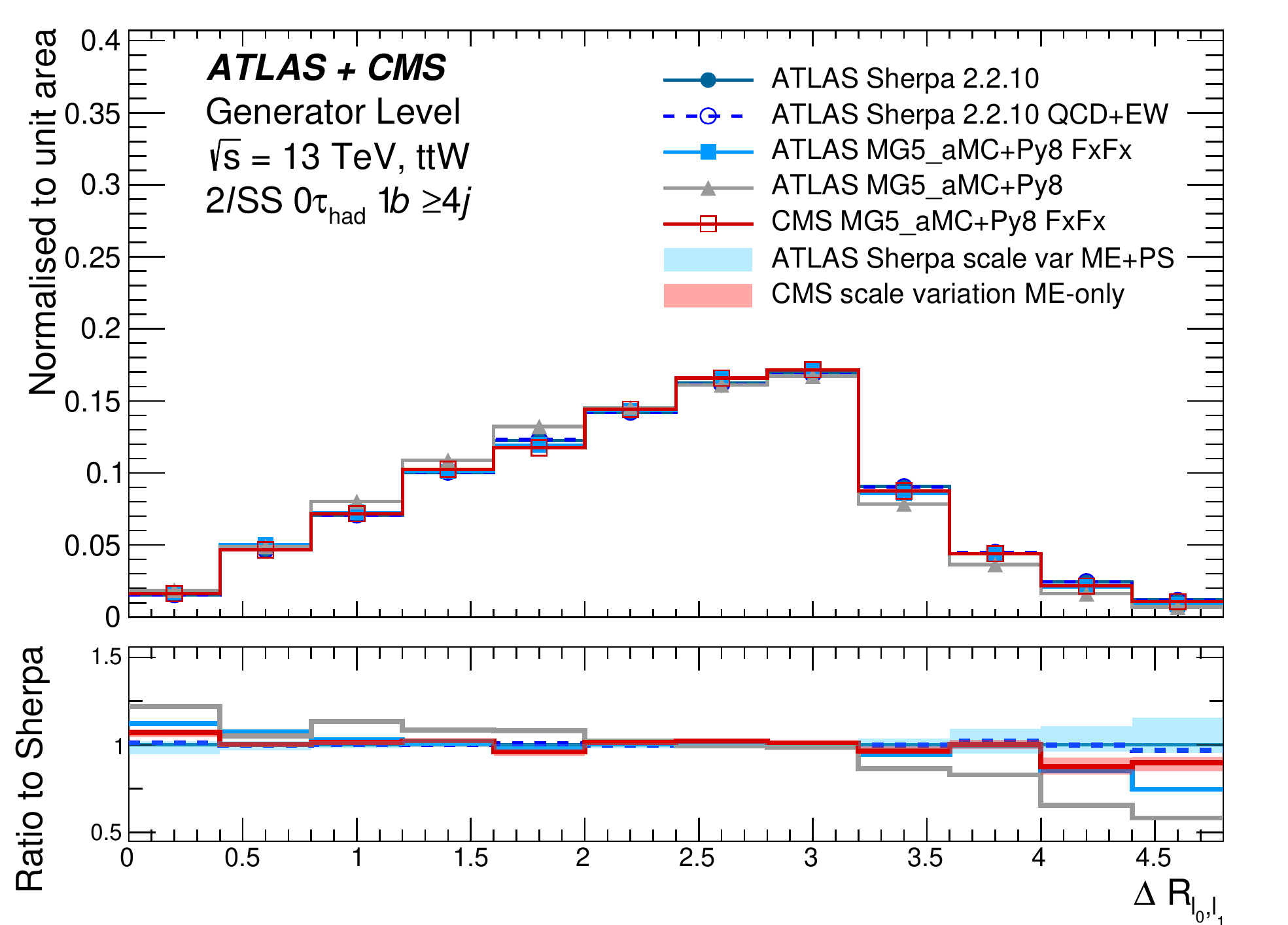}
\includegraphics[width=0.45\textwidth]{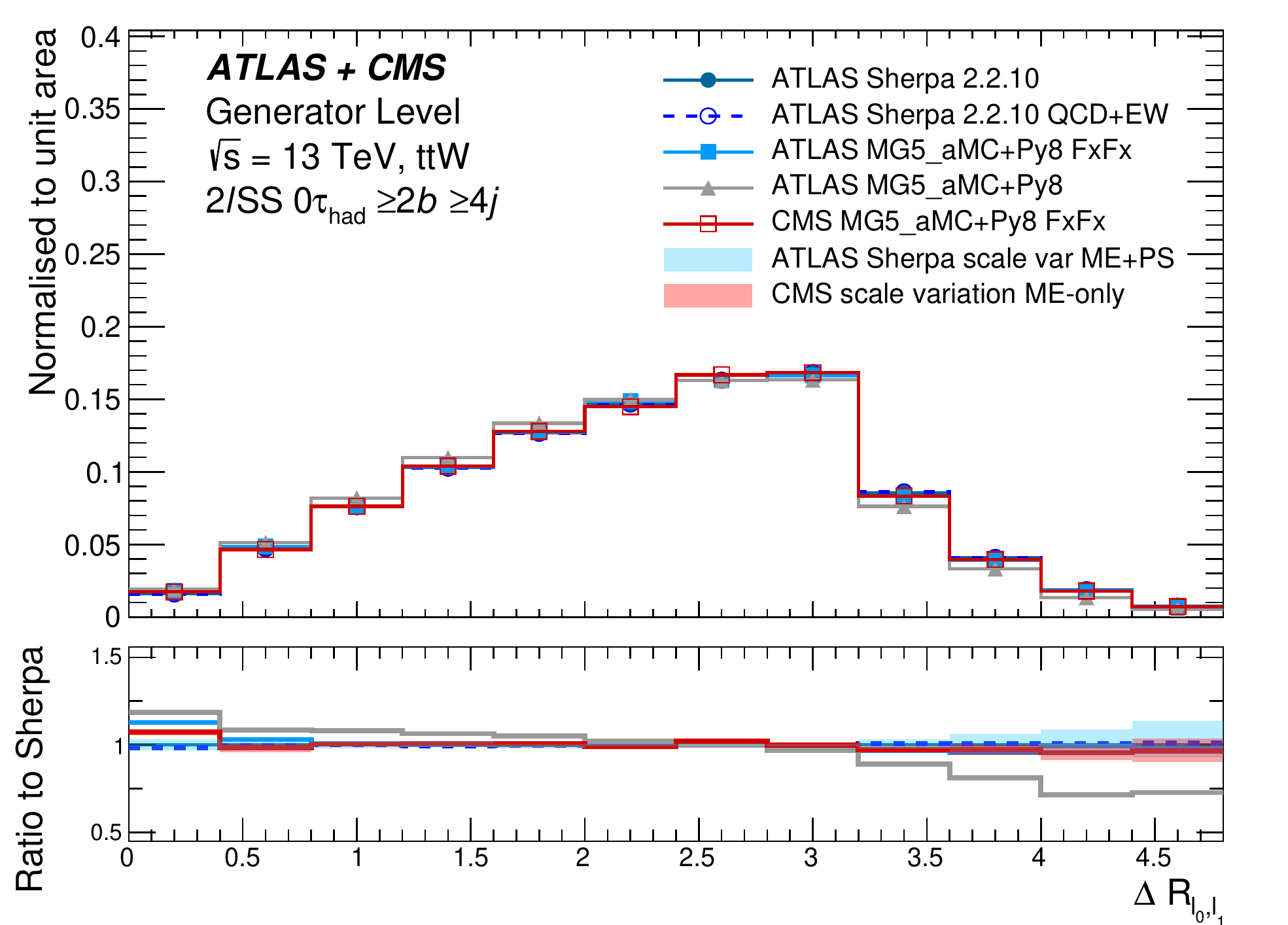}\\
\includegraphics[width=0.45\textwidth]{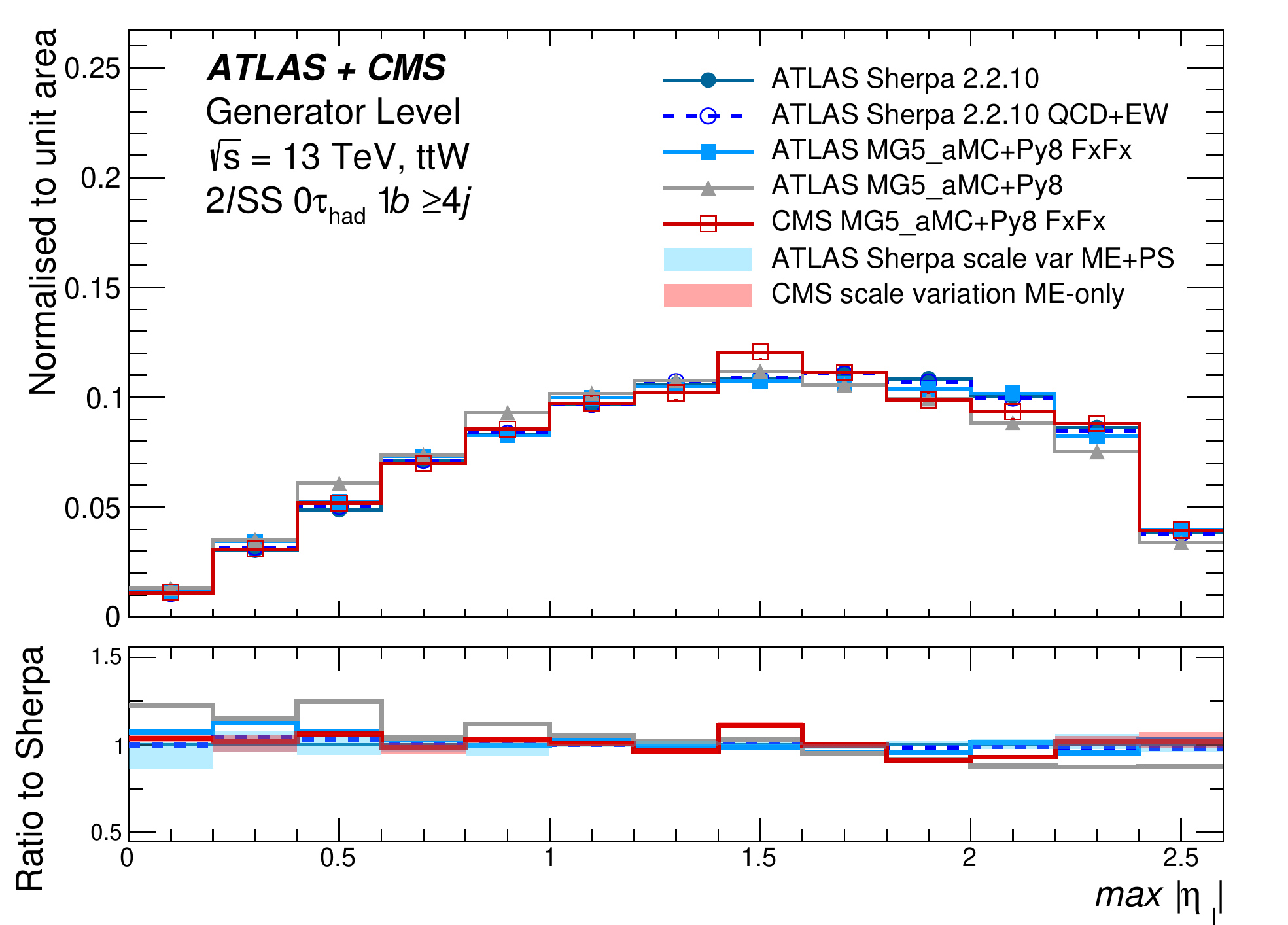}
\includegraphics[width=0.45\textwidth]{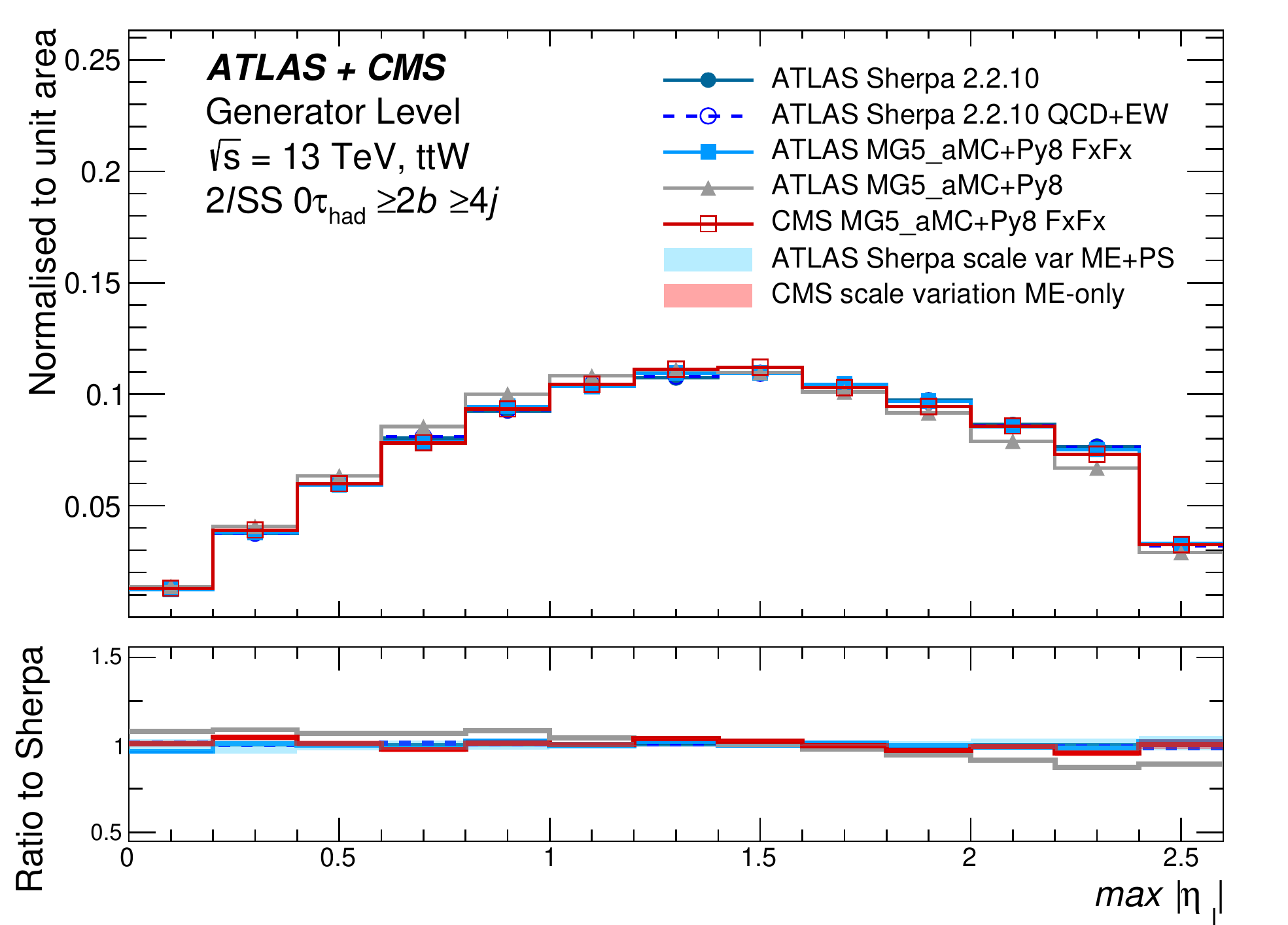}\\
  \caption{Distribution of the angular distance between the two leptons (top), maximum of lepton $|\eta_{\ell 0}|$ and $|\eta_{\ell 1}|$ (bottom) , for the Region 1 with $N_{b-\text{jets}}$=1 (left) and Region 2 with $N_{b-\text{jets}}\geq$2 (right) selection requiring four and more jets.
   \label{ttW:ll_kin_shape}}
\end{figure}

Only ATLAS \amc shows significant differences  for the angular distance between the two leptons and the value of lepton's pseudo-rapidity  as demonstrated in Fig.\,\ref{ttW:lep_kin_shape}. The lepton \pt distributions are similar, but their distance to the closest jet vary at  as seen in Fig.\,\ref{ttW:ll_kin_shape}.

\begin{figure}[!htb]
\centering

\includegraphics[width=0.45\textwidth]{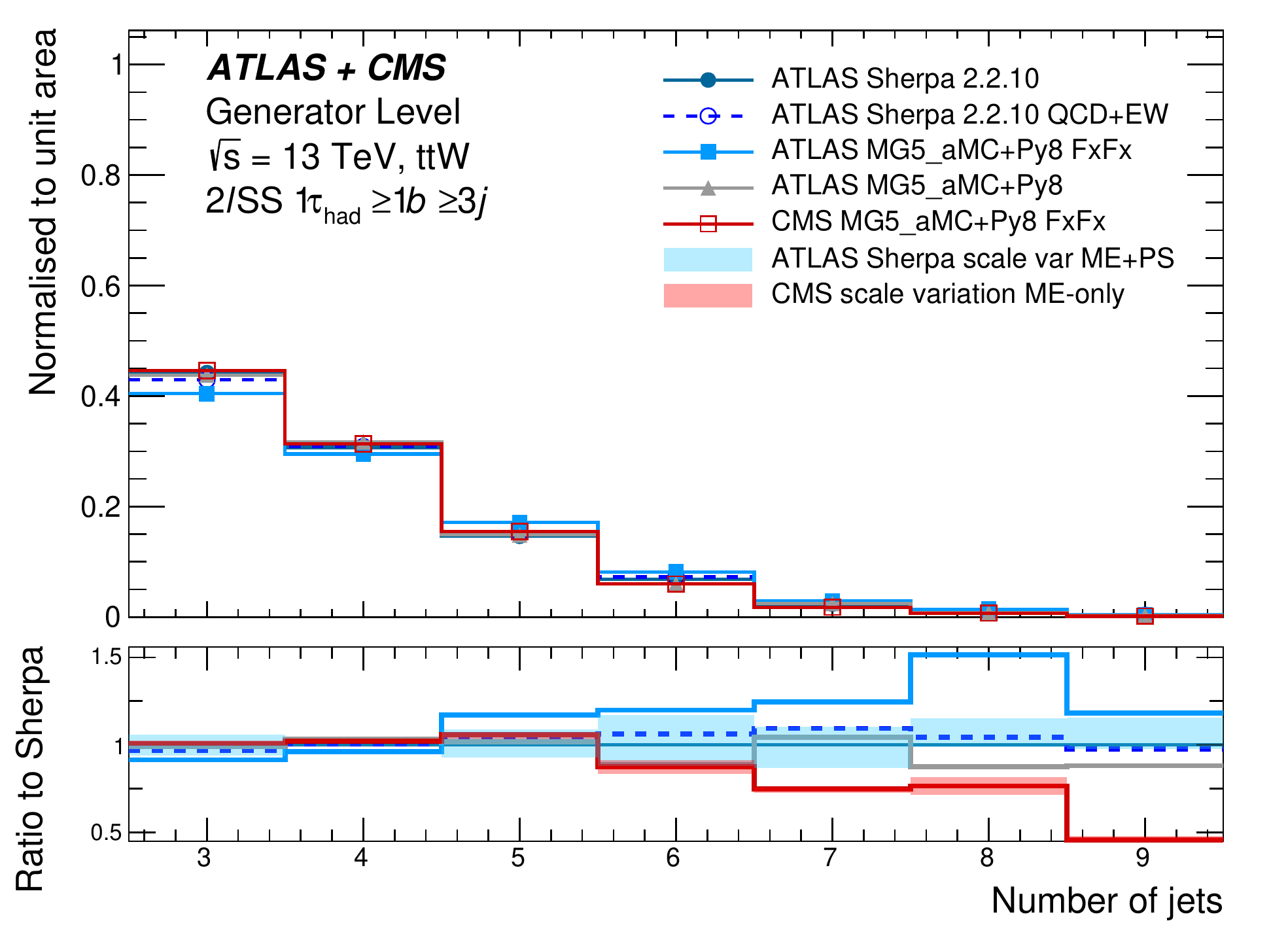}
\includegraphics[width=0.45\textwidth]{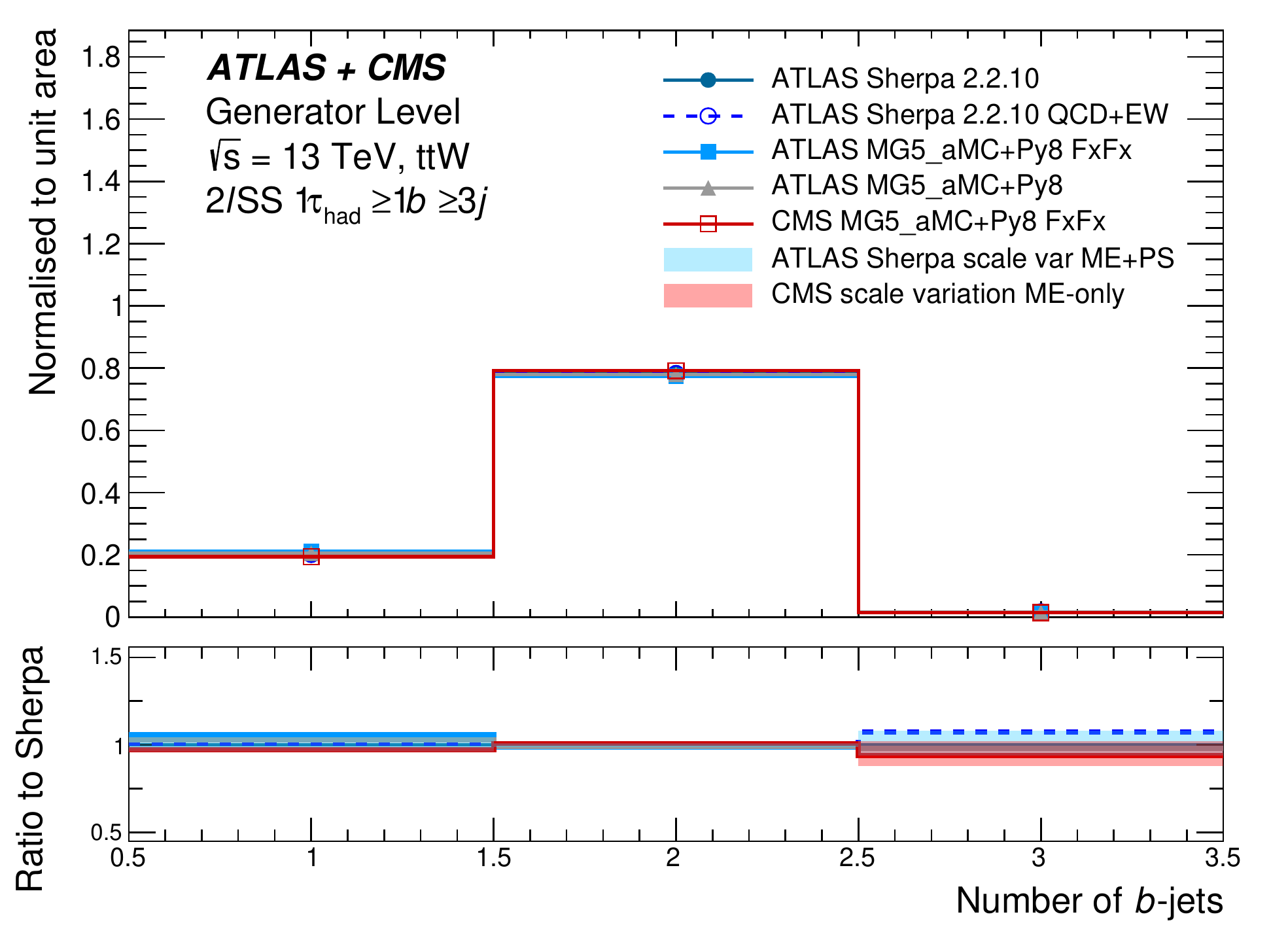}\\
\includegraphics[width=0.45\textwidth]{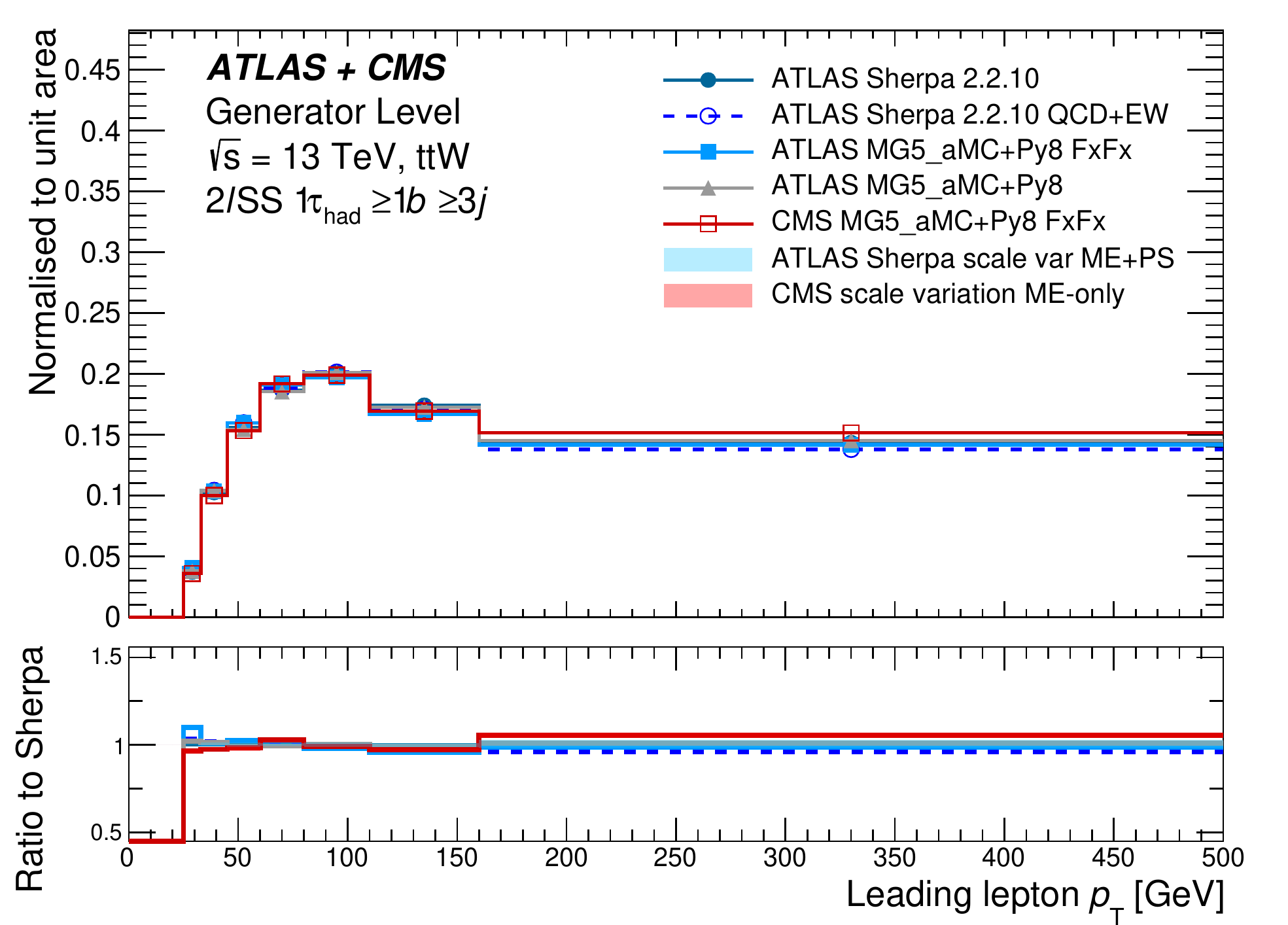}
\includegraphics[width=0.45\textwidth]{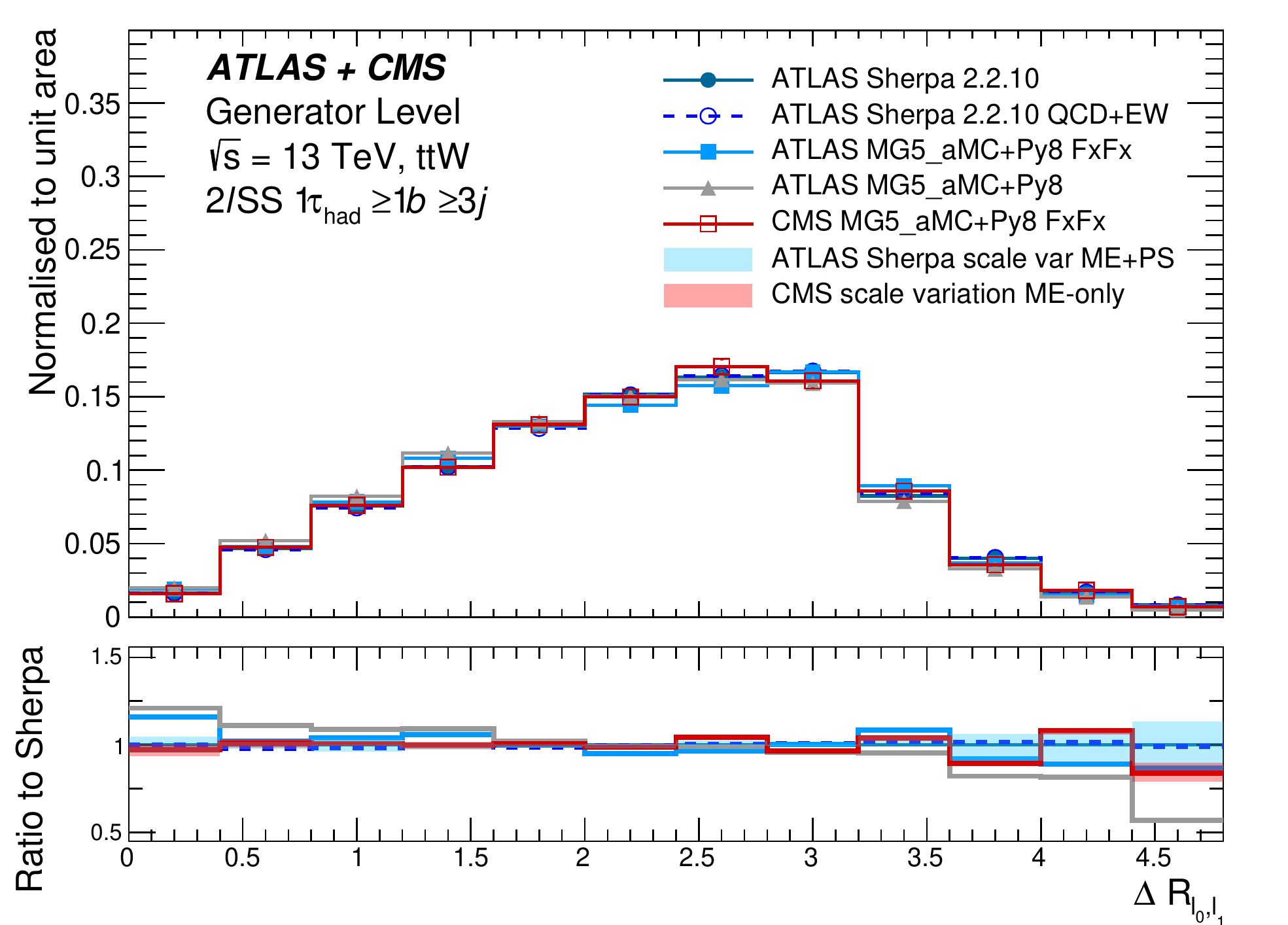}\\
  \caption{Distribution of the the jet multiplicity, number of $b$-jets, the leading lepton transverse momentum and the angular distance between the two leptons  $\Delta R _{\ell \ell }$ for the Region 5 with 1$\tau_{\text{had}}$ selection.
   \label{ttW:tauR_kin_shape}}
\end{figure}

Distributions of the jet multiplicity, number of $b$-jets, the leading lepton transverse momentum and the angular distance between the two leptons  $\Delta R _{\ell 0\ell1 }$ for the Region 5 with $N_{\tau_\text{had}}$ = 1 selection are presented in Fig.\,\ref{ttW:tauR_kin_shape}. The jet multiplicity predictions of MG5\_aMC+Py8~FxFx with the ATLAS and CMS set-ups differ most from the other predictions in this region.

\clearpage
\newpage

\subsubsection{Comparisons of predictions including acceptance effects}
\label{sec:ttW:ttw_gen}

\begin{figure}[!htb]
\centering
\includegraphics[width=0.45\textwidth]{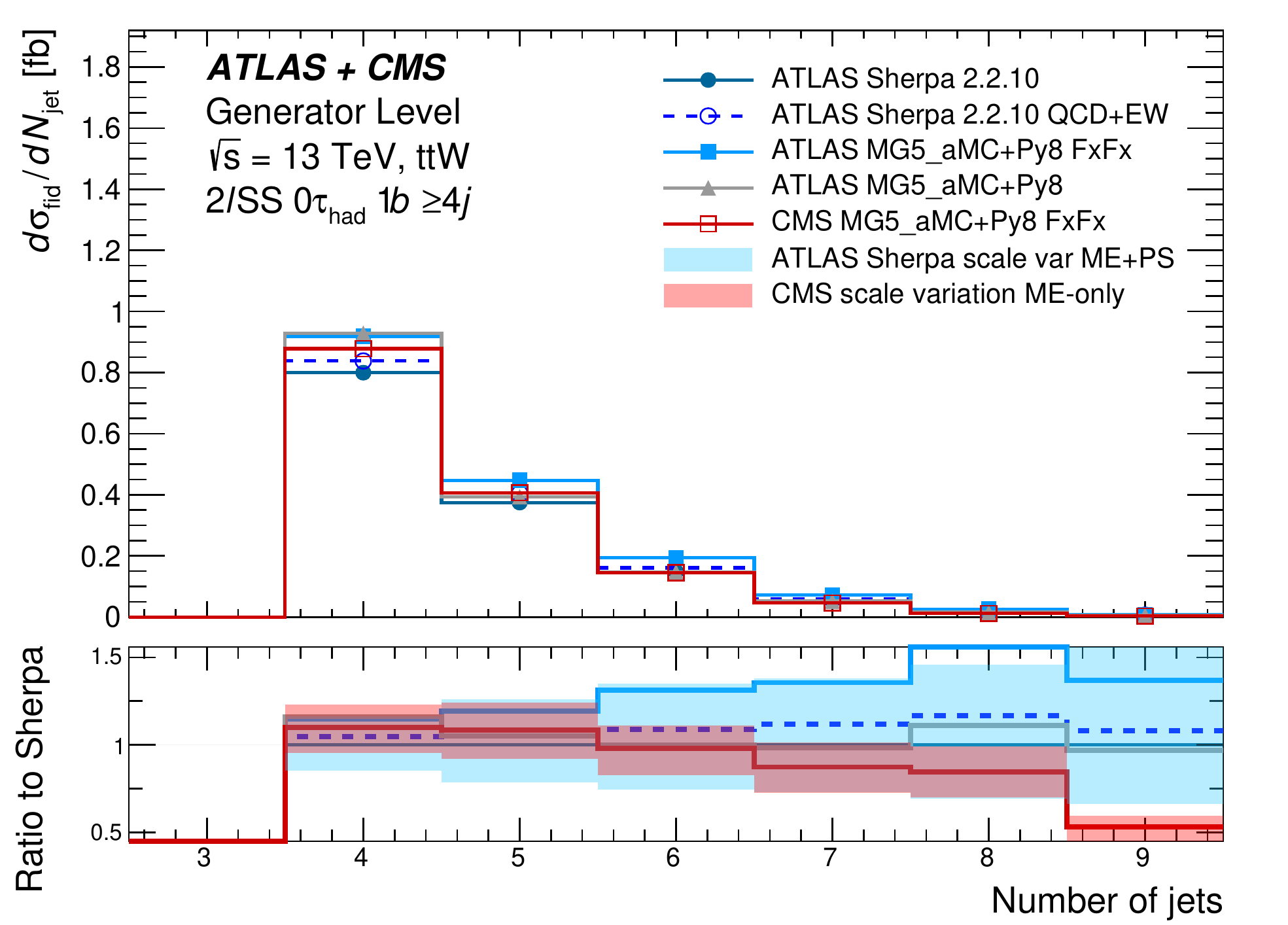}
\includegraphics[width=0.45\textwidth]{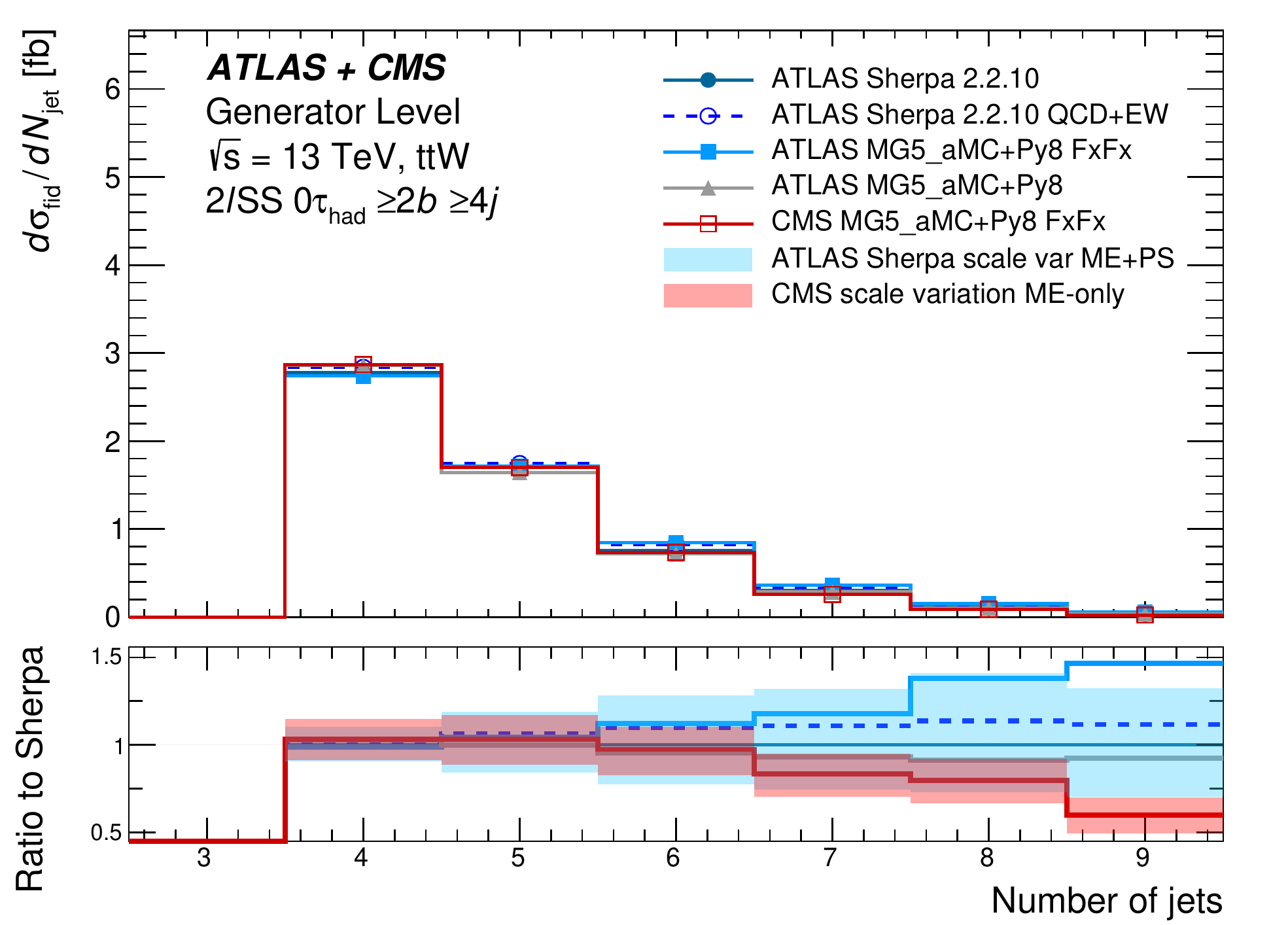}\\
\includegraphics[width=0.45\textwidth]{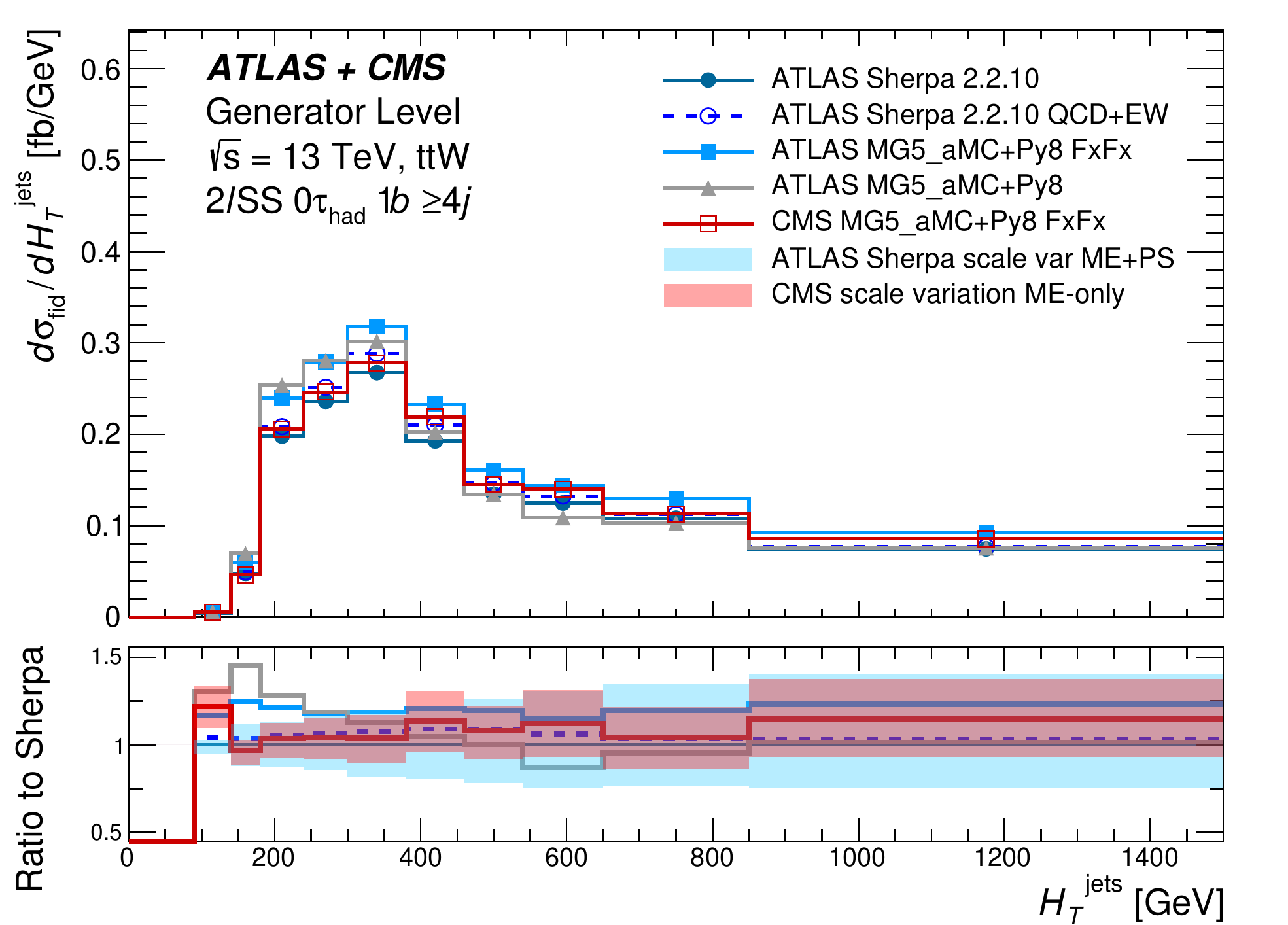}
\includegraphics[width=0.45\textwidth]{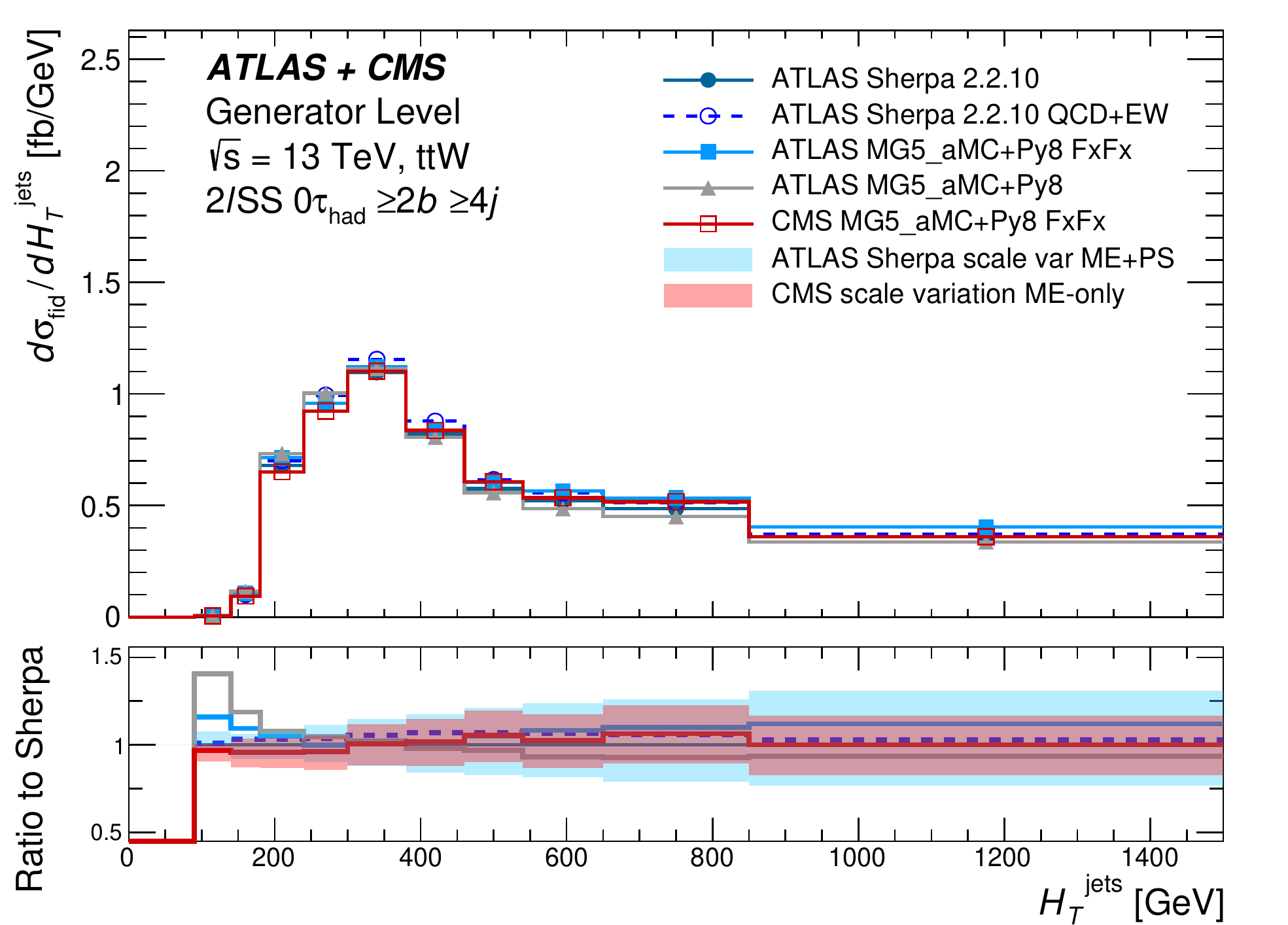}\\
\includegraphics[width=0.45\textwidth]{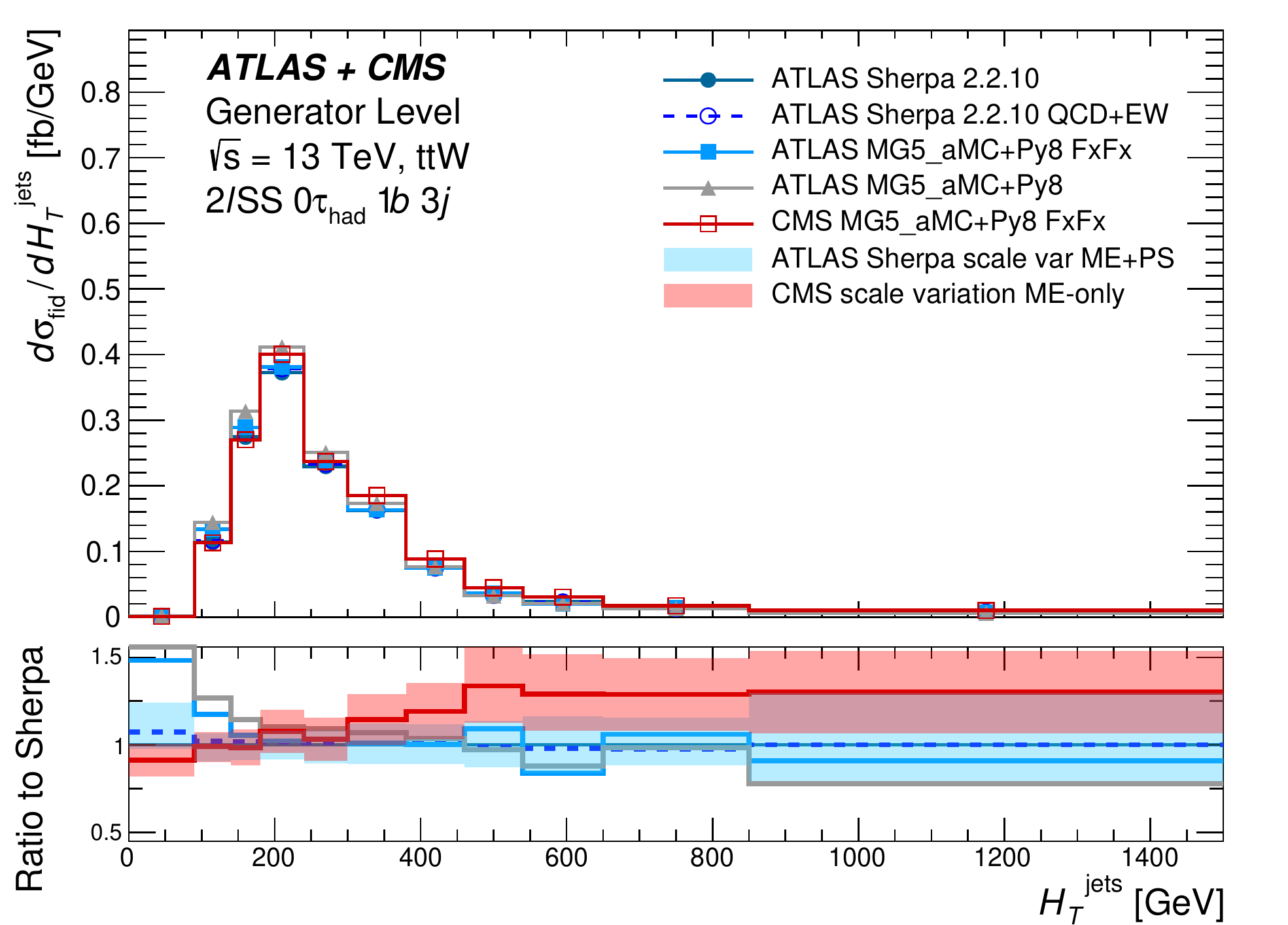}
\includegraphics[width=0.45\textwidth]{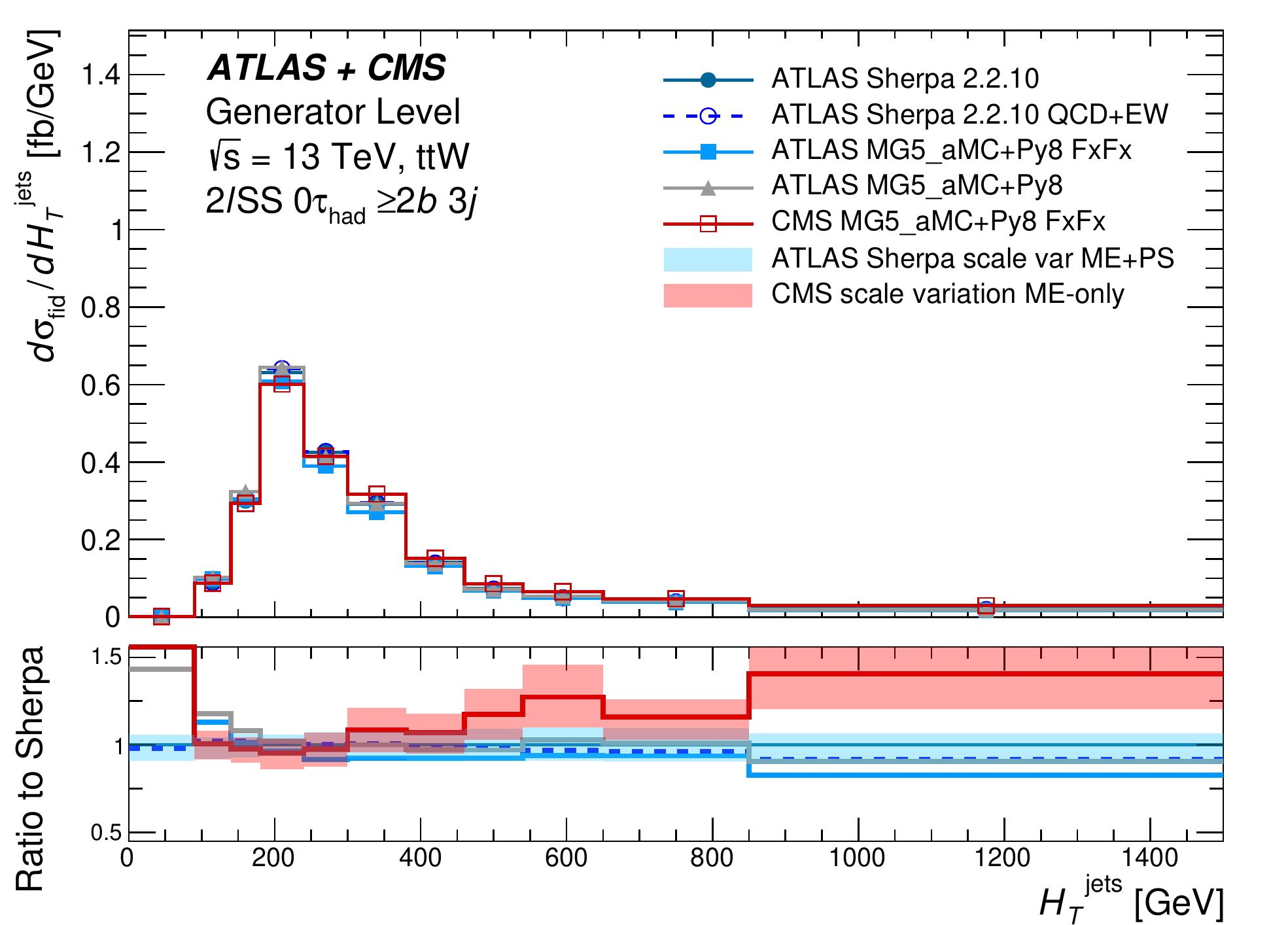}\\
  \caption{Distribution of jet multiplicities (top) and scalar sum of jets transverse momentum, $H_\text{T}^{\text{jets}}$ (middle), for the Region 1 with $N_{b-\text{jets}}$=1 (middle, left) and Region 2 with $N_{b-\text{jets}}\geq$2 (middle, right) selection requiring four and more jets.
   $H_\text{T}^{\text{jets}}$, for the Region 3 with $N_{b-\text{jets}}$=1 (bottom, left) and Region 4 with $N_{b-\text{jets}}\geq$2 (bottom, right) selection requiring exactly three jets.
  All distributions are normalised to the YR4 cross section of \SI{600.8}{\femto\barn}, except \sherpa 2.2.10 QCD+EW which is normalised to \SI{614.7}{\femto\barn}.
  \label{fig:den_3j12b}
  \label{fig:den_4j12b}}
\end{figure}

In the following section, a comparison of the generators will be given in the fiducial phase space, i.e.\ the predicted distributions include acceptance effects.
For this comparison,  all  distributions  are normalised to  a common total cross section value of $\sigma_{\text{tot}}^{\text{YR4}}=\SI{600.8}{\femto\barn}$ as given in the Yellow Report 4~\cite{YR4}, except the distributions of Sherpa~2.2.10~QCD+EW which  is normalised to its generator cross section of \SI{614.7}{\femto\barn}.
The same set of distributions as discussed in Section~\ref{sec:ttW:ttw_shape} are presented. 
In all distributions, a significant increase of scale uncertainties is observed, reaching up to 50\,\% at high jet multiplicity.
The observables related to jet multiplicity and \HT show   similar trends as in the shape comparisons, see Fig.\,\ref{fig:den_3j12b}. Only the discrepancy of the jet multiplicity prediction in MG5\_aMC+Py8~FxFx is significantly enhanced.

\begin{figure}[!htb]
\centering
\includegraphics[width=0.45\textwidth]{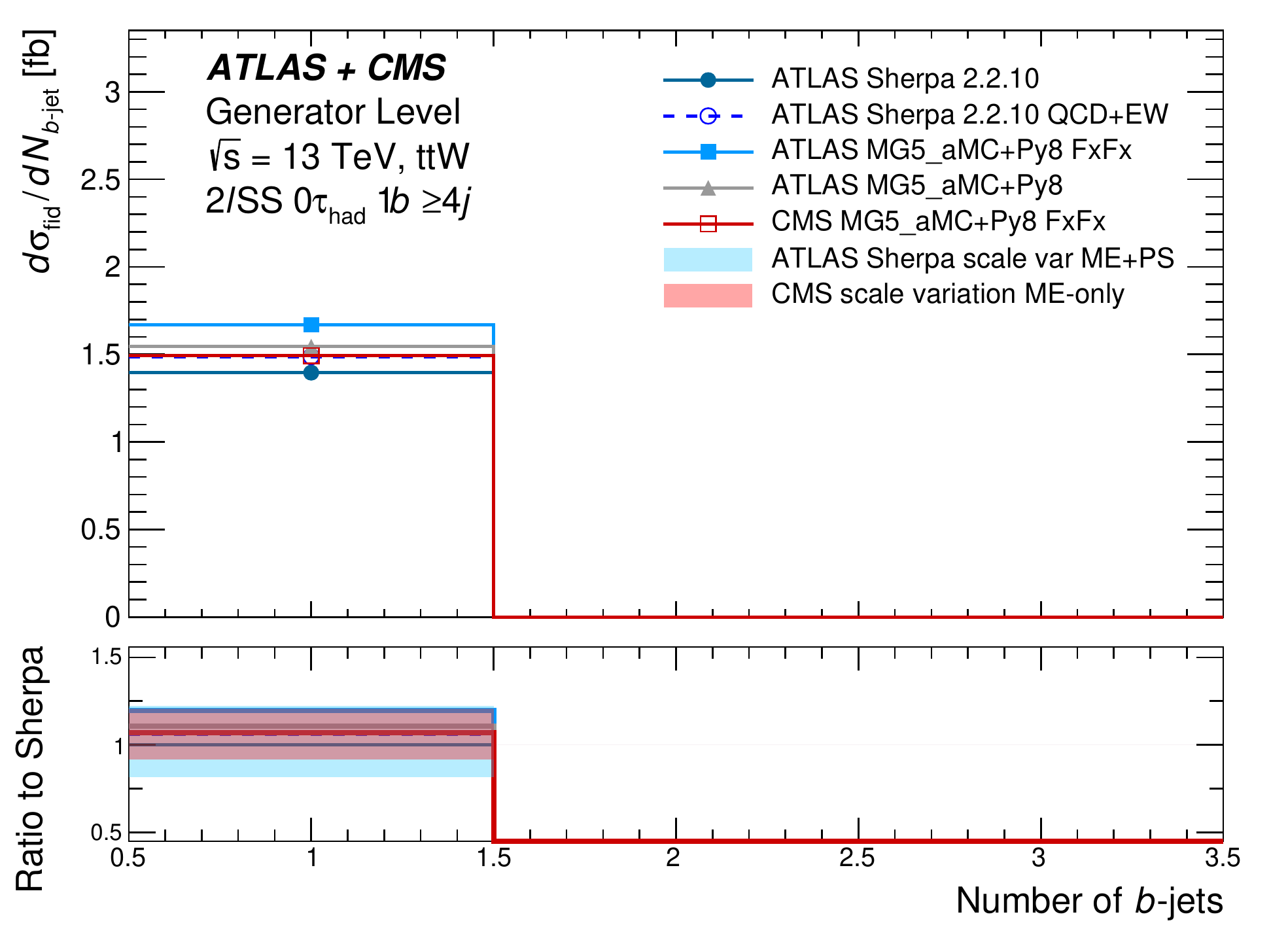}
\includegraphics[width=0.45\textwidth]{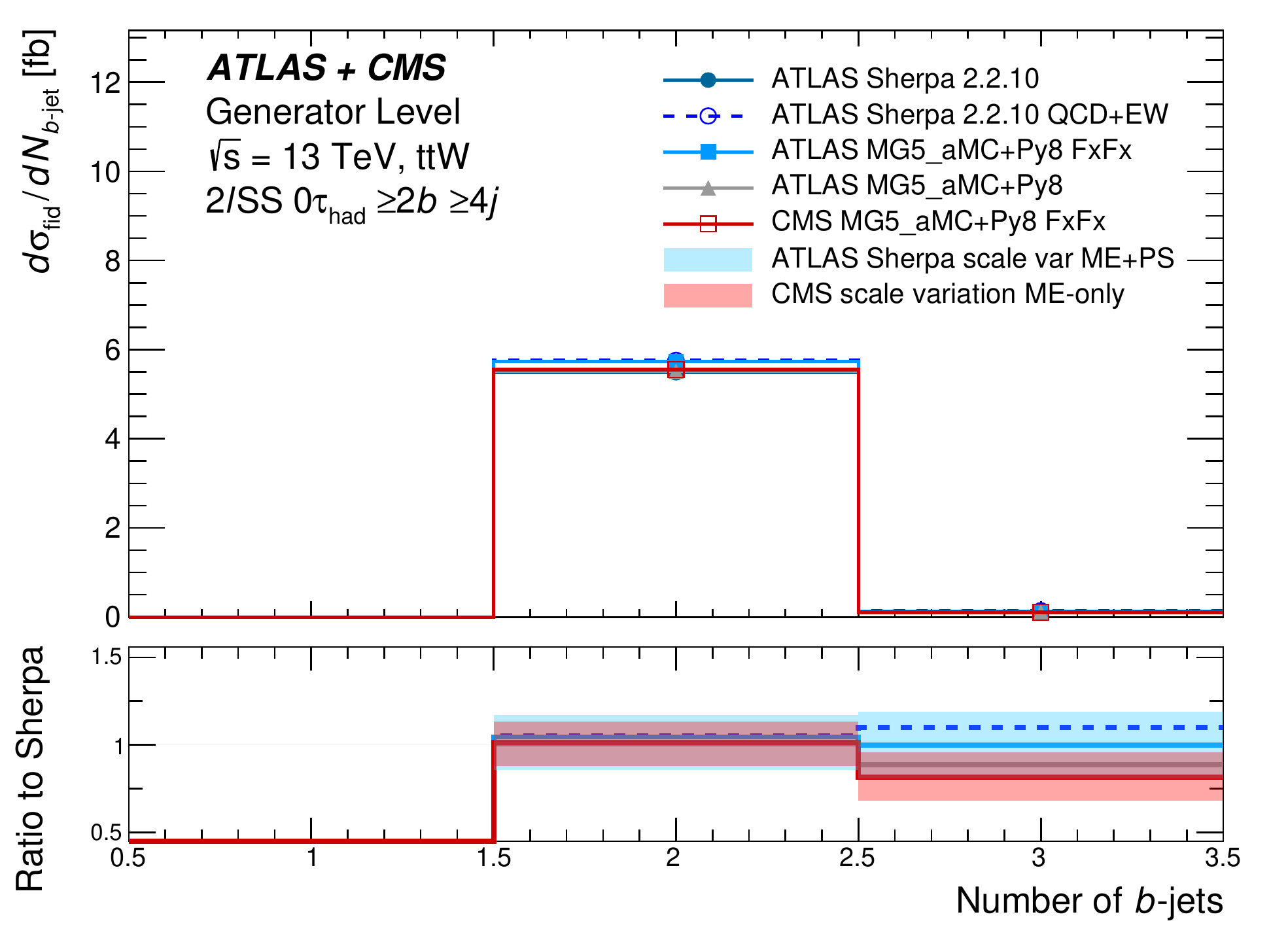}\\
\includegraphics[width=0.45\textwidth]{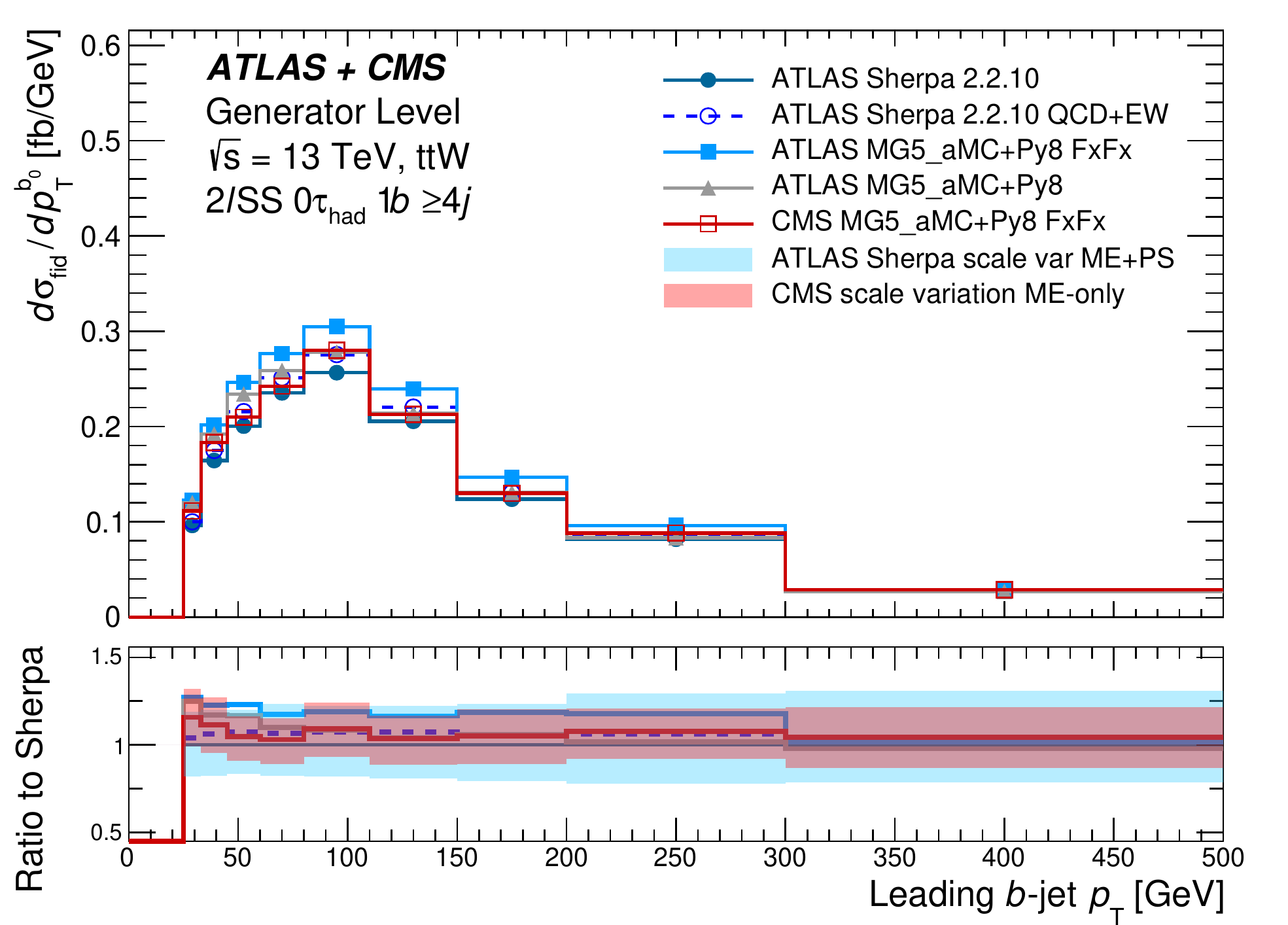}
\includegraphics[width=0.45\textwidth]{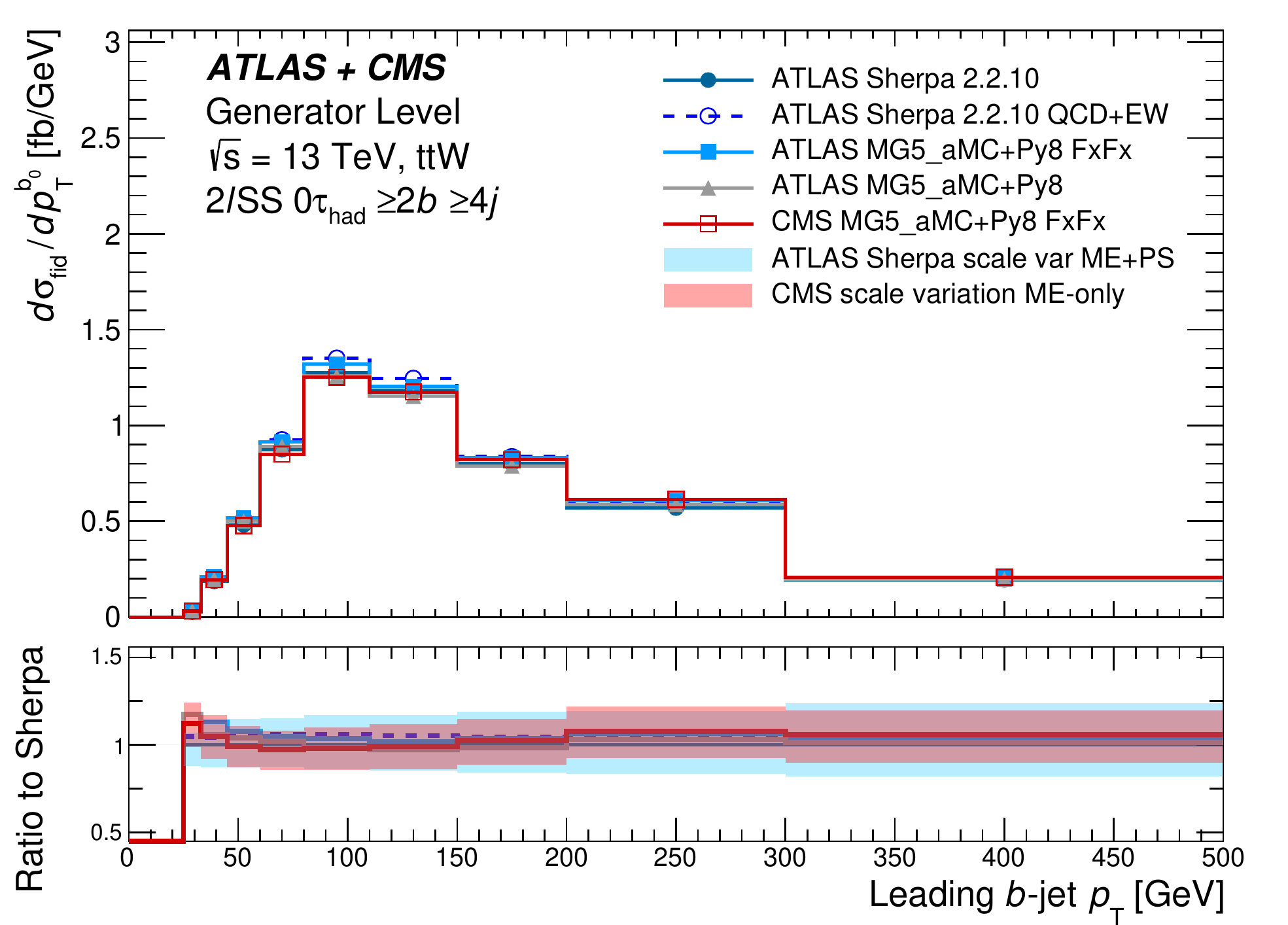}\\
  \caption{Distribution of the $b$-jet multiplicities (top) and the leading $b$-jet transverse momentum (bottom), for the Region 1 with $N_{b-\text{jets}}$=1 (left) and Region 2 with $N_{b-\text{jets}}\geq$2 (right) selection requiring four and more jets. All distributions are normalised to the YR4 cross section of \SI{600.8}{\femto\barn} except \sherpa 2.2.10 QCD+EW which is normalised to \SI{614.7}{\femto\barn}.  \label{ttW:den_4jbinfo}}
\end{figure}

\begin{figure}[!htb]
\centering
\includegraphics[width=0.45\textwidth]{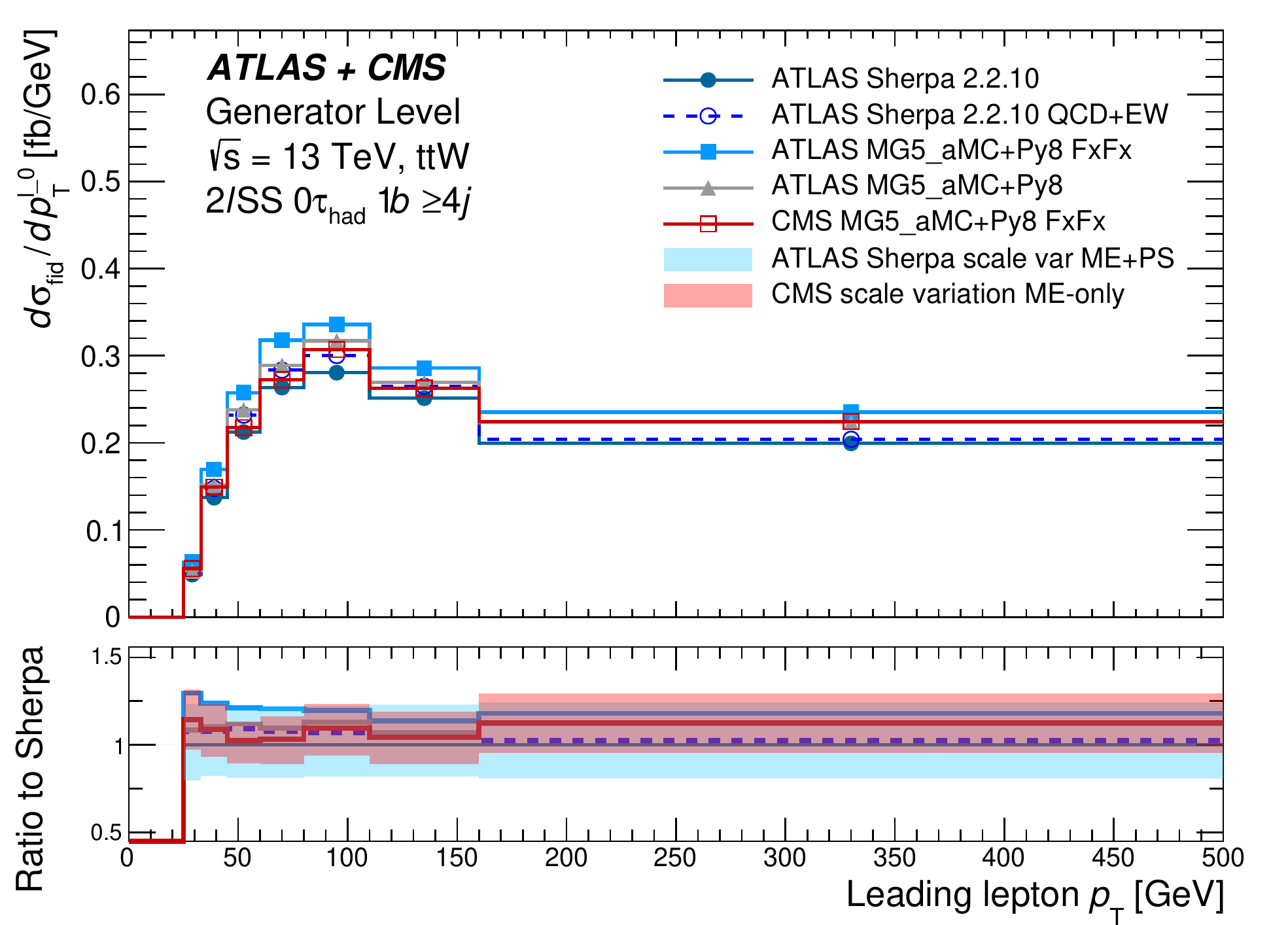}
\includegraphics[width=0.45\textwidth]{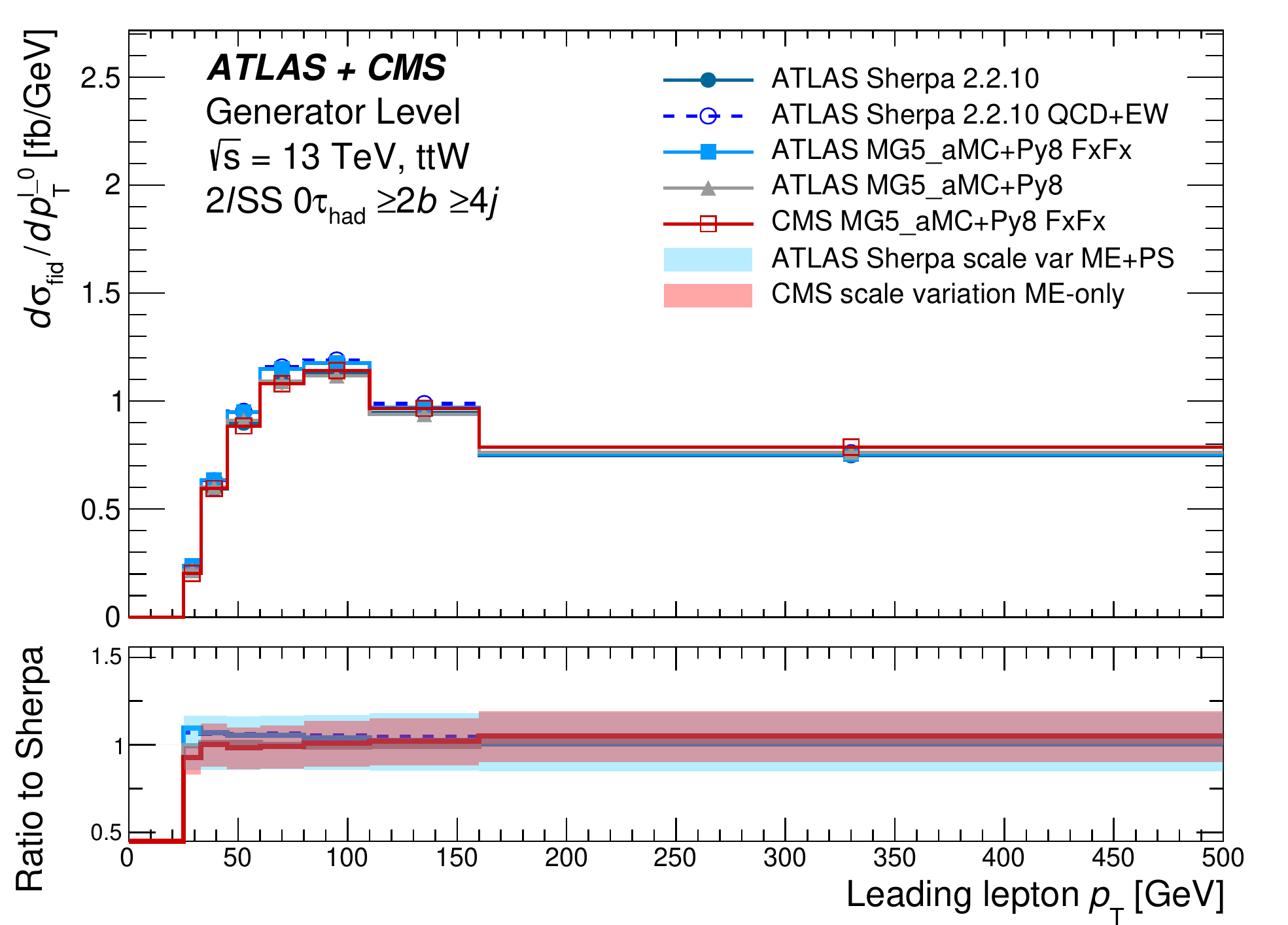}\\
\includegraphics[width=0.45\textwidth]{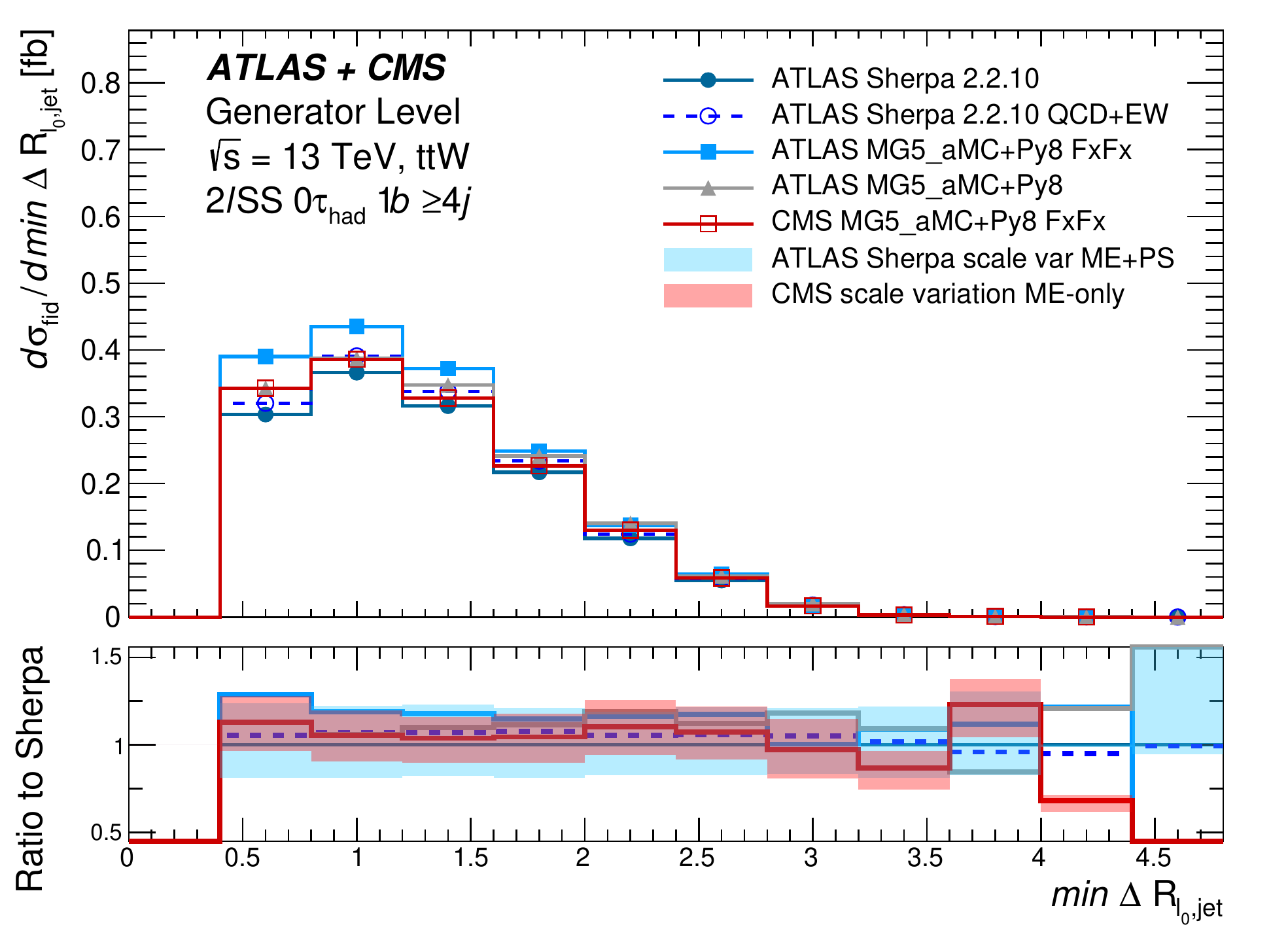}
\includegraphics[width=0.45\textwidth]{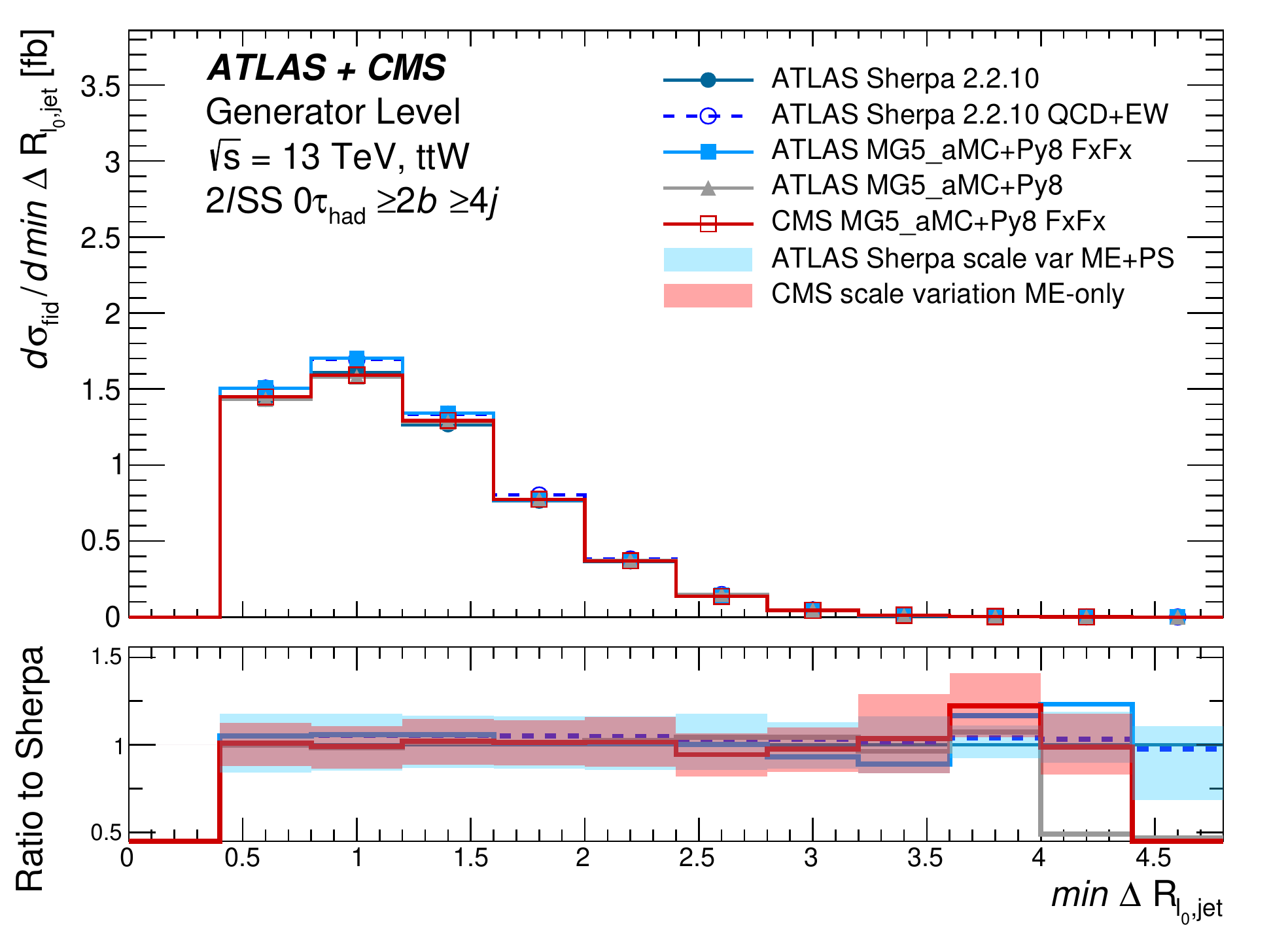}\\
  \caption{Distribution of the leading lepton transverse momentum (top) and the minimum angular separation between the leading lepton and the nearest jet (bottom), for the Region 1 with $N_{b-\text{jets}}$=1 (left) and Region 2 with $N_{b-\text{jets}}\geq$2 (right) selection requiring four and more jets. All distributions are normalised to the YR4 cross section of \SI{600.8}{\femto\barn} except \sherpa 2.2.10 QCD+EW which is normalised to \SI{614.7}{\femto\barn}.
  \label{ttW:den_lep_kin}}
\end{figure}

\begin{figure}[!htb]
\centering
\includegraphics[width=0.45\textwidth]{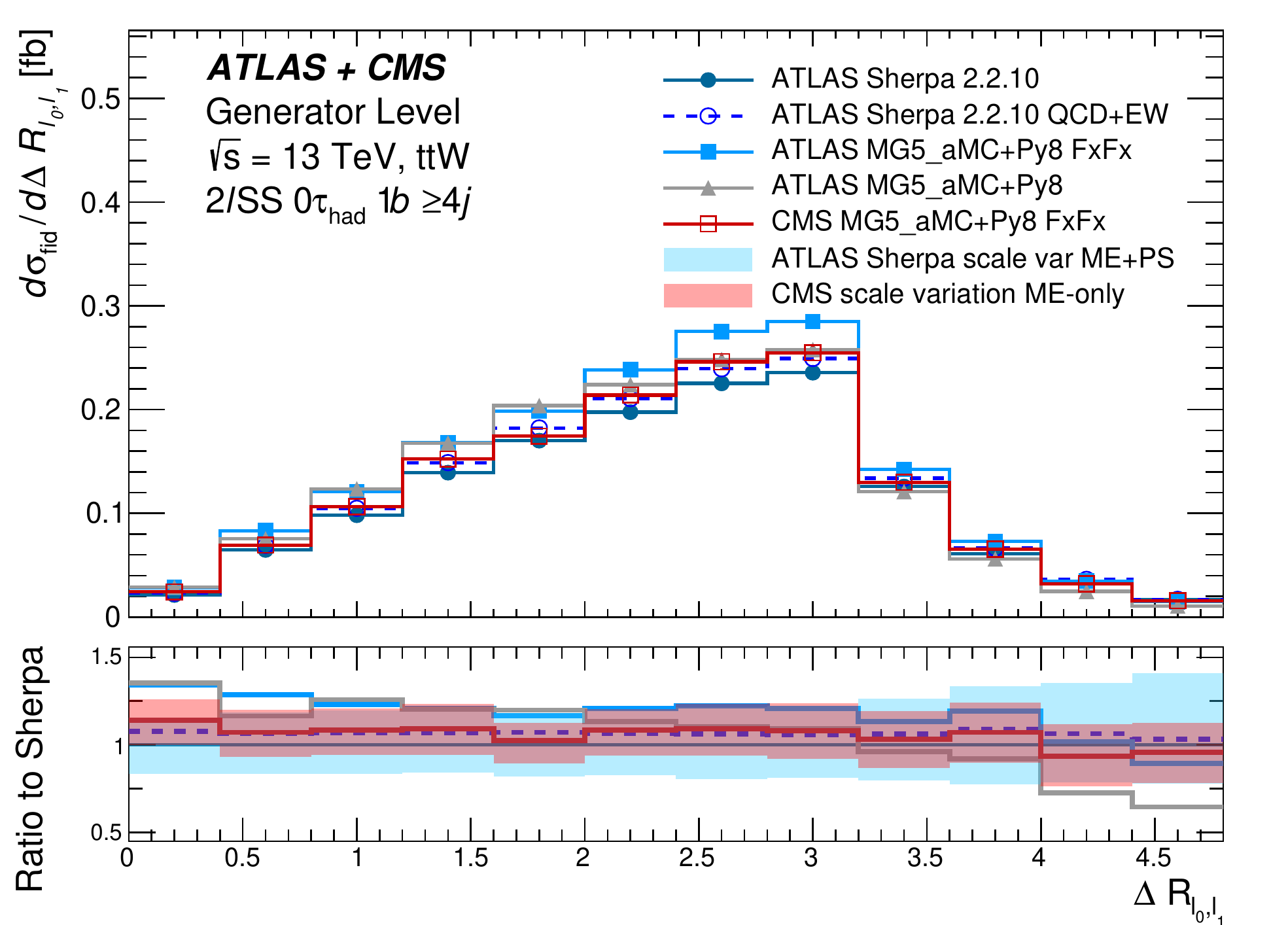}
\includegraphics[width=0.45\textwidth]{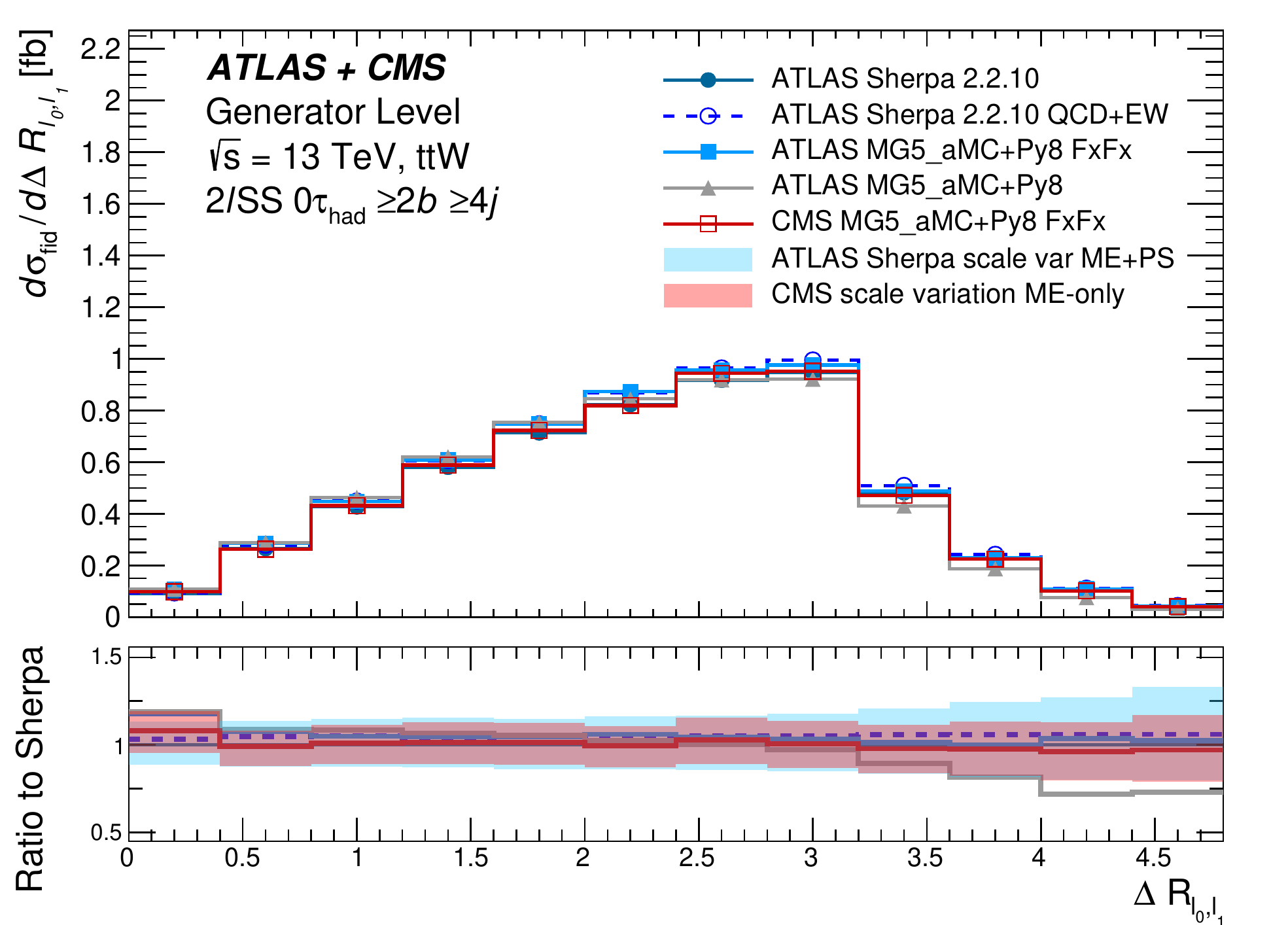}\\
\includegraphics[width=0.45\textwidth]{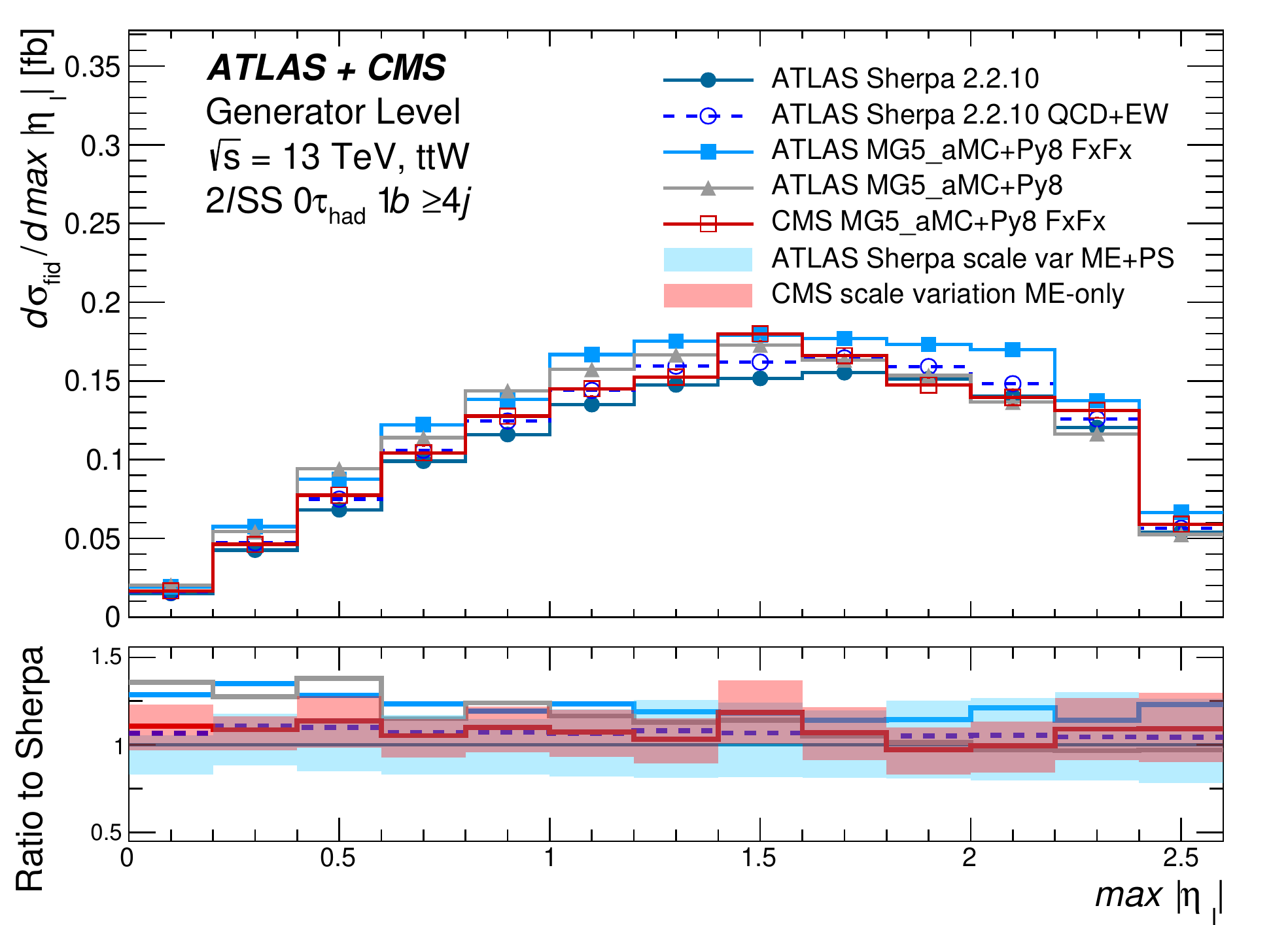}
\includegraphics[width=0.45\textwidth]{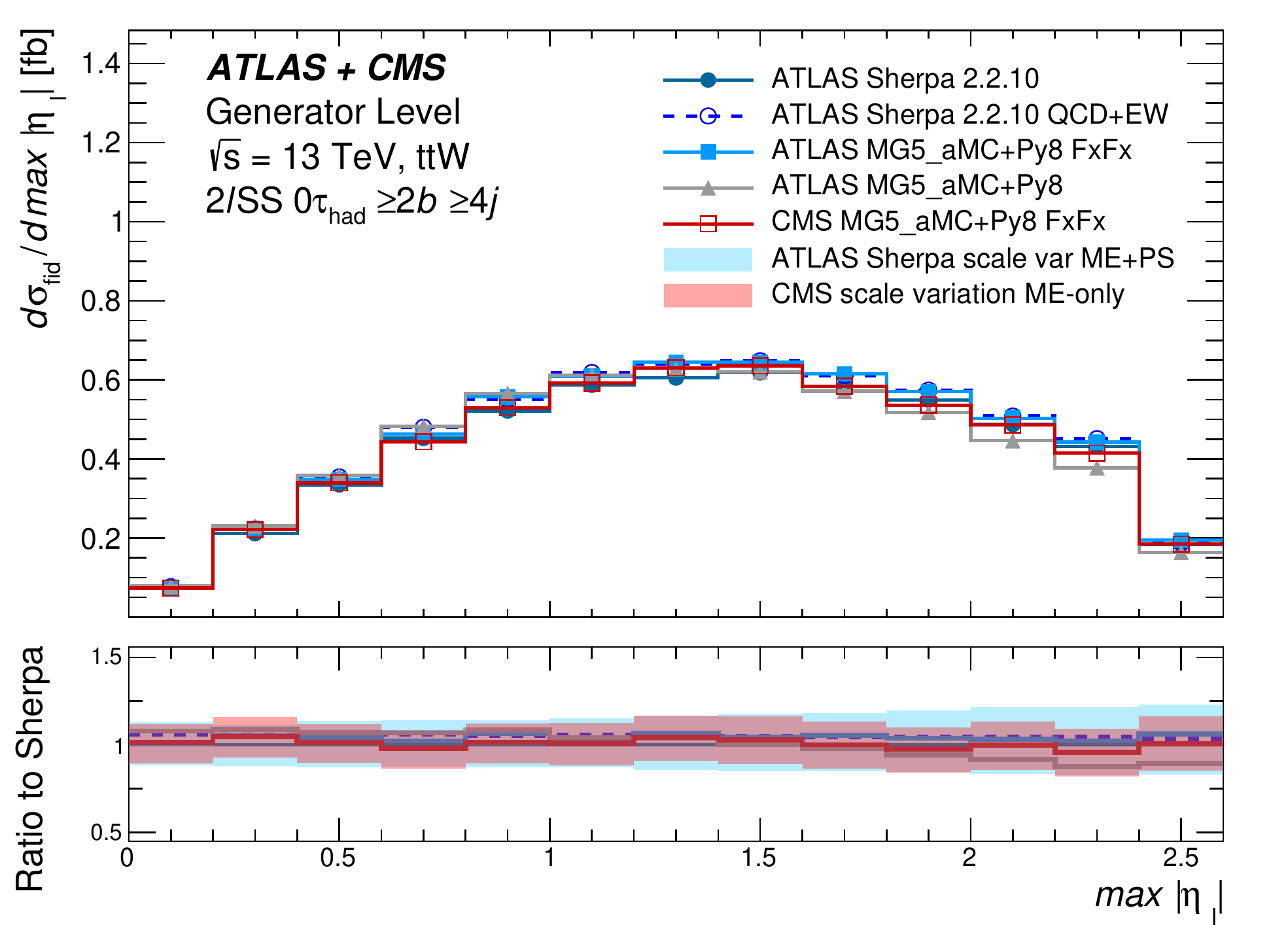}\\
  \caption{Distribution of the angular distance between the two leptons (top), maximum between lepton $|\eta_{\ell 0}|$ and $|\eta_{\ell 1}|$ (centre), 
   for the Region 1 with $N_{b-\text{jets}}$=1 (left) and Region 2 with $N_{b-\text{jets}}\geq$2 (right) selection requiring four and more jets. All distributions are normalised to the YR4 cross section of \SI{600.8}{\femto\barn}  except \sherpa 2.2.10 QCD+EW which is normalised to \SI{614.7}{\femto\barn}.
   \label{ttW:den_ll_kin}}
\end{figure}

\begin{figure}[!htb]
\centering
\includegraphics[width=0.45\textwidth]{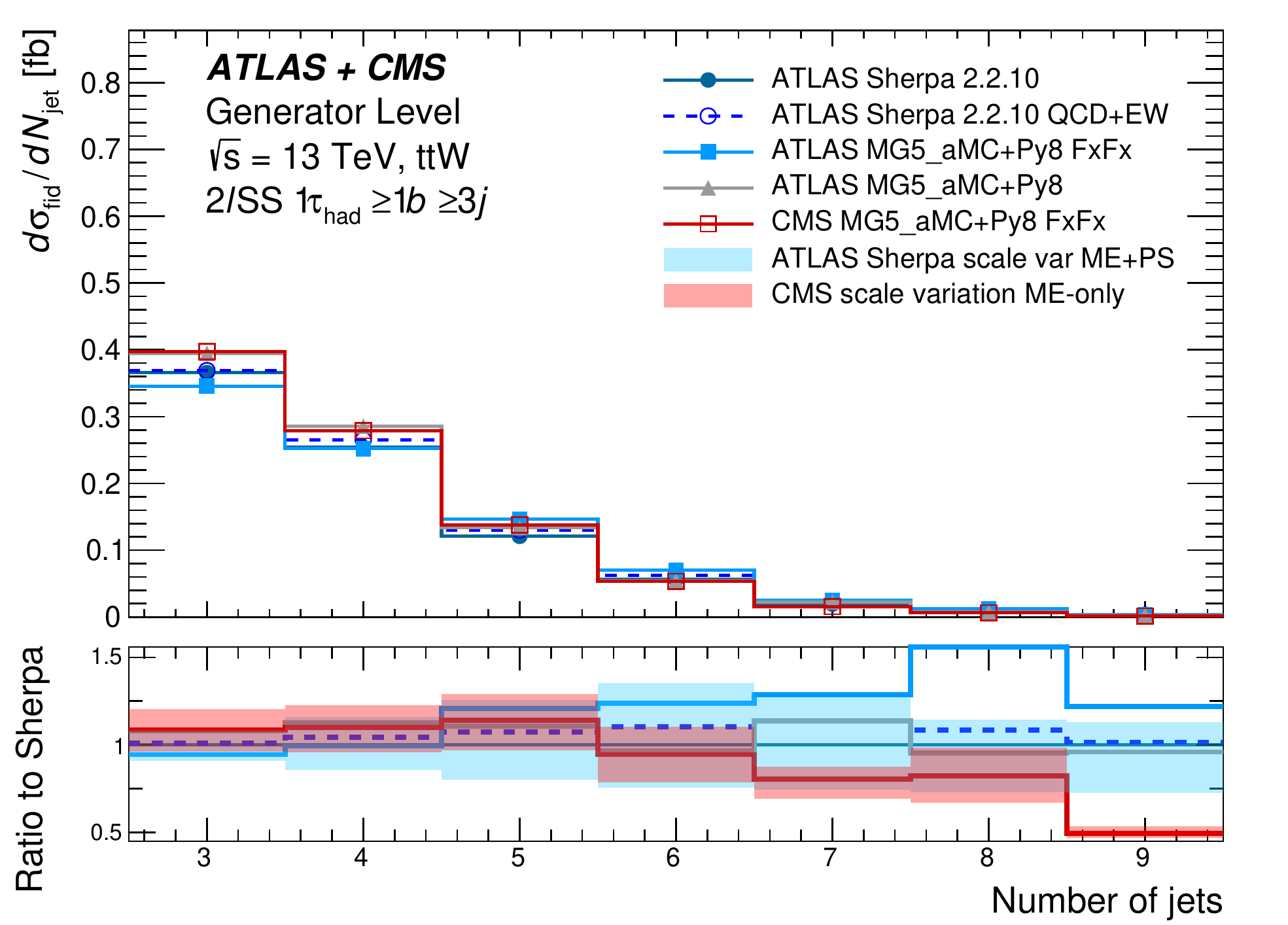}
\includegraphics[width=0.45\textwidth]{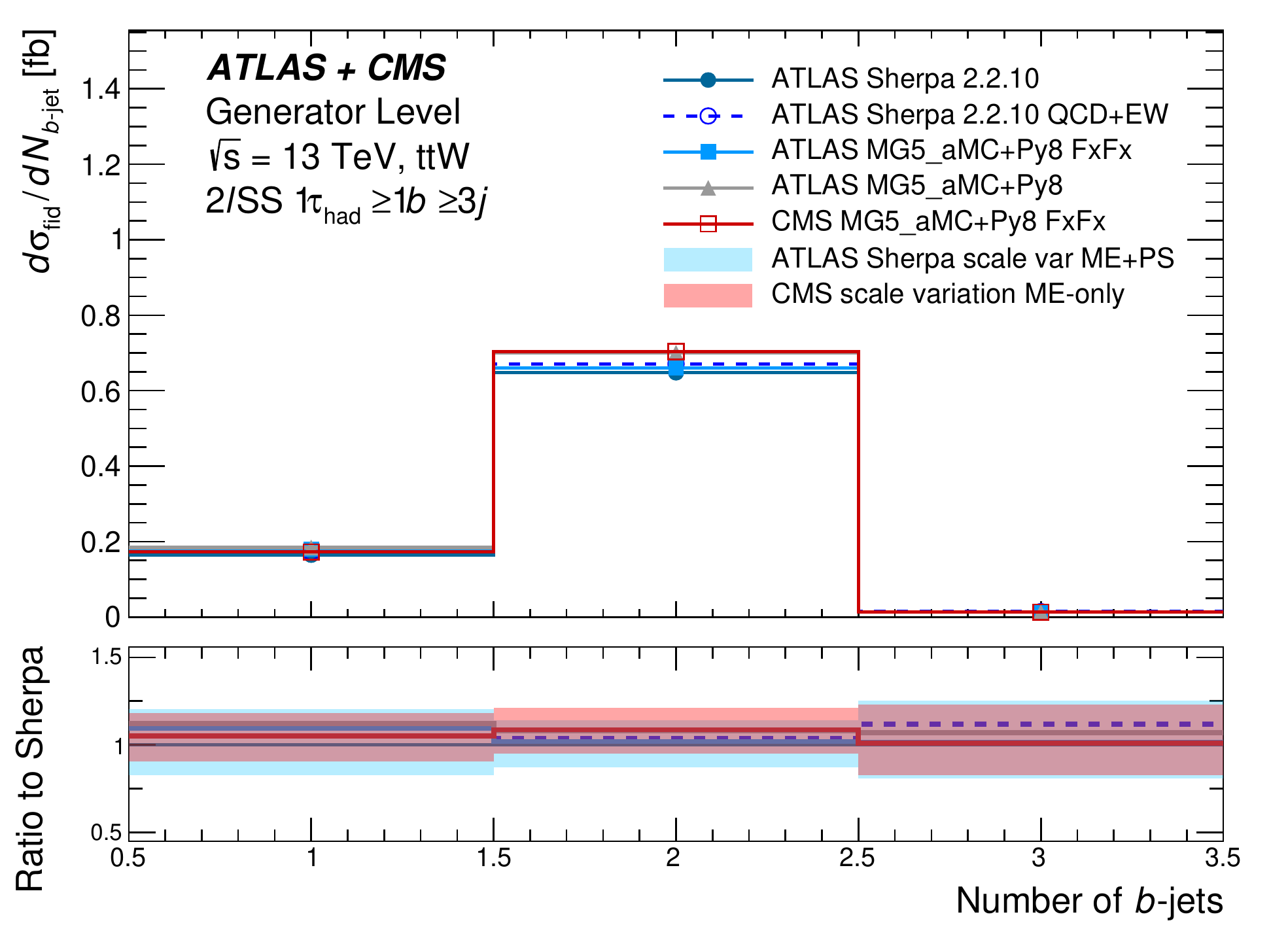}\\
\includegraphics[width=0.45\textwidth]{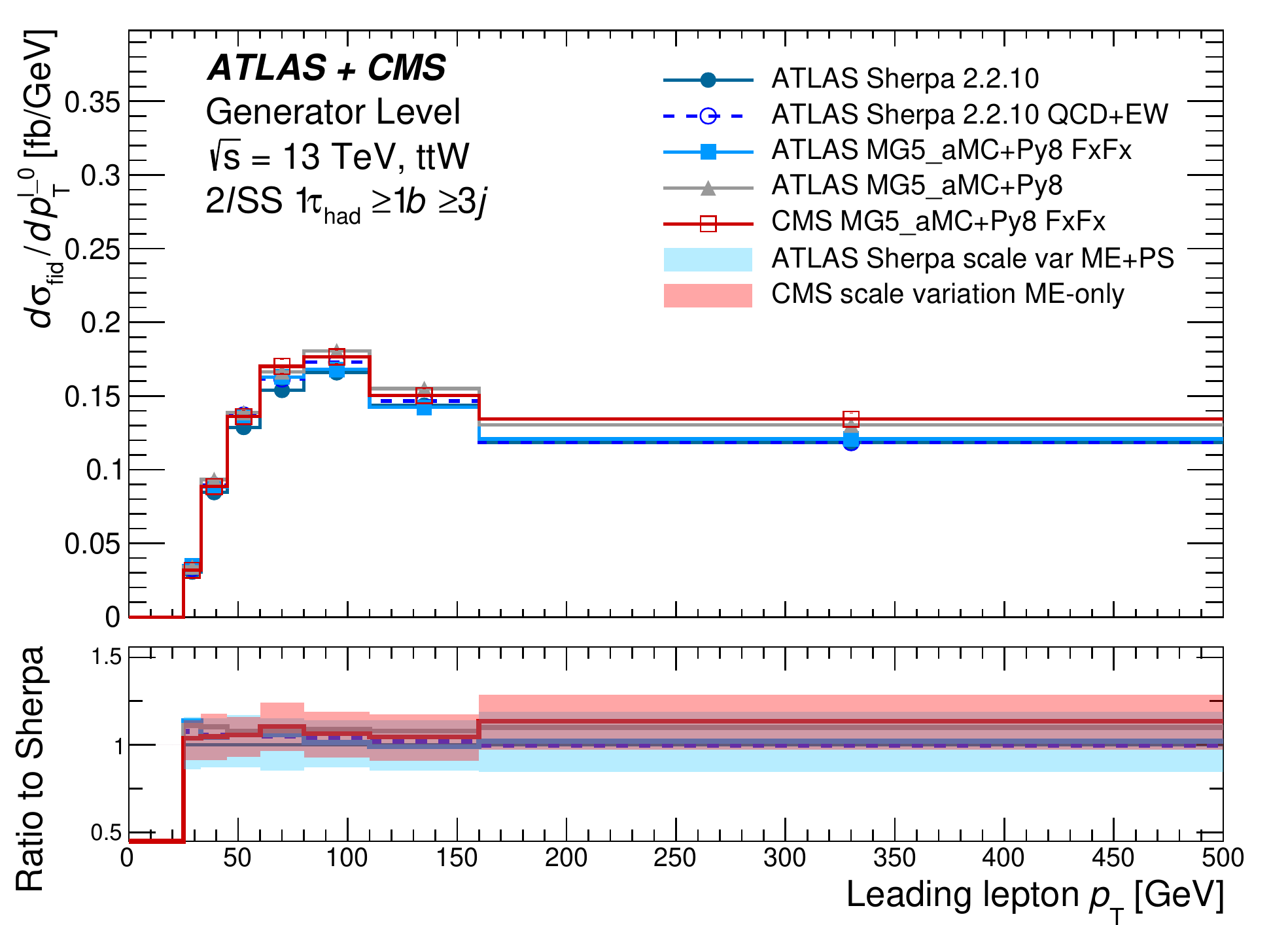}
\includegraphics[width=0.45\textwidth]{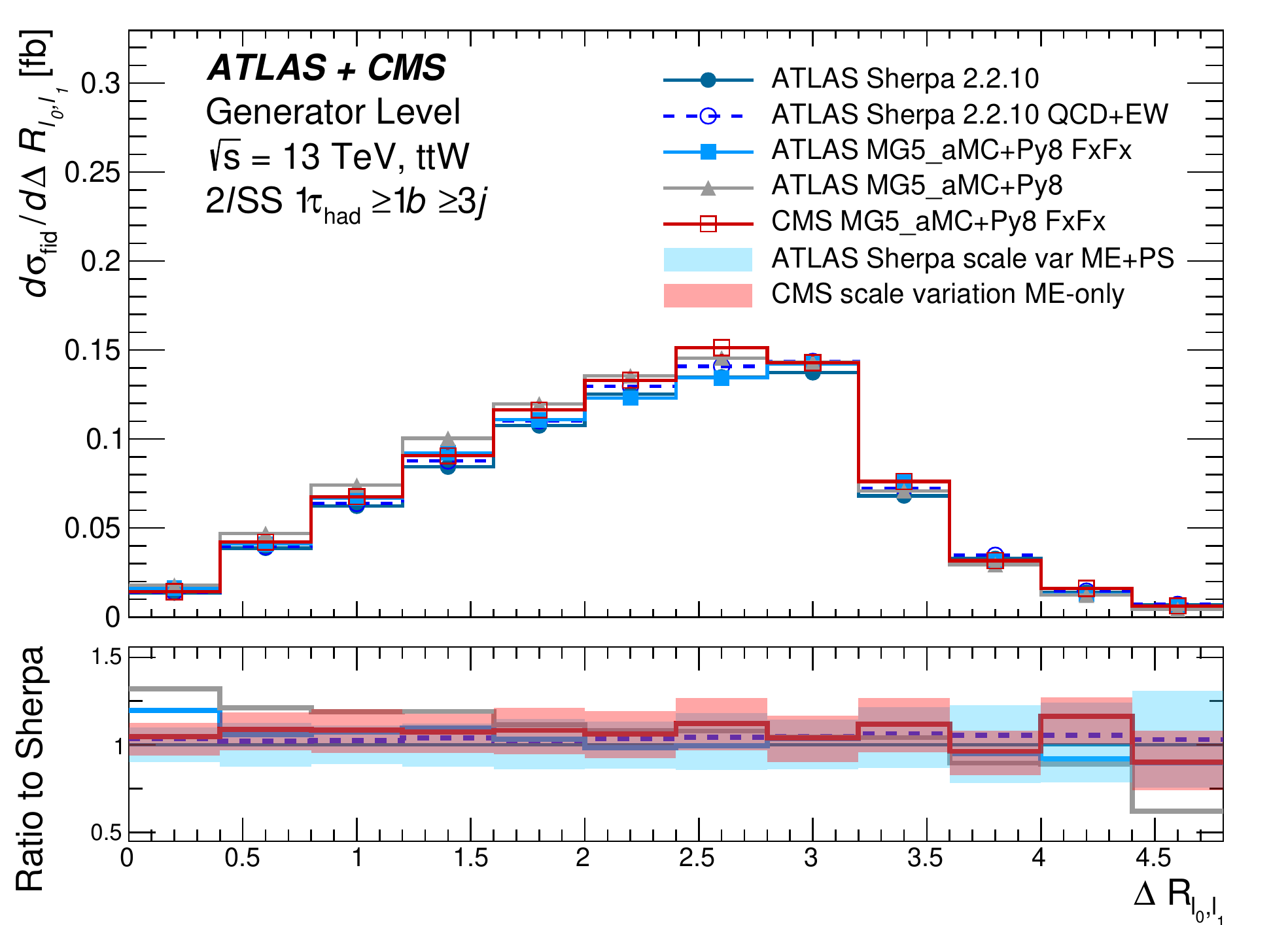}\\
  \caption{Distribution of the the jet multiplicity, number of $b$-jets, the leading lepton transverse momentum and the angular distance between the two leptons  $\Delta R _{\ell \ell }$ for the Region 5 with 1$\tau_{\text{had}}$ selection. All distributions are normalised to the YR4 cross section of \SI{600.8}{\femto\barn}  except \sherpa 2.2.10 QCD+EW which is normalised to \SI{614.7}{\femto\barn}.
   \label{fig:den_tauR_kin}}
\end{figure}
\include{FxFxPlots}

\subsection{Conclusions}\label{sec:ttW:conclusion}
The \ttW preditions of \sherpa and \amc with different settings have been compared  with respect to their inclusive ttW cross section predictions and their differential cross section predictions  in regions and observables relevant for the  measurement of \ttH in the multi-lepton final state.

For the inclusive \ttW cross section slightly different values are predicted \cite{Sherpa, Frederix:2021agh} for calculations with similar theoretical accuracy which is subject to ongoing  theoretical studies. 
Based on the studies presented in this note, additional studies and discussions in the LHC Higgs Working Group and the LHC Top Working Group~\cite{meeting}, ATLAS and CMS agreed to use the inclusive \ttW cross section of $722 ^{+70}_{-78}$  (scale) $\pm7$ (PDF)\,fb \cite{Frederix:2021agh} as a reference inclusive cross section to allow direct comparisons between experiments.

The normalised distributions sensitive to shape differences have  very small scale uncertainties, below 10\,\% in most of the phase space, while these scale uncertainties are significant when the acceptance effects are included, i.e.\ the distributions are normalised to the \ttW cross section. The inclusion of tree-level EW effects only causes minor shape effects but can lead to up to 20\,\% difference in the cross section at high jet multiplicity. As expected,  including the FxFx algorithm into the \amc prediction leads to significant effects in all regions, especially at low \HT.
Significant differences between the \amc FxFx predictions of ATLAS and
CMS are observed, especially in the jet multiplicity. Further studies
are required to investigate the origin of these differences. Given
that both setups consider the same perturbative accuracy such
differences could be attributed to the choice of merging scale value or Pythia8 tune, so this
could be an area of future study.


For many observables the shape differences between the various model
predictions are within the scale uncertainties of each
prediction. Observables relating to jet activity such as the jet
multiplicity and \HT are notable exceptions to this. This is
especially the case for Region 3 where the differences in shape
between predictions for \HT is particlularly large. This region is
important to constrain the interplay between ttW background and
backgrounds arising from ttbar production where at least one lepton is
mis-identified. It represents a phase space where one of the jets in
the ttW decay is not reconstructed or is out of acceptance and is not
expected to be as sensitive to the additional jet modelling as Regions
1 and 2. Therefore, it is somewhat surprising that such large
differences are observed between predictions. This should be
investigated in future studies.

The inclusion of EW corrections shows only a small shape and
normalisation effects for most observables. One place where a notable
effect on the shape of a distribution can be observed is for the jet
multiplicity, however the effect is small enough to be covered by the QCD
scale variations. Future studies could specifically target the
sub-leading EW contribution with cuts related to the rapidity
difference between jets which has been shown~\cite{Dror:2015nkp} to be
different with respect to the central ttW QCD process.

These distributions shall be used as a starting point to derive a
strategy for the theory uncertainty estimates for a combination of the
expected measurement results based on the full Run-2 data set. 
Beyond what has been shown in the comparisons included in this document, this 
strategy is expected to take into consideration the latest developments on the
theoretical models. For example, the NLO+PS calculations provided
in \powheg~\cite{FebresCordero:2021kcc} can act as systematic
variation with respect to the \amc and \sherpa calculations for more
inclusive phase-spaces. They can also be used to understand parton
shower, hadronisation and underlying event effects through interfaces
to \pythia and \herwig. In addition, recent off-shell
calculations~\cite{Bevilacqua:2020pzy,Denner:2020hgg,Bevilacqua:2020srb,Denner:2021hqi}
and in particular the single-resonant contributions could be of 
importance. In the absence of explicit parton shower-matched
calculations, corrections can be applied through the procedure
outlined in Ref.~\cite{Bevilacqua:2021tzp}. It would also be important
to extend existing calculations to additional final states, such as
2$\ell$SS. 
 Finally, given the current discrepancy between ATLAS and CMS,  the strategy 
must address how different model predictions are considered  in addition to the scale uncertainties as part of the theoretical 
uncertainties on the measurement.

\clearpage
\bibliography{bib/ttbb,bib/ttw}{}

\begin{thebibliography}{10}
\bibitem{rivetroutine}
The routine will be released as MC-ttbb routine in the Rivet analysis toolkit
  \cite{rivet}.
\end{thebibliography}
\bibliographystyle{bib/lucas_unsrt}


\end{document}